\documentclass[longauth]{aa}

\usepackage{graphicx}
\usepackage{natbib}
\usepackage{scalerel}
\usepackage{rotating}

\usepackage[table]{xcolor}

\usepackage{txfonts}
\usepackage[pdfencoding=auto,psdextra]{hyperref}
\hypersetup{
    colorlinks=true,
    linkcolor=blue,
    filecolor=magenta,      
    urlcolor=blue,
    citecolor=blue
}
\makeatletter
\renewcommand*\aa@pageof{, page \thepage{} of \pageref*{LastPage}}
\makeatother

\usepackage[utf8]{inputenc}

\usepackage[switch, modulo]{lineno}
              
\renewcommand{\linenumbers}[0]{}

\usepackage{euclid}

\usepackage[nameinlink,capitalise]{cleveref}
\crefname{section}{Sect.}{Sects.}
\Crefname{section}{Section}{Sections}
\crefname{figure}{Fig.}{Figs.}
\Crefname{figure}{Figure}{Figures}
\crefname{equation}{Eq.}{Eqs.}
\Crefname{equation}{Equation}{Equations}
\crefname{table}{Table}{Tables}
\crefname{appendix}{Appendix}{Appendices}

\usepackage[nopostdot, nonumberlist, acronym]{glossaries}
\glsdisablehyper

\usepackage{orcidlink} 
\newcommand{\orcid}[1]{\orcidlink{#1}}
	
\usepackage[nolist,nohyperlinks]{acronym}

\newcommand*{\ectile} {\texttt{ECTile}\xspace}

\newcommand*{\esspt}
{\texttt{ESSPT}\xspace}

\newcommand*{\essot}
{\texttt{ESSOT}\xspace}

\begin{document}

\acrodef{AA}{azimuth or $\alpha$ angle}
\acrodef{APE}{absolute pointing error}
\acrodef{BGS}{blue grism}
\acrodef{CalF} {Calibration Framework}
\acrodef{CCB}{configuration control board}
\acrodef{CCD}{charge-coupled device}
\acrodef{CDFS}{Chandra Deep Field South}
\acrodef{CPC}{completeness and purity calibration}
\acrodef{DES}{Dark Energy Survey}
\acrodef{DR1}{first data release}
\acrodef{DTCP}{daily telemetry communication period}
\acrodef{EAF}{Euclid Auxiliary Field}
\acrodef{EBF}{Euclid Blank Field}
\acrodef{EDF}{Euclid Deep Field}
\acrodef{EDF-F}{Euclid Deep Field Fornax}
\acrodef{EDF-N}{Euclid Deep Field North}
\acrodef{EDF-S}{Euclid Deep Field South}
\acrodef{EDS}{Euclid Deep Survey}
\acrodef{EES}{Euclid Ecliptic Survey}
\acrodef{EGBS}{Euclid Galactic Bulge Survey}
\acrodef{EMDS}{Euclid Medium-Deep Survey}
\acrodef{ERO}{Early Release Observations}
\acrodef{ESA}{European Space Agency}
\acrodef{ESS}{Euclid survey system}
\acrodef{EST}{Euclid Science Team}
\acrodef{EUDF}{Euclid Ultra-Deep Field}
\acrodef{EWS}{Euclid Wide Survey}
\acrodef{FoM}{figure of merit}
\acrodef{FoV}{field of view}
\acrodef{FPA}{focal-plane array}
\acrodef{FWA}{filter-wheel assembly}
\acrodef{GWA}{grism-wheel assembly}
\acrodef{HSC}{Subaru Hyper Suprime Camera}
\acrodef{ICR}{Instrument Commanding Request}
\acrodef{ILS}{Independent Legacy Scientist}
\acrodef{IOT}{Instrument Operation Team}
\acrodef{KiDS}{Kilo-Degree Survey}
\acrodef{LMC}{Large Magellanic Cloud}
\acrodef{MIB}{Mission Information Base}
\acrodef{MOC}{Mission Operations Centre}
\acrodef{NEP}{North ecliptic pole}
\acrodef{NISP}{Near-Infrared Spectrometer and Photometer}
\acrodef{OSS}{operational sky survey}
\acrodef{OU}{Organisation Unit}
\acrodef{PC}{planning cycle}
\acrodef{PDC}{phase-diversity calibration}
\acrodef{PN}{planetary nebula}
\acrodef{PHZ}{PHotometric redshift}
\acrodef{PLM}{payload module}
\acrodef{PSF}{point-spread function}
\acrodef{PV}{performance-verification}
\acrodef{QLA}{quick look analysis}
\acrodef{RGS}{red grism}
\acrodef{RIAR}{RSD ingestion and analysis report}
\acrodef{RoI}{region of interest}
\acrodef{ROS}{reference observing sequence}
\acrodef{RSD}{reference survey definition}
\acrodef{SAA}{Solar aspect angle}
\acrodef{SCG}{Science Coordination Group}
\acrodef{SCS}{SOC Commanding System}
\acrodef{SGS}{Science Ground Segment}
\acrodef{SIR}{NISP Spectroscopy}
\acrodef{SMC}{Small Magellanic Cloud}
\acrodef{SOC}{Science Operations Centre}
\acrodef{SOST}{survey operations support team}
\acrodef{SPE}{NISP SPEctra analysis}
\acrodef{SSO}{Solar System object}
\acrodef{SSR}{Survey Schedule Request}
\acrodef{UNIONS}{Ultraviolet Near-Infrared Optical Northern Survey}
\acrodef{VIS}{Visible instrument}  
\acrodef{XML}{Extensible Markup Language}

%



   
%
%
\title{Euclid Data Release 1 (DR1) } 
 \subtitle{The \Euclid surveys at DR1}
%

%


\renewcommand{\orcid}[1]{} 
\author{Euclid Collaboration: I.~Tereno\orcid{0000-0002-4537-6218}\thanks{\email{iatereno@ciencias.ulisboa.pt}}\inst{\ref{aff1},\ref{aff2}}
\and J.~Dinis\orcid{0000-0001-5075-1601}\inst{\ref{aff1},\ref{aff2}}
\and X.~Dupac\inst{\ref{aff3}}
\and R.~Scaramella\orcid{0000-0003-2229-193X}\inst{\ref{aff4}}
\and B.~Altieri\orcid{0000-0003-3936-0284}\inst{\ref{aff3}}
\and H.~Aussel\orcid{0000-0002-1371-5705}\inst{\ref{aff5}}
\and I.~Baldry\inst{\ref{aff6}}
\and P.~Battaglia\orcid{0000-0002-7337-5909}\inst{\ref{aff7}}
\and G.~Buenadicha\inst{\ref{aff3}}
\and C.~Burigana\orcid{0000-0002-3005-5796}\inst{\ref{aff8},\ref{aff9}}
\and B.~Carry\orcid{0000-0001-5242-3089}\inst{\ref{aff10}}
\and P.~Casenove\orcid{0009-0006-6736-1670}\inst{\ref{aff11}}
\and J.-C.~Cuillandre\orcid{0000-0002-3263-8645}\inst{\ref{aff5}}
\and C.~Dominguez-Tagle\orcid{0000-0002-0463-6212}\inst{\ref{aff12},\ref{aff13}}
\and A.~Franco\orcid{0000-0002-4761-366X}\inst{\ref{aff14},\ref{aff15},\ref{aff16}}
\and B.~R.~Granett\orcid{0000-0003-2694-9284}\inst{\ref{aff17}}
\and M.~Granvik\orcid{0000-0002-5624-1888}\inst{\ref{aff18},\ref{aff19}}
\and A.~Gregorio\orcid{0000-0003-4028-8785}\inst{\ref{aff20},\ref{aff21},\ref{aff22}}
\and B.~Y.~Irureta-Goyena\orcid{0009-0004-5327-8767}\inst{\ref{aff23}}
\and R.~Kohley\inst{\ref{aff3}}
\and K.~Kuijken\orcid{0000-0002-3827-0175}\inst{\ref{aff24}}
\and P.~Liebing\orcid{0009-0005-8043-7150}\inst{\ref{aff25}}
\and E.~L.~Mart\'in\orcid{0000-0002-1208-4833}\inst{\ref{aff12},\ref{aff13}}
\and J.~Mart\'{i}n-Fleitas\orcid{0000-0002-8594-569X}\inst{\ref{aff26}}
\and J.~Mendes\inst{\ref{aff27}}
\and L.~Miller\orcid{0000-0002-3376-6200}\inst{\ref{aff28}}
\and A.~A.~Nucita\inst{\ref{aff14},\ref{aff15},\ref{aff16}}
\and R.~Nakajima\orcid{0009-0009-1213-7040}\inst{\ref{aff29}}
\and K.~Paterson\orcid{0000-0001-8340-3486}\inst{\ref{aff30}}
\and V.~Pettorino\orcid{0000-0002-4203-9320}\inst{\ref{aff31}}
\and M.~P\"ontinen\orcid{0000-0001-5442-2530}\inst{\ref{aff18}}
\and G.~D.~Racca\orcid{0000-0002-9883-8981}\inst{\ref{aff24},\ref{aff31}}
\and E.~Romelli\orcid{0000-0003-3069-9222}\inst{\ref{aff21}}
\and M.~Schirmer\orcid{0000-0003-2568-9994}\inst{\ref{aff30}}
\and M.~Schmidt\inst{\ref{aff27}}
\and T.~Schrabback\orcid{0000-0002-6987-7834}\inst{\ref{aff32}}
\and D.~Scott\orcid{0000-0002-6878-9840}\inst{\ref{aff33}}
\and C.~Sirignano\orcid{0000-0002-0995-7146}\inst{\ref{aff34},\ref{aff35}}
\and R.~Tata\inst{\ref{aff36}}
\and T.~Trombetti\orcid{0000-0001-5166-2467}\inst{\ref{aff8}}
\and A.~Zacchei\orcid{0000-0003-0396-1192}\inst{\ref{aff21},\ref{aff37}}
\and M.~{{\v{Z}erjal}}\orcid{0000-0001-6023-4974}\inst{\ref{aff12},\ref{aff13}}
\and S.~Andreon\orcid{0000-0002-2041-8784}\inst{\ref{aff17}}
\and N.~Auricchio\orcid{0000-0003-4444-8651}\inst{\ref{aff7}}
\and S.~Awan\inst{\ref{aff25}}
\and C.~Baccigalupi\orcid{0000-0002-8211-1630}\inst{\ref{aff37},\ref{aff21},\ref{aff22},\ref{aff38}}
\and M.~Baldi\orcid{0000-0003-4145-1943}\inst{\ref{aff39},\ref{aff7},\ref{aff40}}
\and A.~Balestra\orcid{0000-0002-6967-261X}\inst{\ref{aff41}}
\and S.~Bardelli\orcid{0000-0002-8900-0298}\inst{\ref{aff7}}
\and A.~N.~Belikov\inst{\ref{aff42},\ref{aff43}}
\and R.~Bender\orcid{0000-0001-7179-0626}\inst{\ref{aff44},\ref{aff45}}
\and A.~Biviano\orcid{0000-0002-0857-0732}\inst{\ref{aff21},\ref{aff37}}
\and R.~P.~Blake\inst{\ref{aff46}}
\and M.~Bolzonella\orcid{0000-0003-3278-4607}\inst{\ref{aff7}}
\and A.~Bonchi\orcid{0000-0002-2667-5482}\inst{\ref{aff47}}
\and E.~Branchini\orcid{0000-0002-0808-6908}\inst{\ref{aff48},\ref{aff49},\ref{aff17}}
\and M.~Brescia\orcid{0000-0001-9506-5680}\inst{\ref{aff50},\ref{aff51}}
\and J.~Brinchmann\orcid{0000-0003-4359-8797}\inst{\ref{aff52},\ref{aff53},\ref{aff54}}
\and S.~Camera\orcid{0000-0003-3399-3574}\inst{\ref{aff55},\ref{aff56},\ref{aff57}}
\and G.~Ca\~nas-Herrera\orcid{0000-0003-2796-2149}\inst{\ref{aff24}}
\and V.~Capobianco\orcid{0000-0002-3309-7692}\inst{\ref{aff57}}
\and C.~Carbone\orcid{0000-0003-0125-3563}\inst{\ref{aff58}}
\and V.~F.~Cardone\inst{\ref{aff4},\ref{aff59}}
\and J.~Carretero\orcid{0000-0002-3130-0204}\inst{\ref{aff60},\ref{aff61}}
\and R.~Casas\orcid{0000-0002-8165-5601}\inst{\ref{aff62},\ref{aff63}}
\and F.~J.~Castander\orcid{0000-0001-7316-4573}\inst{\ref{aff63},\ref{aff62}}
\and M.~Castellano\orcid{0000-0001-9875-8263}\inst{\ref{aff4}}
\and G.~Castignani\orcid{0000-0001-6831-0687}\inst{\ref{aff7}}
\and S.~Cavuoti\orcid{0000-0002-3787-4196}\inst{\ref{aff51},\ref{aff64}}
\and P.-Y.~Chabaud\inst{\ref{aff65}}
\and K.~C.~Chambers\orcid{0000-0001-6965-7789}\inst{\ref{aff66}}
\and A.~Cimatti\inst{\ref{aff67}}
\and C.~Colodro-Conde\inst{\ref{aff12}}
\and G.~Congedo\orcid{0000-0003-2508-0046}\inst{\ref{aff46}}
\and C.~J.~Conselice\orcid{0000-0003-1949-7638}\inst{\ref{aff68}}
\and L.~Conversi\orcid{0000-0002-6710-8476}\inst{\ref{aff69},\ref{aff3}}
\and Y.~Copin\orcid{0000-0002-5317-7518}\inst{\ref{aff70}}
\and A.~Costille\inst{\ref{aff65}}
\and F.~Courbin\orcid{0000-0003-0758-6510}\inst{\ref{aff71},\ref{aff72},\ref{aff73}}
\and H.~M.~Courtois\orcid{0000-0003-0509-1776}\inst{\ref{aff74}}
\and M.~Cropper\orcid{0000-0003-4571-9468}\inst{\ref{aff25}}
\and A.~Da~Silva\orcid{0000-0002-6385-1609}\inst{\ref{aff1},\ref{aff2}}
\and H.~Degaudenzi\orcid{0000-0002-5887-6799}\inst{\ref{aff75}}
\and S.~de~la~Torre\inst{\ref{aff65}}
\and G.~De~Lucia\orcid{0000-0002-6220-9104}\inst{\ref{aff21}}
\and H.~Dole\orcid{0000-0002-9767-3839}\inst{\ref{aff76},\ref{aff77}}
\and F.~Dubath\orcid{0000-0002-6533-2810}\inst{\ref{aff75}}
\and S.~Escoffier\orcid{0000-0002-2847-7498}\inst{\ref{aff78}}
\and M.~Farina\orcid{0000-0002-3089-7846}\inst{\ref{aff79}}
\and R.~Farinelli\inst{\ref{aff7}}
\and S.~Farrens\orcid{0000-0002-9594-9387}\inst{\ref{aff5}}
\and F.~Faustini\orcid{0000-0001-6274-5145}\inst{\ref{aff4}}
\and S.~Ferriol\inst{\ref{aff70}}
\and P.~Fosalba\orcid{0000-0002-1510-5214}\inst{\ref{aff62},\ref{aff63}}
\and S.~Fotopoulou\orcid{0000-0002-9686-254X}\inst{\ref{aff80}}
\and N.~Fourmanoit\orcid{0009-0005-6816-6925}\inst{\ref{aff78}}
\and M.~Frailis\orcid{0000-0002-7400-2135}\inst{\ref{aff21}}
\and E.~Franceschi\orcid{0000-0002-0585-6591}\inst{\ref{aff7}}
\and M.~Fumana\orcid{0000-0001-6787-5950}\inst{\ref{aff58}}
\and L.~Gabarra\orcid{0000-0002-8486-8856}\inst{\ref{aff28}}
\and S.~Galeotta\orcid{0000-0002-3748-5115}\inst{\ref{aff21}}
\and K.~George\orcid{0000-0002-1734-8455}\inst{\ref{aff81}}
\and W.~Gillard\orcid{0000-0003-4744-9748}\inst{\ref{aff78}}
\and C.~Giocoli\orcid{0000-0002-9590-7961}\inst{\ref{aff7},\ref{aff40}}
\and J.~Gracia-Carpio\orcid{0000-0003-4689-3134}\inst{\ref{aff44}}
\and A.~Grazian\orcid{0000-0002-5688-0663}\inst{\ref{aff41}}
\and F.~Grupp\inst{\ref{aff44},\ref{aff45}}
\and L.~Guzzo\orcid{0000-0001-8264-5192}\inst{\ref{aff82},\ref{aff17},\ref{aff83}}
\and W.~G.~Hartley\orcid{0000-0002-4563-5101}\inst{\ref{aff75}}
\and S.~V.~H.~Haugan\orcid{0000-0001-9648-7260}\inst{\ref{aff84}}
\and G.~Helou\orcid{0000-0003-3367-3415}\inst{\ref{aff85},\ref{aff86}}
\and S.~Hemmati\orcid{0000-0003-2226-5395}\inst{\ref{aff86}}
\and H.~Hoekstra\orcid{0000-0002-0641-3231}\inst{\ref{aff24}}
\and M.~S.~Holliman\orcid{0009-0006-9850-9770}\inst{\ref{aff46}}
\and W.~Holmes\orcid{0009-0007-8554-4646}\inst{\ref{aff87}}
\and I.~M.~Hook\orcid{0000-0002-2960-978X}\inst{\ref{aff88}}
\and F.~Hormuth\inst{\ref{aff89}}
\and A.~Hornstrup\orcid{0000-0002-3363-0936}\inst{\ref{aff90},\ref{aff91}}
\and K.~Jahnke\orcid{0000-0003-3804-2137}\inst{\ref{aff30}}
\and M.~Jhabvala\inst{\ref{aff92}}
\and S.~Kermiche\orcid{0000-0002-0302-5735}\inst{\ref{aff78}}
\and A.~Kiessling\orcid{0000-0002-2590-1273}\inst{\ref{aff87}}
\and B.~Kubik\orcid{0009-0006-5823-4880}\inst{\ref{aff70}}
\and M.~K\"ummel\orcid{0000-0003-2791-2117}\inst{\ref{aff45}}
\and M.~Kunz\orcid{0000-0002-3052-7394}\inst{\ref{aff93}}
\and H.~Kurki-Suonio\orcid{0000-0002-4618-3063}\inst{\ref{aff18},\ref{aff94}}
\and A.~M.~C.~Le~Brun\orcid{0000-0002-0936-4594}\inst{\ref{aff95}}
\and V.~Le~Brun\orcid{0000-0002-5027-1939}\inst{\ref{aff65}}
\and D.~Le~Mignant\orcid{0000-0002-5339-5515}\inst{\ref{aff65}}
\and S.~Ligori\orcid{0000-0003-4172-4606}\inst{\ref{aff57}}
\and P.~B.~Lilje\orcid{0000-0003-4324-7794}\inst{\ref{aff84}}
\and V.~Lindholm\orcid{0000-0003-2317-5471}\inst{\ref{aff18},\ref{aff94}}
\and I.~Lloro\orcid{0000-0001-5966-1434}\inst{\ref{aff96}}
\and M.~Magliocchetti\orcid{0000-0001-9158-4838}\inst{\ref{aff79}}
\and G.~Mainetti\orcid{0000-0003-2384-2377}\inst{\ref{aff97}}
\and O.~Marggraf\orcid{0000-0001-7242-3852}\inst{\ref{aff29}}
\and M.~Martinelli\orcid{0000-0002-6943-7732}\inst{\ref{aff4},\ref{aff59}}
\and N.~Martinet\orcid{0000-0003-2786-7790}\inst{\ref{aff65}}
\and F.~Marulli\orcid{0000-0002-8850-0303}\inst{\ref{aff98},\ref{aff7},\ref{aff40}}
\and R.~J.~Massey\orcid{0000-0002-6085-3780}\inst{\ref{aff99}}
\and N.~Mauri\orcid{0000-0001-8196-1548}\inst{\ref{aff67},\ref{aff40}}
\and S.~Maurogordato\inst{\ref{aff10}}
\and H.~J.~McCracken\orcid{0000-0002-9489-7765}\inst{\ref{aff100}}
\and C.~J.~R.~McPartland\orcid{0000-0003-0639-025X}\inst{\ref{aff91},\ref{aff101}}
\and E.~Medinaceli\orcid{0000-0002-4040-7783}\inst{\ref{aff7}}
\and S.~Mei\orcid{0000-0002-2849-559X}\inst{\ref{aff102},\ref{aff103}}
\and M.~Meneghetti\orcid{0000-0003-1225-7084}\inst{\ref{aff7},\ref{aff40}}
\and E.~Merlin\orcid{0000-0001-6870-8900}\inst{\ref{aff41}}
\and G.~Meylan\orcid{0000-0001-6503-0209}\inst{\ref{aff23}}
\and P.~Monaco\orcid{0000-0003-2083-7564}\inst{\ref{aff20},\ref{aff21},\ref{aff22},\ref{aff37}}
\and M.~Moresco\orcid{0000-0002-7616-7136}\inst{\ref{aff98},\ref{aff7}}
\and C.~Moretti\orcid{0000-0003-3314-8936}\inst{\ref{aff21},\ref{aff37},\ref{aff22}}
\and L.~Moscardini\orcid{0000-0002-3473-6716}\inst{\ref{aff98},\ref{aff7},\ref{aff40}}
\and E.~Munari\orcid{0000-0002-1751-5946}\inst{\ref{aff21},\ref{aff37}}
\and C.~Neissner\orcid{0000-0001-8524-4968}\inst{\ref{aff104},\ref{aff61}}
\and R.~C.~Nichol\orcid{0000-0003-0939-6518}\inst{\ref{aff105}}
\and S.-M.~Niemi\orcid{0009-0005-0247-0086}\inst{\ref{aff31}}
\and C.~Padilla\orcid{0000-0001-7951-0166}\inst{\ref{aff104}}
\and S.~Paltani\orcid{0000-0002-8108-9179}\inst{\ref{aff75}}
\and F.~Pasian\orcid{0000-0002-4869-3227}\inst{\ref{aff21}}
\and J.~A.~Peacock\orcid{0000-0002-1168-8299}\inst{\ref{aff46}}
\and W.~J.~Percival\orcid{0000-0002-0644-5727}\inst{\ref{aff106},\ref{aff107},\ref{aff108}}
\and S.~Pires\orcid{0000-0002-0249-2104}\inst{\ref{aff5}}
\and G.~Polenta\orcid{0000-0003-4067-9196}\inst{\ref{aff47}}
\and M.~Poncet\inst{\ref{aff11}}
\and L.~A.~Popa\inst{\ref{aff109}}
\and L.~Pozzetti\orcid{0000-0001-7085-0412}\inst{\ref{aff7}}
\and F.~Raison\orcid{0000-0002-7819-6918}\inst{\ref{aff44}}
\and A.~Refregier\inst{\ref{aff110}}
\and A.~Renzi\orcid{0000-0001-9856-1970}\inst{\ref{aff34},\ref{aff35},\ref{aff7}}
\and J.~Rhodes\orcid{0000-0002-4485-8549}\inst{\ref{aff87}}
\and G.~Riccio\inst{\ref{aff51}}
\and I.~Risso\orcid{0000-0003-2525-7761}\inst{\ref{aff48},\ref{aff49},\ref{aff17}}
\and H.-W.~Rix\orcid{0000-0003-4996-9069}\inst{\ref{aff30}}
\and F.~Rizzo\orcid{0000-0002-9407-585X}\inst{\ref{aff21}}
\and M.~Roncarelli\orcid{0000-0001-9587-7822}\inst{\ref{aff7}}
\and C.~Rosset\orcid{0000-0003-0286-2192}\inst{\ref{aff102}}
\and B.~Rusholme\orcid{0000-0001-7648-4142}\inst{\ref{aff86}}
\and R.~Saglia\orcid{0000-0003-0378-7032}\inst{\ref{aff45},\ref{aff44}}
\and Z.~Sakr\orcid{0000-0002-4823-3757}\inst{\ref{aff111},\ref{aff112},\ref{aff113}}
\and A.~G.~S\'anchez\orcid{0000-0003-1198-831X}\inst{\ref{aff44}}
\and D.~Sapone\orcid{0000-0001-7089-4503}\inst{\ref{aff114}}
\and M.~Sauvage\orcid{0000-0002-0809-2574}\inst{\ref{aff5}}
\and P.~Schneider\orcid{0000-0001-8561-2679}\inst{\ref{aff29}}
\and M.~Scodeggio\inst{\ref{aff58}}
\and A.~Secroun\orcid{0000-0003-0505-3710}\inst{\ref{aff78}}
\and E.~Sefusatti\orcid{0000-0003-0473-1567}\inst{\ref{aff21},\ref{aff37},\ref{aff22}}
\and E.~Sihvola\orcid{0000-0003-1804-7715}\inst{\ref{aff18},\ref{aff94}}
\and P.~Simon\inst{\ref{aff29}}
\and G.~Sirri\orcid{0000-0003-2626-2853}\inst{\ref{aff40}}
\and A.~Spurio~Mancini\orcid{0000-0001-5698-0990}\inst{\ref{aff115}}
\and L.~Stanco\orcid{0000-0002-9706-5104}\inst{\ref{aff35}}
\and P.~Tallada-Cresp\'{i}\orcid{0000-0002-1336-8328}\inst{\ref{aff60},\ref{aff61}}
\and A.~N.~Taylor\inst{\ref{aff46}}
\and H.~I.~Teplitz\orcid{0000-0002-7064-5424}\inst{\ref{aff116}}
\and N.~Tessore\orcid{0000-0002-9696-7931}\inst{\ref{aff25}}
\and S.~Toft\orcid{0000-0003-3631-7176}\inst{\ref{aff117},\ref{aff101}}
\and R.~Toledo-Moreo\orcid{0000-0002-2997-4859}\inst{\ref{aff118},\ref{aff119}}
\and F.~Torradeflot\orcid{0000-0003-1160-1517}\inst{\ref{aff61},\ref{aff60}}
\and A.~Tsyganov\inst{\ref{aff120}}
\and I.~Tutusaus\orcid{0000-0002-3199-0399}\inst{\ref{aff63},\ref{aff62},\ref{aff112}}
\and E.~A.~Valentijn\inst{\ref{aff42}}
\and J.~Valiviita\orcid{0000-0001-6225-3693}\inst{\ref{aff18},\ref{aff94}}
\and T.~Vassallo\orcid{0000-0001-6512-6358}\inst{\ref{aff21},\ref{aff81}}
\and G.~Verdoes~Kleijn\orcid{0000-0001-5803-2580}\inst{\ref{aff42}}
\and A.~Veropalumbo\orcid{0000-0003-2387-1194}\inst{\ref{aff17},\ref{aff49},\ref{aff48}}
\and D.~Vibert\orcid{0009-0008-0607-631X}\inst{\ref{aff65}}
\and Y.~Wang\orcid{0000-0002-4749-2984}\inst{\ref{aff86}}
\and J.~Weller\orcid{0000-0002-8282-2010}\inst{\ref{aff45},\ref{aff44}}
\and G.~Zamorani\orcid{0000-0002-2318-301X}\inst{\ref{aff7}}
\and F.~M.~Zerbi\orcid{0000-0002-9996-973X}\inst{\ref{aff17}}
\and E.~Zucca\orcid{0000-0002-5845-8132}\inst{\ref{aff7}}
\and V.~Allevato\orcid{0000-0001-7232-5152}\inst{\ref{aff51}}
\and I.~T.~Andika\orcid{0000-0001-6102-9526}\inst{\ref{aff45}}
\and M.~Ballardini\orcid{0000-0003-4481-3559}\inst{\ref{aff121},\ref{aff122},\ref{aff7}}
\and P.~Bergamini\orcid{0000-0003-1383-9414}\inst{\ref{aff7}}
\and S.~Borgani\orcid{0000-0001-6151-6439}\inst{\ref{aff20},\ref{aff37},\ref{aff21},\ref{aff22},\ref{aff123}}
\and E.~Bozzo\orcid{0000-0002-8201-1525}\inst{\ref{aff75}}
\and M.~Calabrese\orcid{0000-0002-2637-2422}\inst{\ref{aff124},\ref{aff58}}
\and A.~Cappi\inst{\ref{aff10},\ref{aff7}}
\and F.~Caro\orcid{0009-0003-1053-0507}\inst{\ref{aff4}}
\and T.~Castro\orcid{0000-0002-6292-3228}\inst{\ref{aff125},\ref{aff21},\ref{aff37}}
\and S.~Conseil\orcid{0000-0002-3657-4191}\inst{\ref{aff126},\ref{aff70}}
\and G.~Daste\inst{\ref{aff65}}
\and D.~Di~Ferdinando\inst{\ref{aff40}}
\and J.~A.~Escartin~Vigo\inst{\ref{aff44}}
\and G.~Fabbian\orcid{0000-0002-3255-4695}\inst{\ref{aff76}}
\and J.~Garc\'ia-Bellido\orcid{0000-0002-9370-8360}\inst{\ref{aff111}}
\and T.~Gasparetto\orcid{0000-0002-7913-4866}\inst{\ref{aff4}}
\and C.~M.~Gutierrez\orcid{0000-0001-7854-783X}\inst{\ref{aff12},\ref{aff13}}
\and A.~Hall\orcid{0000-0002-3139-8651}\inst{\ref{aff46}}
\and Y.~Kang\orcid{0009-0000-8588-7250}\inst{\ref{aff32}}
\and A.~Loureiro\orcid{0000-0002-4371-0876}\inst{\ref{aff127},\ref{aff128}}
\and J.~F.~Macias-Perez\orcid{0000-0002-5385-2763}\inst{\ref{aff129}}
\and R.~Maoli\orcid{0000-0002-6065-3025}\inst{\ref{aff130},\ref{aff4}}
\and R.~B.~Metcalf\orcid{0000-0003-3167-2574}\inst{\ref{aff98},\ref{aff7}}
\and S.~Nadathur\orcid{0000-0001-9070-3102}\inst{\ref{aff131}}
\and L.~Patrizii\inst{\ref{aff40}}
\and E.~Sarpa\orcid{0000-0002-1256-655X}\inst{\ref{aff21}}
\and V.~Scottez\orcid{0009-0008-3864-940X}\inst{\ref{aff132},\ref{aff133}}
\and M.~Sereno\orcid{0000-0003-0302-0325}\inst{\ref{aff7},\ref{aff40}}
\and K.~Tanidis\orcid{0000-0001-9843-5130}\inst{\ref{aff134}}
\and M.~Tenti\orcid{0000-0002-4254-5901}\inst{\ref{aff40}}
\and A.~Troja\orcid{0000-0003-0239-4595}\inst{\ref{aff21}}
\and M.~Tucci\inst{\ref{aff75}}
\and M.~Viel\orcid{0000-0002-2642-5707}\inst{\ref{aff37},\ref{aff21},\ref{aff38},\ref{aff22},\ref{aff123}}
\and J.~R.~Weaver\orcid{0000-0003-1614-196X}\inst{\ref{aff135}}
\and M.~Wiesmann\orcid{0009-0000-8199-5860}\inst{\ref{aff84}}
\and Abdurro'uf\orcid{0000-0002-5258-8761}\inst{\ref{aff136}}
\and J.~A.~Acevedo~Barroso\orcid{0000-0002-9654-1711}\inst{\ref{aff87}}
\and R.~Adam\orcid{0009-0000-5380-1109}\inst{\ref{aff10}}
\and Y.~Akrami\orcid{0000-0002-2407-7956}\inst{\ref{aff111},\ref{aff137}}
\and G.~Alguero\inst{\ref{aff129}}
\and G.~Angora\orcid{0000-0002-0316-6562}\inst{\ref{aff51},\ref{aff121}}
\and S.~Anselmi\orcid{0000-0002-3579-9583}\inst{\ref{aff35},\ref{aff34},\ref{aff138}}
\and M.~Archidiacono\orcid{0000-0003-4952-9012}\inst{\ref{aff82},\ref{aff83}}
\and G.~Aric\`o\orcid{0000-0002-2802-2928}\inst{\ref{aff40}}
\and F.~Atrio-Barandela\orcid{0000-0002-2130-2513}\inst{\ref{aff139}}
\and M.~Aubert\orcid{0009-0002-7667-8814}\inst{\ref{aff70}}
\and E.~Aubourg\orcid{0000-0002-5592-023X}\inst{\ref{aff102},\ref{aff140}}
\and M.~Baes\orcid{0000-0002-3930-2757}\inst{\ref{aff141}}
\and N.~Bartolo\inst{\ref{aff34},\ref{aff35},\ref{aff41}}
\and L.~Bazzanini\orcid{0000-0003-0727-0137}\inst{\ref{aff8},\ref{aff121}}
\and J.~R.~Bermejo-Climent\inst{\ref{aff142},\ref{aff143}}
\and D.~Bertacca\orcid{0000-0002-2490-7139}\inst{\ref{aff34},\ref{aff41},\ref{aff35}}
\and M.~Bethermin\orcid{0000-0002-3915-2015}\inst{\ref{aff144}}
\and F.~Beutler\orcid{0000-0003-0467-5438}\inst{\ref{aff46}}
\and A.~Blanchard\orcid{0000-0001-8555-9003}\inst{\ref{aff112}}
\and H.~B\"ohringer\orcid{0000-0001-8241-4204}\inst{\ref{aff44},\ref{aff81},\ref{aff145}}
\and M.~Bonici\orcid{0000-0002-8430-126X}\inst{\ref{aff106},\ref{aff58}}
\and F.~Bouch\`e\orcid{0000-0002-4663-1786}\inst{\ref{aff146},\ref{aff64}}
\and M.~L.~Brown\orcid{0000-0002-0370-8077}\inst{\ref{aff68}}
\and S.~Bruton\orcid{0000-0002-6503-5218}\inst{\ref{aff85}}
\and F.~Buitrago\orcid{0000-0002-2861-9812}\inst{\ref{aff147},\ref{aff148},\ref{aff149}}
\and A.~Calabro\orcid{0000-0003-2536-1614}\inst{\ref{aff4}}
\and P.~Carrilho\orcid{0000-0003-1339-0194}\inst{\ref{aff150}}
\and C.~S.~Carvalho\inst{\ref{aff149}}
\and N.~Chartab\orcid{0000-0003-3691-937X}\inst{\ref{aff86}}
\and F.~Cogato\orcid{0000-0003-4632-6113}\inst{\ref{aff7}}
\and T.~E.~Collett\orcid{0000-0001-5564-3140}\inst{\ref{aff131}}
\and T.~Contini\orcid{0000-0003-0275-938X}\inst{\ref{aff112}}
\and P.~Corcho-Caballero\orcid{0000-0001-6327-7080}\inst{\ref{aff42}}
\and M.~Correnti\orcid{0000-0001-6464-3257}\inst{\ref{aff4},\ref{aff47}}
\and M.~Costanzi\orcid{0000-0001-8158-1449}\inst{\ref{aff20},\ref{aff21},\ref{aff37}}
\and B.~Csizi\orcid{0000-0003-3227-6581}\inst{\ref{aff32}}
\and O.~Cucciati\orcid{0000-0002-9336-7551}\inst{\ref{aff7}}
\and H.~Dannerbauer\orcid{0000-0001-7147-3575}\inst{\ref{aff12},\ref{aff13}}
\and T.~de~Boer\orcid{0000-0001-5486-2747}\inst{\ref{aff66}}
\and F.~De~Paolis\orcid{0000-0001-6460-7563}\inst{\ref{aff14},\ref{aff15},\ref{aff16}}
\and G.~Desprez\orcid{0000-0001-8325-1742}\inst{\ref{aff42}}
\and A.~D\'iaz-S\'anchez\orcid{0000-0003-0748-4768}\inst{\ref{aff151}}
\and S.~Di~Domizio\orcid{0000-0003-2863-5895}\inst{\ref{aff48},\ref{aff49}}
\and J.~M.~Diego\orcid{0000-0001-9065-3926}\inst{\ref{aff152}}
\and P.~Dimauro\orcid{0000-0001-7399-2854}\inst{\ref{aff153},\ref{aff4}}
\and P.-A.~Duc\orcid{0000-0003-3343-6284}\inst{\ref{aff144}}
\and F.~Dufresne\orcid{0009-0009-0580-7886}\inst{\ref{aff65}}
\and V.~Duret\orcid{0009-0009-0383-4960}\inst{\ref{aff154}}
\and A.~Eggemeier\orcid{0000-0002-1841-8910}\inst{\ref{aff29}}
\and A.~Enia\orcid{0000-0002-0200-2857}\inst{\ref{aff155},\ref{aff7}}
\and Y.~Fang\orcid{0000-0002-0334-6950}\inst{\ref{aff45}}
\and A.~Farina\orcid{0009-0000-3420-929X}\inst{\ref{aff17},\ref{aff49}}
\and A.~M.~N.~Ferguson\orcid{0000-0001-7934-1278}\inst{\ref{aff46}}
\and P.~G.~Ferreira\orcid{0000-0002-3021-2851}\inst{\ref{aff28}}
\and A.~Finoguenov\orcid{0000-0002-4606-5403}\inst{\ref{aff18}}
\and A.~Fontana\orcid{0000-0003-3820-2823}\inst{\ref{aff4}}
\and F.~Fontanot\orcid{0000-0003-4744-0188}\inst{\ref{aff21},\ref{aff37}}
\and N.~Frusciante\orcid{0000-0002-7375-1230}\inst{\ref{aff156},\ref{aff157}}
\and A.~Fumagalli\orcid{0009-0004-0300-2535}\inst{\ref{aff21}}
\and Y.~Fu\orcid{0000-0002-0759-0504}\inst{\ref{aff24},\ref{aff42}}
\and K.~Ganga\orcid{0000-0001-8159-8208}\inst{\ref{aff102}}
\and R.~Gavazzi\orcid{0000-0002-5540-6935}\inst{\ref{aff65},\ref{aff100}}
\and E.~Gaztanaga\orcid{0000-0001-9632-0815}\inst{\ref{aff63},\ref{aff62},\ref{aff131}}
\and Z.~Ghaffari\orcid{0000-0002-6467-8078}\inst{\ref{aff21},\ref{aff37}}
\and F.~Giacomini\orcid{0000-0002-3129-2814}\inst{\ref{aff40}}
\and F.~Gianotti\orcid{0000-0003-4666-119X}\inst{\ref{aff7}}
\and G.~Gozaliasl\orcid{0000-0002-0236-919X}\inst{\ref{aff158},\ref{aff18}}
\and A.~Gruppuso\orcid{0000-0001-9272-5292}\inst{\ref{aff7},\ref{aff40}}
\and M.~Guidi\orcid{0000-0001-9408-1101}\inst{\ref{aff39},\ref{aff7}}
\and N.~A.~Hatch\orcid{0000-0001-5600-0534}\inst{\ref{aff159}}
\and C.~Hern\'andez-Monteagudo\orcid{0000-0001-5471-9166}\inst{\ref{aff13},\ref{aff12}}
\and T.~Heuberger\orcid{0009-0009-3474-8467}\inst{\ref{aff160}}
\and H.~Hildebrandt\orcid{0000-0002-9814-3338}\inst{\ref{aff161}}
\and J.~Hjorth\orcid{0000-0002-4571-2306}\inst{\ref{aff162}}
\and M.~Jauzac\orcid{0000-0003-1974-8732}\inst{\ref{aff163},\ref{aff99},\ref{aff164},\ref{aff165},\ref{aff112}}
\and S.~Joudaki\orcid{0000-0001-8820-673X}\inst{\ref{aff60},\ref{aff131}}
\and J.~J.~E.~Kajava\orcid{0000-0002-3010-8333}\inst{\ref{aff166},\ref{aff167},\ref{aff168}}
\and D.~Karagiannis\orcid{0000-0002-4927-0816}\inst{\ref{aff169},\ref{aff121},\ref{aff170}}
\and J.~Kim\orcid{0000-0003-2776-2761}\inst{\ref{aff28}}
\and C.~C.~Kirkpatrick\inst{\ref{aff171}}
\and A.~Kov\'acs\orcid{0000-0002-5825-579X}\inst{\ref{aff143},\ref{aff142}}
\and I.~Kova{\v{c}}i{\'{c}}\orcid{0000-0001-6751-3263}\inst{\ref{aff141}}
\and K.~Koyama\orcid{0000-0001-6727-6915}\inst{\ref{aff131}}
\and S.~Kruk\orcid{0000-0001-8010-8879}\inst{\ref{aff3}}
\and M.~C.~Lam\orcid{0000-0002-9347-2298}\inst{\ref{aff46}}
\and M.~Lattanzi\orcid{0000-0003-1059-2532}\inst{\ref{aff122}}
\and F.~Leclercq\orcid{0000-0002-9339-1404}\inst{\ref{aff100}}
\and S.~Lee\orcid{0000-0002-8289-740X}\inst{\ref{aff36},\ref{aff87}}
\and J.~Le~Graet\orcid{0000-0001-6523-7971}\inst{\ref{aff78}}
\and M.~Lembo\orcid{0000-0002-5271-5070}\inst{\ref{aff100}}
\and M.~D.~Lepinzan\orcid{0000-0003-1287-9801}\inst{\ref{aff21}}
\and G.~F.~Lesci\orcid{0000-0002-4607-2830}\inst{\ref{aff8}}
\and J.~Lesgourgues\orcid{0000-0001-7627-353X}\inst{\ref{aff172}}
\and T.~I.~Liaudat\orcid{0000-0002-9104-314X}\inst{\ref{aff140}}
\and L.~Linke\orcid{0000-0002-2622-8113}\inst{\ref{aff32}}
\and A.~Manj\'on-Garc\'ia\orcid{0000-0002-7413-8825}\inst{\ref{aff151}}
\and F.~Mannucci\orcid{0000-0002-4803-2381}\inst{\ref{aff173}}
\and M.~Marinucci\orcid{0000-0003-1159-3756}\inst{\ref{aff174}}
\and F.~R.~Marleau\orcid{0000-0002-1442-2947}\inst{\ref{aff32}}
\and C.~J.~A.~P.~Martins\orcid{0000-0002-4886-9261}\inst{\ref{aff175},\ref{aff53}}
\and M.~Migliaccio\inst{\ref{aff176},\ref{aff177}}
\and M.~Miluzio\inst{\ref{aff3},\ref{aff178}}
\and G.~Morgante\inst{\ref{aff7}}
\and C.~Murray\orcid{0000-0002-4668-1273}\inst{\ref{aff5}}
\and K.~Naidoo\orcid{0000-0002-9182-1802}\inst{\ref{aff131},\ref{aff30}}
\and P.~Natoli\orcid{0000-0003-0126-9100}\inst{\ref{aff121},\ref{aff122}}
\and A.~Navarro-Alsina\orcid{0000-0002-3173-2592}\inst{\ref{aff29}}
\and A.~Nersesian\orcid{0000-0001-6843-409X}\inst{\ref{aff179},\ref{aff141}}
\and S.~Nesseris\orcid{0000-0002-0567-0324}\inst{\ref{aff111}}
\and L.~Nicastro\orcid{0000-0001-8534-6788}\inst{\ref{aff7}}
\and M.~Oguri\orcid{0000-0003-3484-399X}\inst{\ref{aff180},\ref{aff181}}
\and F.~Oppizzi\orcid{0000-0003-3904-8370}\inst{\ref{aff49},\ref{aff35},\ref{aff34}}
\and F.~Pace\orcid{0000-0001-8039-0480}\inst{\ref{aff55},\ref{aff56},\ref{aff57},\ref{aff2}}
\and M.~J.~Page\orcid{0000-0002-6689-6271}\inst{\ref{aff25}}
\and D.~Paoletti\orcid{0000-0003-4761-6147}\inst{\ref{aff7},\ref{aff9}}
\and G.~Papini\orcid{0000-0002-5810-318X}\inst{\ref{aff7},\ref{aff98}}
\and G.~Parimbelli\orcid{0000-0002-2539-2472}\inst{\ref{aff63},\ref{aff38}}
\and F.~Passalacqua\orcid{0000-0002-8606-4093}\inst{\ref{aff35}}
\and C.~Pattison\orcid{0000-0003-3272-2617}\inst{\ref{aff131}}
\and A.~Pisani\orcid{0000-0002-6146-4437}\inst{\ref{aff78}}
\and L.~Pizzuti\orcid{0000-0001-5654-7580}\inst{\ref{aff182}}
\and J.~J.~Popp\orcid{0000-0002-3724-1727}\inst{\ref{aff183}}
\and D.~Potter\orcid{0000-0002-0757-5195}\inst{\ref{aff184}}
\and M.~Poulain\orcid{0000-0002-7664-4510}\inst{\ref{aff185}}
\and G.~W.~Pratt\inst{\ref{aff5}}
\and S.~Quai\orcid{0000-0002-0449-8163}\inst{\ref{aff98},\ref{aff7}}
\and M.~Radovich\orcid{0000-0002-3585-866X}\inst{\ref{aff41}}
\and F.~Ricci\orcid{0000-0001-5742-5980}\inst{\ref{aff186},\ref{aff4}}
\and G.~Rodighiero\orcid{0000-0002-9415-2296}\inst{\ref{aff34},\ref{aff41}}
\and W.~Roster\orcid{0000-0002-9149-6528}\inst{\ref{aff44}}
\and M.~Sahl\'en\orcid{0000-0003-0973-4804}\inst{\ref{aff187}}
\and D.~B.~Sanders\orcid{0000-0002-1233-9998}\inst{\ref{aff66}}
\and C.~Scarlata\orcid{0000-0002-9136-8876}\inst{\ref{aff188}}
\and J.~Schaye\orcid{0000-0002-0668-5560}\inst{\ref{aff24}}
\and A.~Schneider\orcid{0000-0001-7055-8104}\inst{\ref{aff184}}
\and D.~Sciotti\orcid{0009-0008-4519-2620}\inst{\ref{aff4},\ref{aff59}}
\and E.~Sellentin\orcid{0009-0002-2655-3458}\inst{\ref{aff189},\ref{aff24}}
\and S.~Serjeant\orcid{0000-0002-0517-7943}\inst{\ref{aff183}}
\and F.~Shankar\orcid{0000-0001-8973-5051}\inst{\ref{aff190}}
\and L.~C.~Smith\orcid{0000-0002-3259-2771}\inst{\ref{aff191}}
\and F.~Soldano\inst{\ref{aff100}}
\and S.~A.~Stanford\orcid{0000-0003-0122-0841}\inst{\ref{aff192}}
\and I.~Szapudi\orcid{0000-0003-2274-0301}\inst{\ref{aff66}}
\and M.~Talia\orcid{0000-0003-4352-2063}\inst{\ref{aff98},\ref{aff7}}
\and F.~Tarsitano\orcid{0000-0002-5919-0238}\inst{\ref{aff193},\ref{aff110}}
\and G.~Testera\orcid{0000-0003-2970-766X}\inst{\ref{aff49}}
\and R.~Teyssier\orcid{0000-0001-7689-0933}\inst{\ref{aff194}}
\and S.~Tosi\orcid{0000-0002-7275-9193}\inst{\ref{aff48},\ref{aff17},\ref{aff49}}
\and C.~Uhlemann\orcid{0000-0001-7831-1579}\inst{\ref{aff195}}
\and M.~Urbano\orcid{0000-0001-5640-0650}\inst{\ref{aff159},\ref{aff144}}
\and C.~Valieri\inst{\ref{aff40}}
\and A.~Venhola\orcid{0000-0001-6071-4564}\inst{\ref{aff185}}
\and D.~Vergani\orcid{0000-0003-0898-2216}\inst{\ref{aff7}}
\and G.~Vernardos\orcid{0000-0001-8554-7248}\inst{\ref{aff196},\ref{aff197}}
\and A.~Viitanen\orcid{0000-0001-9383-786X}\inst{\ref{aff75},\ref{aff171},\ref{aff4}}
\and E.~Vilenius\orcid{0000-0002-6184-7681}\inst{\ref{aff25}}
\and S.~Vinciguerra\orcid{0009-0005-4018-3184}\inst{\ref{aff65}}
\and M.~von~Wietersheim-Kramsta\orcid{0000-0003-4986-5091}\inst{\ref{aff99},\ref{aff163}}
\and N.~A.~Walton\orcid{0000-0003-3983-8778}\inst{\ref{aff191}}
\and L.~Wang\orcid{0000-0002-6736-9158}\inst{\ref{aff198},\ref{aff42}}
\and A.~H.~Wright\orcid{0000-0001-7363-7932}\inst{\ref{aff161}}}
										   
\institute{Departamento de F\'isica, Faculdade de Ci\^encias, Universidade de Lisboa, Edif\'icio C8, Campo Grande, PT1749-016 Lisboa, Portugal\label{aff1}
\and
Instituto de Astrof\'isica e Ci\^encias do Espa\c{c}o, Faculdade de Ci\^encias, Universidade de Lisboa, Campo Grande, 1749-016 Lisboa, Portugal\label{aff2}
\and
ESAC/ESA, Camino Bajo del Castillo, s/n., Urb. Villafranca del Castillo, 28692 Villanueva de la Ca\~nada, Madrid, Spain\label{aff3}
\and
INAF-Osservatorio Astronomico di Roma, Via Frascati 33, 00078 Monteporzio Catone, Italy\label{aff4}
\and
Universit\'e Paris-Saclay, Universit\'e Paris Cit\'e, CEA, CNRS, AIM, 91191, Gif-sur-Yvette, France\label{aff5}
\and
Astrophysics Research Institute, Liverpool John Moores University, 146 Brownlow Hill, Liverpool L3 5RF, UK\label{aff6}
\and
INAF-Osservatorio di Astrofisica e Scienza dello Spazio di Bologna, Via Piero Gobetti 93/3, 40129 Bologna, Italy\label{aff7}
\and
INAF, Istituto di Radioastronomia, Via Piero Gobetti 101, 40129 Bologna, Italy\label{aff8}
\and
INFN-Bologna, Via Irnerio 46, 40126 Bologna, Italy\label{aff9}
\and
Universit\'e C\^{o}te d'Azur, Observatoire de la C\^{o}te d'Azur, CNRS, Laboratoire Lagrange, Bd de l'Observatoire, CS 34229, 06304 Nice cedex 4, France\label{aff10}
\and
Centre National d'Etudes Spatiales -- Centre spatial de Toulouse, 18 avenue Edouard Belin, 31401 Toulouse Cedex 9, France\label{aff11}
\and
Instituto de Astrof\'{\i}sica de Canarias, E-38205 La Laguna, Tenerife, Spain\label{aff12}
\and
Universidad de La Laguna, Dpto. Astrof\'\i sica, E-38206 La Laguna, Tenerife, Spain\label{aff13}
\and
Department of Mathematics and Physics E. De Giorgi, University of Salento, Via per Arnesano, CP-I93, 73100, Lecce, Italy\label{aff14}
\and
INFN, Sezione di Lecce, Via per Arnesano, CP-193, 73100, Lecce, Italy\label{aff15}
\and
INAF-Sezione di Lecce, c/o Dipartimento Matematica e Fisica, Via per Arnesano, 73100, Lecce, Italy\label{aff16}
\and
INAF-Osservatorio Astronomico di Brera, Via Brera 28, 20122 Milano, Italy\label{aff17}
\and
Department of Physics, P.O. Box 64, University of Helsinki, 00014 Helsinki, Finland\label{aff18}
\and
Asteroid Engineering Laboratory, Lule\aa{} University of Technology, Box 848, 98128 Kiruna, Sweden\label{aff19}
\and
Dipartimento di Fisica - Sezione di Astronomia, Universit\`a di Trieste, Via Tiepolo 11, 34131 Trieste, Italy\label{aff20}
\and
INAF-Osservatorio Astronomico di Trieste, Via G. B. Tiepolo 11, 34143 Trieste, Italy\label{aff21}
\and
INFN, Sezione di Trieste, Via Valerio 2, 34127 Trieste TS, Italy\label{aff22}
\and
Institute of Physics, Laboratory of Astrophysics, Ecole Polytechnique F\'ed\'erale de Lausanne (EPFL), Observatoire de Sauverny, 1290 Versoix, Switzerland\label{aff23}
\and
Leiden Observatory, Leiden University, Einsteinweg 55, 2333 CC Leiden, The Netherlands\label{aff24}
\and
Mullard Space Science Laboratory, University College London, Holmbury St Mary, Dorking, Surrey RH5 6NT, UK\label{aff25}
\and
Aurora Technology for European Space Agency (ESA), Camino bajo del Castillo, s/n, Urbanizacion Villafranca del Castillo, Villanueva de la Ca\~nada, 28692 Madrid, Spain\label{aff26}
\and
European Space Agency/ESOC, Robert-Bosch-Str. 5, 64293 Darmstadt, Germany\label{aff27}
\and
Department of Physics, University of Oxford, Keble Road, Oxford OX1 3RH, UK\label{aff28}
\and
Universit\"at Bonn, Argelander-Institut f\"ur Astronomie, Auf dem H\"ugel 71, 53121 Bonn, Germany\label{aff29}
\and
Max-Planck-Institut f\"ur Astronomie, K\"onigstuhl 17, 69117 Heidelberg, Germany\label{aff30}
\and
European Space Agency/ESTEC, Keplerlaan 1, 2201 AZ Noordwijk, The Netherlands\label{aff31}
\and
Universit\"at Innsbruck, Institut f\"ur Astro- und Teilchenphysik, Technikerstr. 25/8, 6020 Innsbruck, Austria\label{aff32}
\and
Department of Physics and Astronomy, University of British Columbia, Vancouver, BC V6T 1Z1, Canada\label{aff33}
\and
Dipartimento di Fisica e Astronomia "G. Galilei", Universit\`a di Padova, Via Marzolo 8, 35131 Padova, Italy\label{aff34}
\and
INFN-Padova, Via Marzolo 8, 35131 Padova, Italy\label{aff35}
\and
Ohio University, Physics \& Astronomy Department,1 Ohio University, Athens, OH 45701, USA\label{aff36}
\and
IFPU, Institute for Fundamental Physics of the Universe, via Beirut 2, 34151 Trieste, Italy\label{aff37}
\and
SISSA, International School for Advanced Studies, Via Bonomea 265, 34136 Trieste TS, Italy\label{aff38}
\and
Dipartimento di Fisica e Astronomia, Universit\`a di Bologna, Via Gobetti 93/2, 40129 Bologna, Italy\label{aff39}
\and
INFN-Sezione di Bologna, Viale Berti Pichat 6/2, 40127 Bologna, Italy\label{aff40}
\and
INAF-Osservatorio Astronomico di Padova, Via dell'Osservatorio 5, 35122 Padova, Italy\label{aff41}
\and
Kapteyn Astronomical Institute, University of Groningen, PO Box 800, 9700 AV Groningen, The Netherlands\label{aff42}
\and
ATG Europe BV, Huygensstraat 34, 2201 DK Noordwijk, The Netherlands\label{aff43}
\and
Max Planck Institute for Extraterrestrial Physics, Giessenbachstr. 1, 85748 Garching, Germany\label{aff44}
\and
Universit\"ats-Sternwarte M\"unchen, Fakult\"at f\"ur Physik, Ludwig-Maximilians-Universit\"at M\"unchen, Scheinerstr.~1, 81679 M\"unchen, Germany\label{aff45}
\and
Institute for Astronomy, University of Edinburgh, Royal Observatory, Blackford Hill, Edinburgh EH9 3HJ, UK\label{aff46}
\and
Space Science Data Center, Italian Space Agency, via del Politecnico snc, 00133 Roma, Italy\label{aff47}
\and
Dipartimento di Fisica, Universit\`a di Genova, Via Dodecaneso 33, 16146, Genova, Italy\label{aff48}
\and
INFN-Sezione di Genova, Via Dodecaneso 33, 16146, Genova, Italy\label{aff49}
\and
Department of Physics "E. Pancini", University Federico II, Via Cinthia 6, 80126, Napoli, Italy\label{aff50}
\and
INAF-Osservatorio Astronomico di Capodimonte, Via Moiariello 16, 80131 Napoli, Italy\label{aff51}
\and
European Southern Observatory, Karl-Schwarzschild-Str.~2, 85748 Garching, Germany\label{aff52}
\and
Instituto de Astrof\'isica e Ci\^encias do Espa\c{c}o, Universidade do Porto, CAUP, Rua das Estrelas, PT4150-762 Porto, Portugal\label{aff53}
\and
Faculdade de Ci\^encias da Universidade do Porto, Rua do Campo de Alegre, 4150-007 Porto, Portugal\label{aff54}
\and
Dipartimento di Fisica, Universit\`a degli Studi di Torino, Via P. Giuria 1, 10125 Torino, Italy\label{aff55}
\and
INFN-Sezione di Torino, Via P. Giuria 1, 10125 Torino, Italy\label{aff56}
\and
INAF-Osservatorio Astrofisico di Torino, Via Osservatorio 20, 10025 Pino Torinese (TO), Italy\label{aff57}
\and
INAF-IASF Milano, Via Alfonso Corti 12, 20133 Milano, Italy\label{aff58}
\and
INFN-Sezione di Roma, Piazzale Aldo Moro, 2 - c/o Dipartimento di Fisica, Edificio G. Marconi, 00185 Roma, Italy\label{aff59}
\and
Centro de Investigaciones Energ\'eticas, Medioambientales y Tecnol\'ogicas (CIEMAT), Avenida Complutense 40, 28040 Madrid, Spain\label{aff60}
\and
Port d'Informaci\'{o} Cient\'{i}fica, Campus UAB, C. Albareda s/n, 08193 Bellaterra (Barcelona), Spain\label{aff61}
\and
Institut d'Estudis Espacials de Catalunya (IEEC),  Edifici RDIT, Campus UPC, 08860 Castelldefels, Barcelona, Spain\label{aff62}
\and
Institute of Space Sciences (ICE, CSIC), Campus UAB, Carrer de Can Magrans, s/n, 08193 Barcelona, Spain\label{aff63}
\and
INFN -- Sezione di Napoli, Via Cinthia 6, 80126, Napoli, Italy\label{aff64}
\and
Aix-Marseille Universit\'e, CNRS, CNES, LAM, Marseille, France\label{aff65}
\and
Institute for Astronomy, University of Hawaii, 2680 Woodlawn Drive, Honolulu, HI 96822, USA\label{aff66}
\and
Dipartimento di Fisica e Astronomia "Augusto Righi" - Alma Mater Studiorum Universit\`a di Bologna, Viale Berti Pichat 6/2, 40127 Bologna, Italy\label{aff67}
\and
Jodrell Bank Centre for Astrophysics, Department of Physics and Astronomy, University of Manchester, Oxford Road, Manchester M13 9PL, UK\label{aff68}
\and
European Space Agency/ESRIN, Largo Galileo Galilei 1, 00044 Frascati, Roma, Italy\label{aff69}
\and
Universit\'e Lyon 1, CNRS/IN2P3, IP2I Lyon, Villeurbanne, France\label{aff70}
\and
Institut de Ci\`{e}ncies del Cosmos (ICCUB), Universitat de Barcelona (IEEC-UB), Mart\'{i} i Franqu\`{e}s 1, 08028 Barcelona, Spain\label{aff71}
\and
Instituci\'o Catalana de Recerca i Estudis Avan\c{c}ats (ICREA), Passeig de Llu\'{\i}s Companys 23, 08010 Barcelona, Spain\label{aff72}
\and
Institut de Ciencies de l'Espai (IEEC-CSIC), Campus UAB, Carrer de Can Magrans, s/n Cerdanyola del Vall\'es, 08193 Barcelona, Spain\label{aff73}
\and
Universit\'e Lyon 1, CNRS/IN2P3, IUF, IP2I Lyon, Villeurbanne, France \label{aff74}
\and
Department of Astronomy, University of Geneva, ch. d'Ecogia 16, 1290 Versoix, Switzerland\label{aff75}
\and
Universit\'e Paris-Saclay, CNRS, Institut d'astrophysique spatiale, 91405, Orsay, France\label{aff76}
\and
Institut universitaire de France (IUF), 1 rue Descartes, 75231 PARIS CEDEX 05, France\label{aff77}
\and
Aix-Marseille Universit\'e, CNRS/IN2P3, CPPM, Marseille, France\label{aff78}
\and
INAF-Istituto di Astrofisica e Planetologia Spaziali, via del Fosso del Cavaliere, 100, 00100 Roma, Italy\label{aff79}
\and
School of Physics, HH Wills Physics Laboratory, University of Bristol, Tyndall Avenue, Bristol, BS8 1TL, UK\label{aff80}
\and
University Observatory, LMU Faculty of Physics, Scheinerstr.~1, 81679 Munich, Germany\label{aff81}
\and
Dipartimento di Fisica "Aldo Pontremoli", Universit\`a degli Studi di Milano, Via Celoria 16, 20133 Milano, Italy\label{aff82}
\and
INFN-Sezione di Milano, Via Celoria 16, 20133 Milano, Italy\label{aff83}
\and
Institute of Theoretical Astrophysics, University of Oslo, P.O. Box 1029 Blindern, 0315 Oslo, Norway\label{aff84}
\and
California Institute of Technology, 1200 E California Blvd, Pasadena, CA 91125, USA\label{aff85}
\and
Caltech/IPAC, 1200 E. California Blvd., Pasadena, CA 91125, USA\label{aff86}
\and
Jet Propulsion Laboratory, California Institute of Technology, 4800 Oak Grove Drive, Pasadena, CA, 91109, USA\label{aff87}
\and
School of Physics and Astronomy, Lancaster University, Lancaster, LA1 4YB, UK\label{aff88}
\and
Felix Hormuth Engineering, Goethestr. 17, 69181 Leimen, Germany\label{aff89}
\and
Technical University of Denmark, Elektrovej 327, 2800 Kgs. Lyngby, Denmark\label{aff90}
\and
Cosmic Dawn Center (DAWN), Denmark\label{aff91}
\and
NASA Goddard Space Flight Center, Greenbelt, MD 20771, USA\label{aff92}
\and
Universit\'e de Gen\`eve, D\'epartement de Physique Th\'eorique and Centre for Astroparticle Physics, 24 quai Ernest-Ansermet, CH-1211 Gen\`eve 4, Switzerland\label{aff93}
\and
Helsinki Institute of Physics, Gustaf H{\"a}llstr{\"o}min katu 2, University of Helsinki, 00014 Helsinki, Finland\label{aff94}
\and
Laboratoire d'etude de l'Univers et des phenomenes eXtremes, Observatoire de Paris, Universit\'e PSL, Sorbonne Universit\'e, CNRS, 92190 Meudon, France\label{aff95}
\and
SKAO, Jodrell Bank, Lower Withington, Macclesfield SK11 9FT, UK\label{aff96}
\and
Centre de Calcul de l'IN2P3/CNRS, 21 avenue Pierre de Coubertin 69627 Villeurbanne Cedex, France\label{aff97}
\and
Dipartimento di Fisica e Astronomia "Augusto Righi" - Alma Mater Studiorum Universit\`a di Bologna, via Piero Gobetti 93/2, 40129 Bologna, Italy\label{aff98}
\and
Department of Physics, Institute for Computational Cosmology, Durham University, South Road, Durham, DH1 3LE, UK\label{aff99}
\and
Institut d'Astrophysique de Paris, UMR 7095, CNRS, and Sorbonne Universit\'e, 98 bis boulevard Arago, 75014 Paris, France\label{aff100}
\and
Niels Bohr Institute, University of Copenhagen, Jagtvej 128, 2200 Copenhagen, Denmark\label{aff101}
\and
Universit\'e Paris Cit\'e, CNRS, Astroparticule et Cosmologie, 75013 Paris, France\label{aff102}
\and
CNRS-UCB International Research Laboratory, Centre Pierre Bin\'etruy, IRL2007, CPB-IN2P3, Berkeley, USA\label{aff103}
\and
Institut de F\'{i}sica d'Altes Energies (IFAE), The Barcelona Institute of Science and Technology, Campus UAB, 08193 Bellaterra (Barcelona), Spain\label{aff104}
\and
School of Mathematics and Physics, University of Surrey, Guildford, Surrey, GU2 7XH, UK\label{aff105}
\and
Waterloo Centre for Astrophysics, University of Waterloo, Waterloo, Ontario N2L 3G1, Canada\label{aff106}
\and
Department of Physics and Astronomy, University of Waterloo, Waterloo, Ontario N2L 3G1, Canada\label{aff107}
\and
Perimeter Institute for Theoretical Physics, Waterloo, Ontario N2L 2Y5, Canada\label{aff108}
\and
Institute of Space Science, Str. Atomistilor, nr. 409 M\u{a}gurele, Ilfov, 077125, Romania\label{aff109}
\and
Institute for Particle Physics and Astrophysics, Dept. of Physics, ETH Zurich, Wolfgang-Pauli-Strasse 27, 8093 Zurich, Switzerland\label{aff110}
\and
Instituto de F\'isica Te\'orica UAM-CSIC, Campus de Cantoblanco, 28049 Madrid, Spain\label{aff111}
\and
Institut de Recherche en Astrophysique et Plan\'etologie (IRAP), Universit\'e de Toulouse, CNRS, UPS, CNES, 14 Av. Edouard Belin, 31400 Toulouse, France\label{aff112}
\and
Universit\'e St Joseph; Faculty of Sciences, Beirut, Lebanon\label{aff113}
\and
Departamento de F\'isica, FCFM, Universidad de Chile, Blanco Encalada 2008, Santiago, Chile\label{aff114}
\and
Department of Physics, Royal Holloway, University of London, Surrey TW20 0EX, UK\label{aff115}
\and
Infrared Processing and Analysis Center, California Institute of Technology, Pasadena, CA 91125, USA\label{aff116}
\and
Cosmic Dawn Center (DAWN)\label{aff117}
\and
Universidad Polit\'ecnica de Cartagena, Departamento de Electr\'onica y Tecnolog\'ia de Computadoras,  Plaza del Hospital 1, 30202 Cartagena, Spain\label{aff118}
\and
European University of Technology EUt+, European Union\label{aff119}
\and
Centre for Information Technology, University of Groningen, P.O. Box 11044, 9700 CA Groningen, The Netherlands\label{aff120}
\and
Dipartimento di Fisica e Scienze della Terra, Universit\`a degli Studi di Ferrara, Via Giuseppe Saragat 1, 44122 Ferrara, Italy\label{aff121}
\and
Istituto Nazionale di Fisica Nucleare, Sezione di Ferrara, Via Giuseppe Saragat 1, 44122 Ferrara, Italy\label{aff122}
\and
ICSC - Centro Nazionale di Ricerca in High Performance Computing, Big Data e Quantum Computing, Via Magnanelli 2, Bologna, Italy\label{aff123}
\and
Astronomical Observatory of the Autonomous Region of the Aosta Valley (OAVdA), Loc. Lignan 39, I-11020, Nus (Aosta Valley), Italy\label{aff124}
\and
Department of Mathematical Physics, Institute of Physics, University of S\~ao Paulo, R. do Mat\~ao 1371, 05508-090, S\~ao Paulo, SP, Brazil\label{aff125}
\and
Centre de Recherche Astrophysique de Lyon, UMR5574, CNRS, Universit\'e Claude Bernard Lyon 1, ENS de Lyon, 69230, Saint-Genis-Laval, France\label{aff126}
\and
Oskar Klein Centre for Cosmoparticle Physics, Department of Physics, Stockholm University, Stockholm, SE-106 91, Sweden\label{aff127}
\and
Astrophysics Group, Blackett Laboratory, Imperial College London, London SW7 2AZ, UK\label{aff128}
\and
Univ. Grenoble Alpes, CNRS, Grenoble INP, LPSC-IN2P3, 53, Avenue des Martyrs, 38000 Grenoble, France\label{aff129}
\and
Dipartimento di Fisica, Sapienza Universit\`a di Roma, Piazzale Aldo Moro 2, 00185 Roma, Italy\label{aff130}
\and
Institute of Cosmology and Gravitation, University of Portsmouth, Portsmouth PO1 3FX, UK\label{aff131}
\and
Institut d'Astrophysique de Paris, 98bis Boulevard Arago, 75014, Paris, France\label{aff132}
\and
ICL, Junia, Universit\'e Catholique de Lille, LITL, 59000 Lille, France\label{aff133}
\and
Center for Astrophysics and Cosmology, University of Nova Gorica, Nova Gorica, Slovenia\label{aff134}
\and
MIT Kavli Institute for Astrophysics and Space Research, Massachusetts Institute of Technology, Cambridge, MA 02139, USA\label{aff135}
\and
Department of Astronomy, Indiana University, 727 East Third Street, Bloomington, IN 47405, USA\label{aff136}
\and
CERCA/ISO, Department of Physics, Case Western Reserve University, 10900 Euclid Avenue, Cleveland, OH 44106, USA\label{aff137}
\and
Laboratoire Univers et Th\'eorie, Observatoire de Paris, Universit\'e PSL, Universit\'e Paris Cit\'e, CNRS, 92190 Meudon, France\label{aff138}
\and
Departamento de F{\'\i}sica Fundamental. Universidad de Salamanca. Plaza de la Merced s/n. 37008 Salamanca, Spain\label{aff139}
\and
IRFU, CEA, Universit\'e Paris-Saclay 91191 Gif-sur-Yvette Cedex, France\label{aff140}
\and
Universiteit Gent, Department of Physics and Astronomy, Proeftuinstraat 86 N3, 9000 Ghent, Belgium
\label{aff141}
\and
Konkoly Observatory, HUN-REN CSFK, MTA Centre of Excellence, Budapest, Konkoly Thege Mikl\'os {\'u}t 15-17. H-1121, Hungary\label{aff142}
\and
MTA-CSFK Lend\"ulet Large-Scale Structure Research Group, Konkoly-Thege Mikl\'os \'ut 15-17, H-1121 Budapest, Hungary\label{aff143}
\and
Universit\'e de Strasbourg, CNRS, Observatoire astronomique de Strasbourg, UMR 7550, 67000 Strasbourg, France\label{aff144}
\and
Max-Planck-Institut f\"ur Physik, Boltzmannstr. 8, 85748 Garching, Germany\label{aff145}
\and
Scuola Superiore Meridionale, Via Mezzocannone 4, 80138, Napoli, Italy\label{aff146}
\and
Departamento de F\'{i}sica Te\'{o}rica, At\'{o}mica y \'{O}ptica, Universidad de Valladolid, 47011 Valladolid, Spain\label{aff147}
\and
Laboratory for Disruptive Interdisciplinary Science (LaDIS), Universidad de Valladolid, 47011 Valladolid, Spain\label{aff148}
\and
Instituto de Astrof\'isica e Ci\^encias do Espa\c{c}o, Faculdade de Ci\^encias, Universidade de Lisboa, Tapada da Ajuda, 1349-018 Lisboa, Portugal\label{aff149}
\and
Department of Physics, Astronomy and Mathematics, University of Hertfordshire, College Lane, Hatfield AL10 9AB, UK\label{aff150}
\and
Departamento F\'isica Aplicada, Universidad Polit\'ecnica de Cartagena, Campus Muralla del Mar, 30202 Cartagena, Murcia, Spain\label{aff151}
\and
Instituto de F\'isica de Cantabria, Edificio Juan Jord\'a, Avenida de los Castros, 39005 Santander, Spain\label{aff152}
\and
Observatorio Nacional, Rua General Jose Cristino, 77-Bairro Imperial de Sao Cristovao, Rio de Janeiro, 20921-400, Brazil\label{aff153}
\and
Univ. Grenoble Alpes, CNRS, IPAG, 38000 Grenoble, France\label{aff154}
\and
CEA Saclay, DFR/IRFU, Service d'Astrophysique, Bat. 709, 91191 Gif-sur-Yvette, France\label{aff155}
\and
Dipartimento di Fisica "E. Pancini", Universita degli Studi di Napoli Federico II, Via Cinthia 6, 80126, Napoli, Italy\label{aff156}
\and
INFN Sezione di Napoli, Universita degli Studi di Napoli "Federico II", Compl. Univ. di Monte S. Angelo, Edificio G, Via Cinthia, I-80126, Napoli, Italy\label{aff157}
\and
Department of Computer Science, Aalto University, PO Box 15400, Espoo, FI-00 076, Finland\label{aff158}
\and
School of Physics and Astronomy, University of Nottingham, University Park, Nottingham NG7 2RD, UK\label{aff159}
\and
University of Applied Sciences and Arts of Northwestern Switzerland, School of Computer Science, 5210 Windisch, Switzerland\label{aff160}
\and
Ruhr University Bochum, Faculty of Physics and Astronomy, Astronomical Institute (AIRUB), German Centre for Cosmological Lensing (GCCL), 44780 Bochum, Germany\label{aff161}
\and
DARK, Niels Bohr Institute, University of Copenhagen, Jagtvej 155, 2200 Copenhagen, Denmark\label{aff162}
\and
Department of Physics, Centre for Extragalactic Astronomy, Durham University, South Road, Durham, DH1 3LE, UK\label{aff163}
\and
Astrophysics Research Centre, University of KwaZulu-Natal, Westville Campus, Durban 4041, South Africa\label{aff164}
\and
School of Mathematics, Statistics \& Computer Science, University of KwaZulu-Natal, Westville Campus, Durban 4041, South Africa\label{aff165}
\and
Department of Physics and Astronomy, Vesilinnantie 5, University of Turku, 20014 Turku, Finland\label{aff166}
\and
Finnish Centre for Astronomy with ESO (FINCA), Quantum, Vesilinnantie 5, University of Turku, 20014 Turku, Finland\label{aff167}
\and
Serco for European Space Agency (ESA), Camino bajo del Castillo, s/n, Urbanizacion Villafranca del Castillo, Villanueva de la Ca\~nada, 28692 Madrid, Spain\label{aff168}
\and
Van Swinderen Institute, University of Groningen, Nijenborgh 3, 9747 AG Groningen, The Netherlands\label{aff169}
\and
Department of Physics and Astronomy, University of the Western Cape, Bellville, Cape Town, 7535, South Africa\label{aff170}
\and
Department of Physics and Helsinki Institute of Physics, Gustaf H\"allstr\"omin katu 2, University of Helsinki, 00014 Helsinki, Finland\label{aff171}
\and
Institute for Theoretical Particle Physics and Cosmology (TTK), RWTH Aachen University, 52056 Aachen, Germany\label{aff172}
\and
INAF-Osservatorio Astrofisico di Arcetri, Largo E. Fermi 5, 50125, Firenze, Italy\label{aff173}
\and
Institute for Theoretical Physics, ETH Zurich, Wolfgang-Pauli-Str. 27, 8093 Zurich, Switzerland\label{aff174}
\and
Centro de Astrof\'{\i}sica da Universidade do Porto, Rua das Estrelas, 4150-762 Porto, Portugal\label{aff175}
\and
Dipartimento di Fisica, Universit\`a di Roma Tor Vergata, Via della Ricerca Scientifica 1, Roma, Italy\label{aff176}
\and
INFN, Sezione di Roma 2, Via della Ricerca Scientifica 1, Roma, Italy\label{aff177}
\and
HE Space for European Space Agency (ESA), Camino bajo del Castillo, s/n, Urbanizacion Villafranca del Castillo, Villanueva de la Ca\~nada, 28692 Madrid, Spain\label{aff178}
\and
STAR Institute, University of Li{\`e}ge, Quartier Agora, All\'ee du six Ao\^ut 19c, 4000 Li\`ege, Belgium\label{aff179}
\and
Center for Frontier Science, Chiba University, 1-33 Yayoi-cho, Inage-ku, Chiba 263-8522, Japan\label{aff180}
\and
Department of Physics, Graduate School of Science, Chiba University, 1-33 Yayoi-Cho, Inage-Ku, Chiba 263-8522, Japan\label{aff181}
\and
Dipartimento di Fisica ``G. Occhialini", Universit\`a degli Studi di Milano Bicocca, Piazza della Scienza 3, 20126 Milano, Italy\label{aff182}
\and
School of Physical Sciences, The Open University, Milton Keynes, MK7 6AA, UK\label{aff183}
\and
Department of Astrophysics, University of Zurich, Winterthurerstrasse 190, 8057 Zurich, Switzerland\label{aff184}
\and
Space physics and astronomy research unit, University of Oulu, Pentti Kaiteran katu 1, FI-90014 Oulu, Finland\label{aff185}
\and
Department of Mathematics and Physics, Roma Tre University, Via della Vasca Navale 84, 00146 Rome, Italy\label{aff186}
\and
Theoretical astrophysics, Department of Physics and Astronomy, Uppsala University, Box 524, 751 20 Uppsala, Sweden\label{aff187}
\and
Minnesota Institute for Astrophysics, University of Minnesota, 116 Church St SE, Minneapolis, MN 55455, USA\label{aff188}
\and
Mathematical Institute, University of Leiden, Einsteinweg 55, 2333 CA Leiden, The Netherlands\label{aff189}
\and
School of Physics \& Astronomy, University of Southampton, Highfield Campus, Southampton SO17 1BJ, UK\label{aff190}
\and
Institute of Astronomy, University of Cambridge, Madingley Road, Cambridge CB3 0HA, UK\label{aff191}
\and
Department of Physics and Astronomy, University of California, Davis, CA 95616, USA\label{aff192}
\and
Kobayashi-Maskawa Institute for the Origin of Particles and the Universe, Nagoya University, Chikusa-ku, Nagoya, 464-8602, Japan\label{aff193}
\and
Department of Astrophysical Sciences, Peyton Hall, Princeton University, Princeton, NJ 08544, USA\label{aff194}
\and
Fakult\"at f\"ur Physik, Universit\"at Bielefeld, Postfach 100131, 33501 Bielefeld, Germany\label{aff195}
\and
Department of Physics and Astronomy, Lehman College of the CUNY, Bronx, NY 10468, USA\label{aff196}
\and
American Museum of Natural History, Department of Astrophysics, New York, NY 10024, USA\label{aff197}
\and
SRON Netherlands Institute for Space Research, Landleven 12, 9747 AD, Groningen, The Netherlands\label{aff198}}    
   
%
%
\abstract{
\Euclid is a surveying mission that is progressively imaging and spectroscopically mapping a wide area of extragalactic sky in visible and near infrared bands. The \Euclid 
mission is, however, not restricted to the Euclid Wide Survey, but it includes data from various other survey components. This paper presents the Data Release~1 (DR1) status of the multiple \Euclid surveys and fields, including the Euclid Deep Survey, Euclid Auxiliary Fields, calibration fields, the Euclid Medium-Deep Survey, the Euclid Ecliptic Survey, and the Independent Legacy Scientists' field, in addition to the Euclid Wide Survey. Some of these fields were observed during the initial Performance-Verification phase, others during the Phase-Diversity calibration campaign, and the majority during the subsequent science operations phase. This paper describes the different procedures of survey implementation applied in the various phases of the mission, from the set of requirements, constraints, and limitations, to the final operational products used to operate the spacecraft and execute the science surveys. The resulting main component of the \Euclid surveys, the Euclid Wide Survey, covers at DR1 an area of 2100$\,{\rm deg}^2$, in two separate regions, of which 1900$\,{\rm deg}^2$ corresponds to data validated in higher-level processing.}
%
%
%
%

  \titlerunning{Euclid DR1: The \Euclid surveys}
   \authorrunning{Euclid Collaboration: I.~Tereno et al.}
   
   \maketitle
%
%
%
%
   
\section{\label{sc:Intro}Introduction}

\Euclid \citep{EuclidSkyOverview} is a survey mission. One of its fundamental requirements is to carry out the \ac{EWS} to conduct its two primary scientific experiments, galaxy clustering and weak gravitational lensing, covering at least 14\,000\,deg$^2$ of the extragalactic sky. \ac{EWS} observations started on 14 February 2024 and will continue until the nominal end of the mission on 31 December 2029. At the same time, \Euclid also performs the \ac{EDS}, consisting of repeated observations of the three \acp{EDF}, namely the North (EDF-N), South (EDF-S) and Fornax (EDF-F) fields. The EDS will accumulate greater depth than that of the \ac{EWS}, up to a gain of about 2 magnitudes. 
It also observes six so-called auxiliary fields, 
supporting instrument calibration and characterising the source population, and acquires data on specific fields at specific cadences to perform calibration activities necessary to process and analyse the \ac{EWS} data. The observations of these four components of the \Euclid survey, namely the \ac{EWS}, the  \ac{EDS}, the \acp{EAF}, and calibrations,  are scheduled in a time-ordered sequence of pointings and spacecraft slews, known as a \ac{RSD}.

The \Euclid  survey is however not restricted to the \ac{RSD} components. During the period of 7.5 months between launch on 1 July 2023 and the start of the \ac{RSD} execution, other observations were made over 3 phases, namely the spacecraft commissioning, the \ac{PV} phase, and the \ac{PDC} campaign. The commissioning of the \Euclid spacecraft was completed on 4 August 2023. It was followed by the \ac{PV} phase \citep{DR1-TP026} that lasted until the beginning of December 2023, longer than originally planned to include investigation, testing, and validation activities associated with spacecraft anomalies detected during commissioning \citep{Laureijs2024}. 
The \ac{PDC} campaign, started on 3 December 2023, is a first component of the strategy to build and calibrate a \ac{PSF} model (Euclid Collaboration: Miller et al. in prep.), and data were required to be taken under very stable thermal conditions. It was therefore necessary to allocate thermal stabilisation periods during which the telescope kept a fixed solar attitude. Three scientific filler programmes were executed during these periods to make the best possible use of \Euclid time. These observations resulted in three additional \Euclid surveys, namely the Euclid Medium-Deep survey, the \ac{ILS} field, and the Euclid ecliptic survey.

The \ac{DR1} of the \Euclid mission contains data from the following six survey components, or `\Euclid surveys', with data acquired over three periods (PV, PDC, RSD).

\begin{itemize}
\item Euclid Wide Survey (RSD);
\item Euclid Deep Survey (RSD, PV);
\item Euclid Auxiliary fields (RSD, PV);
\item Calibration fields (RSD, PV, PDC);
\item PV science fields (PV);
\item PDC surveys (PDC).
\end{itemize}

The contents of DR1 \citep{DR1cite} are 
defined in a memo\footnote{\url{https://www.cosmos.esa.int/documents/10647/12245842/DR1-memo\_260922.pdf/3d879dc6-d5e0-9125-54a7-f68d1b4971d8?t=1790090418174}} approved by the \ac{EST}
and are detailed in the \ac{DR1} overview paper \citep{DR1-TP001}. DR1 includes all 
complete Level~1 (LE1) data acquired since launch until 21 May 2025. Complete LE1 data are \Euclid raw data generated by ESA's \Euclid \ac{SOC} and ingested in the Data Processing System. LE1 products are fully processed into Level~2 (LE2) products by the various \ac{SGS} processing functions (PF) of the Euclid Consortium, as described in the SGS overview paper \citep{DR1-TP033} and in the PF-specific DR1 processing papers describing VIS \citep{DR1-TP002}, NIR \citep{DR1-TP003}, EXT \citep{DR1-TP006}, MER \citep{DR1-TP009}, SIR \citep{DR1-TP023}, SPE \citep{DR1-TP025} and PHZ \citep{DR1-TP011}. We note that not all scheduled observations were successfully executed or correspond to complete LE1 data products, and not all LE1 data originated valid and complete LE2 products. There is thus a reduction in coverage from scheduled observations to LE1 data, and from LE1 to LE2 data. These differences will be mentioned throughout the paper when appropriate.

For completeness, it is relevant to mention that there is one additional component of the \Euclid survey that is not part of DR1 or described in this paper. It is the \ac{EGBS} performed on 23 and 24 March 2025, which forms the content of the Quick Data Release~2 (Q2) and is presented in \cite{Q2-TP002}. On the other hand, the Quick Data Release~1 (Q1) data, consisting of one visit to each of the \acp{EDF}, are included in DR1, as are the data from the 17 fields observed in the PV phase for the six science programmes \citep{EROData} that were the subject of \Euclid’s \ac{ERO} data release. We also note that data from Commissioning are not part of DR1 because they lack the metadata required to generate LE1 complete data products.

The \ac{RSD} is computed with the dedicated \ectile software by the \ac{SOST}, a Euclid Consortium group created for the mission's operations phase that inherited the responsibilities of the ECSURV group that was active during the mission's development phase.
In the development phase, prior to launch, a large number of \acp{RSD} were produced to evaluate survey scenarios and implementation strategies, against the mission  requirements. During operations, the survey is re-planned a few times per year, with the computation of a new \ac{RSD}. The \acp{RSD} follow a naming convention consisting of the year its first new observations are executed and an alphabetical sequence that runs through each calendar year. \Euclid's \ac{DR1} RSD data were acquired with observations scheduled in \acp{RSD}~2024A, 2024B, 2024C, and 2025A. Each \ac{RSD} is always computed from the starting date of 14 February 2024 to the nominal end of the mission, with each version being a copy of the previous one up to a restart point; therefore \ac{RSD}~2025A contains the relevant information for DR1. 

The \ac{RSD} is delivered to \Euclid's \ac{SOC} that generates the \ac{OSS} and the operational products required for the execution of the survey by \Euclid's \ac{MOC}. In the other phases of operations, the \ac{PV} phase and the \ac{PDC} campaign, the scheduling was built by \ac{SOC} using a different set of tools, and a manual assembly of calibration blocks.

This paper proceeds in \cref{sc:surveystrategy} with a presentation of the process of \ac{RSD} computation with the \ectile software,
describing relevant inputs and constraints and addressing optimisation strategies and re-planning methods. The resulting DR1 \ac{RSD}~2025A scheduled survey properties of the \ac{EWS}, \acp{EAF}, \acp{EDF}, and calibration fields are presented in 
\cref{sc:surveyresults}, where we  also detail deviations of the DR1 content from the scheduled one. \Cref{sc:socsurvey} describes the \ac{SOC} procedures to generate an \ac{OSS} from the \ac{RSD} and the production of the operational products sent to \Euclid's \ac{MOC} for the execution of the survey.
The implementation of the \ac{PV} phase observations and a list of the resulting PV fields is presented in \cref{sc:pv}. The 
\ac{PDC} campaign is addressed in \cref{sc:PDC}, 
with a description of the resulting calibration fields and of the three surveys performed during its thermal stabilisation periods.
We conclude in \cref{sc:conclusion} with a summary table of the DR1 \Euclid surveys and a full-sky map showing their locations.
The appendices include exposure maps of the DR1 observations of the \acp{EDF}, \acp{EAF}, main calibration fields, and PDC surveys. A list of acronyms used in the paper with a glossary of terms is also provided in the Appendix. 

\section{\label{sc:surveystrategy} RSD computation }

The \ac{RSD} is built using the \ectile software. \ectile contains a set of tools developed to design, implement and analyse \Euclid surveys. The tools relevant for the present description are three: {\tt plan}, {\tt wide}, and {\tt surv}. The computation of an \ac{RSD} is carried out in three stages \citep{Scaramella-EP1}. In a preparatory Stage~0, a set of input parameters is defined. After that, in Stage~1, a skeleton of the schedule of the observations of the calibration fields, \acp{EDF}, and \acp{EAF} is manually set and their exact observing dates are computed by \ectile. The locations of the \acp{EDF} and \acp{EAF} were carefully selected before launch \cite{EuclidSkyOverview}. The number of visits, i.e., the number of passes over the full field, needed to reach the required depth depends on the zodiacal background at its ecliptic latitude \citep{Scaramella-EP1}. 
The core of the procedure is Stage~2, where the \ac{EWS} is computed in a predefined tessellation.

\begin{figure*}
\includegraphics[width=1.0\linewidth]{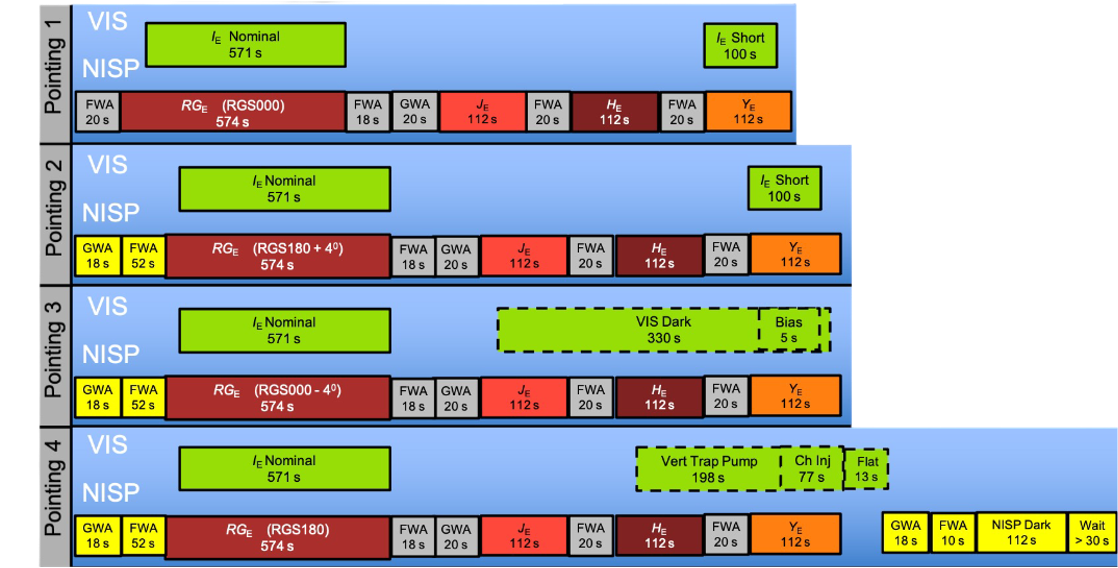}
\caption{ROS overview. Each of the four panels shows the timelines of the VIS and NISP activities that are carried out during one dither. VIS exposures are shown in green and NISP ones in shades of red. The times given for filter wheel (FWA) and grism wheel (GWA) include stabilisation times. The times given in VIS boxes include overheads such as shutter motion, flush, and surface inversion. Dashed boxes cycle through the various VIS inline calibrations indicated.
Yellow indicate that the activities are performed during slews. The final box labelled `Wait' has a variable time, depending on the duration of the slew to the next field. Readout, data
compression and transfer activities are not included in the figure.}
\label{fig:ros}
\end{figure*}

\subsection{\label{sc:ROS} RSD observing sequences}

Each \ac{EWS}, \ac{EDF}, and \ac{EAF} observation is made with the \Euclid \ac{ROS}, depicted in Fig.~\ref{fig:ros}, and contains four dithered pointings (usually called `dithers'). \Euclid performs, in each dither, a VIS exposure \citep{EuclidSkyVIS} in parallel with \ac{NISP} spectroscopy \citep{EuclidSkyNISP} and proceeds with a sequence of observations in the three \ac{NISP} photometric bands. In each dither, the spectroscopic exposure is taken with a different rotation of one of the two \acp{RGS}, RGS000 and RGS180. The first and second dithers also include a \ac{VIS} short science exposure during the NISP \YE-band observation. Prior to each NISP observation, the filter wheel (FWA) or grism wheel (GWA) assemblies are actuated and time is allocated for stabilisation at the end of the movement. In the second, third, and fourth pointings, the first activation of the wheels is done during the dither slew, allowing for a larger stabilisation time than in the first pointing. The last dither is longer, ending with a NISP dark exposure taken during the slew to the next field. 
The third and fourth dithers contain time allocated for VIS inline calibrations, both during NISP observations and during the final field slew. These calibrations are implemented as a fixed sequence over a 20-ROS block, consisting of alternating darks and biases (in the third dither), and alternating flats, charge injection, and vertical trap pumping (in the fourth dither). In a 20-ROS block, six are allocated for flats, six for vertical trap pumping, and eight for charge injection.
Each of these calibration types then cycles over different parameter settings over a longer baseline, such that a coherent set of calibration products is generated over approximately 30 days of ROS execution. The parameter settings and the calibration types are scheduled in a way that similar cases are  spread out over the 30 days, to be resilient against the loss of consecutive observations.

The duration of the first three dithers is 996\,s, excluding the 20\,s lead time and the 70\,s dither slews, while the fourth dither is 1006\,s long. The ROS thus lasts 4204\,s, excluding lead time, dither slews and the final field slew that has a minimum duration of 170\,s. This implies that \Euclid makes roughly 19 ROS observations per day.

Besides the ROS, there is also a `ROS-blue' variant, where the \ac{BGS}, BGS000 is used in all four dithers and always in the nominal position. This variant is used in some of the visits to the \acp{EDF} and \acp{EAF}.

Calibration observations are made with special non-ROS sequences defined \citep[see][]{DR1-TP026} in \Euclid's \ac{CalF}. Those sequences are in general very different from the ROS, one exception being the ROS-D1 sequence that consists of the first dither of the ROS.

\subsection{\label{sc:surveystrategy0} Stage~0: Configuration file}

A long set of input parameters describing system constraints, targets properties, and survey optimisation strategy is defined in the configuration file {\tt {ectile.cfg}}. The main inputs are the following.
\begin{itemize}
\item \ac{RSD} start date and duration.
\item Orbit file: \Euclid's orbit around L2 provided by the \ac{SOC}. The orbit, i.e., the set of future positions of the \Euclid satellite, is regularly updated as the orbit drifts and station-keeping maneuvers are performed.
\item Slew thresholds and durations: there are three different types of slews (dither, field, and large) depending on the total rotation executed between two pointings. The total rotation, also known as `eigen-slew' includes the angular distance between the pointings and the change of telescope attitude from the initial to the final position. 
The threshold values from dither to field slews and from field to large slews are $130\farcs0$ and $\ang{3.5;;}$, respectively, with the caveat that for specific calibrations field slews are allowed to be as small as $73\farcs0$.
The duration of a dither slew is 70\,s, the field slews duration vary from 170\,s to 283\,s and the duration of a large slew varies from 5.8 minutes to around 34 minutes for eigen-slews of $\ang{180;;}$.
\item Dither pattern: the geometry of the \ac{ROS} follows the shape of an `S’, with defined angular step sizes \citep{Scaramella-EP1} and a $\ang{180;;}$ 
symmetry to ensure the same footprint whether an observation is made looking forward or backwards with respect to the direction of the orbital movement.
\item Pointing constraints: spacecraft 
manoeuvrability is conditioned by two parameters, \ac{AA} and \ac{SAA}. For a given time of observation, these two angles uniquely define the spacecraft attitude. Roughly, \ac{AA} defines a rotation around the telescope boresight and \ac{SAA} defines the angle between the telescope boresight and the Sun. 
\Euclid is allowed to point away from the transit meridian by tilting up to $\ang{3;;}$ towards the Sun and up to $\ang{30;;}$ away from the Sun, maximising thermo-optical stability; that is, the \ac{SAA} range is restricted to $[\ang{87;;},\ang{120;;}]$. The \ac{AA} is restricted to the asymmetric range $[\ang{-2.9;;},\ang{-8.5;;}]$, only containing negative values, which is a significant reduction of the originally intended configuration. This modification is needed  to mitigate solar stray light found during commissioning; its impact on the survey is appreciable, leading to a redesign of the survey strategy, as discussed in \citet[][see also the tessellation item below]{EuclidSkyOverview}.
\item Region of Interest (RoI): the \ac{EWS} is implemented within a \ac{RoI}, defined by the intersection of threshold lines of three sky backgrounds, namely zodiacal light, Galactic extinction, and stellar density. There are a few custom-made carvings on the thresholds to exclude the vicinity of the Galactic bulge, the \ac{LMC}, the Taurus/Orion clouds, and a circle of $\ang{10;;}$ radius around the northern equatorial pole that is out of reach of the complementary ground-based telescopes. The \ac{RoI} must ensure that the required average signal-to-noise ratios are achieved in all \Euclid bands. The current \Euclid \ac{RoI}, introduced for \ac{RSD}~2023A, has an area of \num{16200}\,$\deg^2$ distributed in four quadrants (in the shape of two large regions known as the `mainlands' and two small regions dubbed the `islands').
\item Field of view (FoV): the geometries of the FoVs of the 
\ac{VIS} and \ac{NISP} are off-centred and do not intersect each other completely, leaving a \ac{NISP} gap \citep{EuclidSkyOverview}.  
The angular extent of both FoVs and their corresponding physical dimensions in the focal plane, including the total size of their sensors (i.e,  the mosaic of detectors), the size of the detectors (CCDs), and the size of the inter-detector gaps, are combined to compute a spherical rectangle observing the common FoV (rectangular area observed by both instruments). The common FoV has an area of $0.525\,{\rm deg}^2$. 
\item Tessellation: the tessellation covers the \ac{RoI} with tiles, small latitude-longitude rectangles, forming one compact covering per quadrant, leaving holes only in the bright stars exclusion regions. Each tile is sized to be fully observed by a FoV. For this, the tile is constructed to be larger than the common FoV, including the NISP angular gap sizes and a safety margin slightly larger than the absolute pointing error. The tile has an area of $0.551\,{\rm deg}^2$. 
Due to the convergence towards the poles, the number of tiles per row decreases with latitude, and moreover they start to acquire some overlap. The natural overlap of the tessellation is around $4\,\%$. One \ac{ROS} observation of the \ac{EWS} corresponds to placing the \ac{VIS} and \ac{NISP} FoVs on one tile of the tessellated sky and moving it along the dither S-pattern. However, the modified \ac{AA} range introduced to avoid stray light forces us to observe with the telescope rotated from its natural position of ${\rm AA}=\ang{0;;}$, resulting in a FoV on the sky with a different orientation than the underlying tile. From a given observing position in the orbit, there is only a very reduced area of the sky where an AA counter-rotation can be applied within the allowed range to align the FoV with the tessellation, making it impossible to build a survey (Sect.~\ref{sc:surveystrategy2}). The solution found to solve this potential show-stopper was to accept tilted observations up to a specified threshold, as a new survey strategy. The maximum misalignment tolerated between tile and FoV position angles is a compromise between visibility and field-to-field overlap. We thus redesigned the tessellation to allow for a maximum FoV-tile misalignment of $\ang{3.5;;}$. The main impact in the tessellation is that the tiles need to be  placed closer together to avoid gaps due to the tilt. Through an analysis of the number of counts at pixel level, we determined the optimal distance between the centres of the tessellation tiles. The resulting tessellation has a mean tile overlap of $14\,\%$, a significant increase from the original $4\,\%$, which causes a significant decrease in surveying efficiency and therefore in the total area at the nominal end of the mission (although it also means that the data are generally a little deeper than planned). This tessellation was first used in \ac{RSD}~2024A.
\item Polar caps: due to the tessellation convergence at the ecliptic poles, high-latitude tiles cannot be observed in a way that is compliant with the pointing constraints. Local tessellations are thus introduced in the polar caps, defined by the two thresholding ecliptic latitudes of $\pm \ang{77.8;;}$. The two polar caps are covered with 22 sectors, each with its own local tessellation. 
\item Bright stars avoidance: a catalogue of stars brighter than 4\,AB magnitudes in any of the \Euclid photometric bands, together with exclusion radii to be applied to avoid harming the detectors, and information on proper motions.
\item Planet avoidance: observations must avoid pointing too close to the outer planets. Angular avoidance radii are specified in the mission requirements.
\item Observing timings: defined for all elements of each observing sequence, including the \ac{ROS} and special sequences for calibration fields. 
\item Targets table: including, for each field, the location, position angle, pattern geometry, number of visits, type of pointing smearing applied between visits, ratio of blue/red grism visits and observing sequence.
\item Scheduling optimisation: several parameters controlling various aspects of the scheduling, such as patch margins, \ac{RoI} clipping, quadrant corners, or patch selector, can be defined (these parameters will be described in Sect.~\ref{sc:surveystrategy2}).
\item Ground-based surveys: the footprints of the surveys providing Euclid DR1 complementary data, which are the \ac{DES} in the southern sky  \citep{Abbott:2021}  and the \ac{UNIONS} in the north \citep {Gwyn:2025}. 
\item Special operations (SOPS): periodic interruptions of the survey observations are scheduled to enable non-survey operational activities, such as orbit adjustment, reaction-wheel momentum offloading, or instrument maintenance.
SOPS used to require 12 hours, but in \ac{RSD}~2025A, applicable to DR1, the duration was reduced to a 6 hour time-slot for SOPS every 4 weeks on Mondays at 12:00\,UTC (within a margin of 2\,h). More recently, the cadence decreased and \ac{RSD}~2026A schedules the SOPS 6 hour interruptions only every 8 weeks on Wednesdays at 15\,h\,UTC (with a $-2$\,h, $+11$\,h margin).
\end{itemize}

\begin{figure}
\includegraphics[width=\columnwidth]{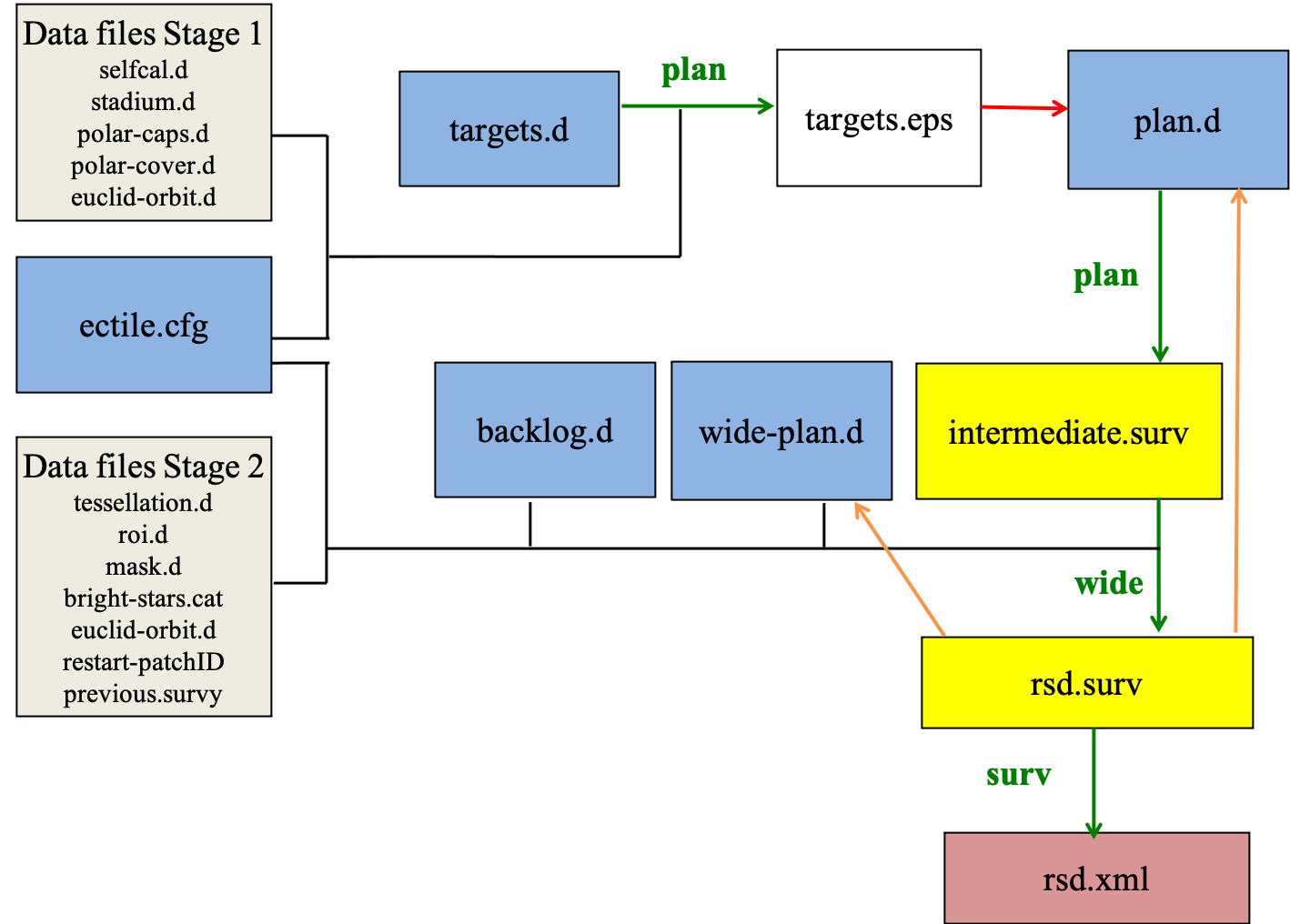}
\caption{Schematic representation of the \ectile pipeline for computing an \ac{RSD}, showing the main tools and files. The configuration file and input data needed in Stage 1 and Stage 2 are shown in the left. The procedure flows with the green arrows. An \ectile tool named in green next to an arrow is applied to an input file, and uses information from the configuration and data files connected with black lines to produce an output. Output binary files are shown in yellow and output plots in white. A red arrow indicates that the output is manually constructed using information of the previous file and not automatically computed by an \ectile tool. Blue boxes are the five core input files that are manually modified by the mission planner and are the stepping stones of the \ac{RSD} computation. The orange arrows indicate the optimisation loop: the monitoring on-screen of the step by step computation of the \ac{RSD} guides the mission planner in the manual modification of the files {\tt plan.d} and {\tt wide-plan.d}. At the end of the process, the final \ac{RSD} file is generated in \ac{XML} format (brown box). See text for a description.}
\label{fig:ectile_diagram}
\end{figure}

\subsection{\label{sc:surveystrategy1} Stage~1: Scheduling targets}

The next step in computing a reference survey is to design a schedule for all observations that are not part of the regular \ac{EWS}. These include all the \ac{EDF} and \ac{EAF} observations, all of the calibrations with the exception of the PSF block defined below, and all the polar cap sectors. The detailed description of the process given in \cite{Scaramella-EP1} is still valid, with the caveat that there have been a few modifications in calibration specifications and the introduction of  new calibrations.

The complete overview of the DR1 calibrations is given in \cite{DR1-TP026}. It contains 13 calibration observation blocks, labelled CALBLOCK-F-001 and 007 through 018. Each calibration block fulfills various calibration products for one or both instruments. F-001 (self-calibration) is observed with a single-dither ROS (ROS-D1), while the others are observed with special sequences. Observations with similar cadence requirements are scheduled together, forming monthly blocks (F-001, F-018, F-007), biannual blocks (F-009, F-016, F-011), and PSF blocks (F-012, F-015, F-013, F-008). 
CALBLOCK-F-002 through F-006 are reserved for ROS observations: F-002 for \acp{EDF}, F-003 for the \acp{CPC} are made with observations in the EDF-N and EDF-S at fixed cadences to obtain a range of observing directions for the spectroscopic data; F-004 and 005 for observations of the \acp{EAF} used, respectively, for colour gradient and photometric redshift calibrations; and F-006 for the \ac{EWS}.

Stage~1 is a process that is mostly manual and is done in two steps. Summarising the description of \cite{Scaramella-EP1} and referring also to Fig.~\ref{fig:ectile_diagram}, in the first step composite targets are assembled by combining one or more targets. The list of composite targets is stored in the file {\tt targets.d}. The composite targets defined in this file form the set of all possibilities that will be available for scheduling and their definition is an optimisation choice. The \ectile tool {\tt plan} computes a visual chart ({\tt targets.eps}) of the visibility window of each composite target given as an interval of the Sun's ecliptic longitude, showing also the duration of each observation within the interval. From this information the mission planner moves to the second step, manually constructing an observing plan ({\tt plan.d}), known as the calibration plan, by placing each composite target in six ordered sequences (one per year) according to the Sun's longitude of their visibility windows. As an optimisation strategy, various composite targets are grouped together in a block to be observed back to back. In this case, one of the targets is assigned the role of anchor and the observing longitudes of the neighbouring targets are shifted (within their visibility windows) to ensure no time gaps exist between the targets of a block. The placement of the composite targets is a core aspect of strategic planning of an \ac{RSD}. It is made with two goals in mind: (i) respecting requirements on cadences and priorities; and (ii) building a plan that enables an optimal implementation of the \ac{EWS} in the next stage, avoiding leaving survey windows that are too small and ensuring that the \ac{EWS} coverage in the first year (where many calibrations need to be scheduled) is capable of reaching the expected area for DR1. The calibration plan is best populated iteratively, adding the highest priority first and then adding the rest in decreasing order of difficulty, clustering them whenever possible. Here, `difficulty' refers to the width and availability of the visibility windows. The most difficult targets have a single possible visibility window. Most targets have two visibility windows but, in some cases, these can be very short, hindering their clustering. The plan is then executed by the \ectile tool plan that generates the intermediate survey file ({\tt intermediate.surv}), where the longitudes are converted to time assuming the given starting date. The result does not follow {\tt plan.d} strictly but performs optimisation steps by automatically making small shifts in time and merging neighbouring blocks of targets according to optimisation parameters defined in {\tt ectile.cfg}.

\subsection{\label{sc:surveystrategy2} Stage~2: Computing the EWS}

The survey windows available to schedule the observations of the \ac{EWS} correspond to the time intervals left unscheduled in Stage~1 between two consecutive target observations. The process of constructing the \ac{EWS} schedule is described in detail in \cite{Scaramella-EP1}. In the following, we provide an overview of the process, also referring to Fig.~\ref{fig:ectile_diagram}. 

The method proceeds by gradually scheduling each survey window in chronological order. The \ectile tool {\tt wide} takes the {\tt intermediate.surv} file as input and computes the 
visibility windows for all  orbital positions of the spacecraft during the dates of the first survey window.
The set of visibility windows shows the tiles that are within reach in a given survey window. 
The limitations on the range of the two solar angles, SAA and AA, together with the need to keep the fields aligned with the tessellation (within the tilting margin, cf. Sect.~\ref{sc:surveystrategy0}) determine the shape of the visibility window. 

\begin{figure*}
    \includegraphics[width=0.3\textwidth]{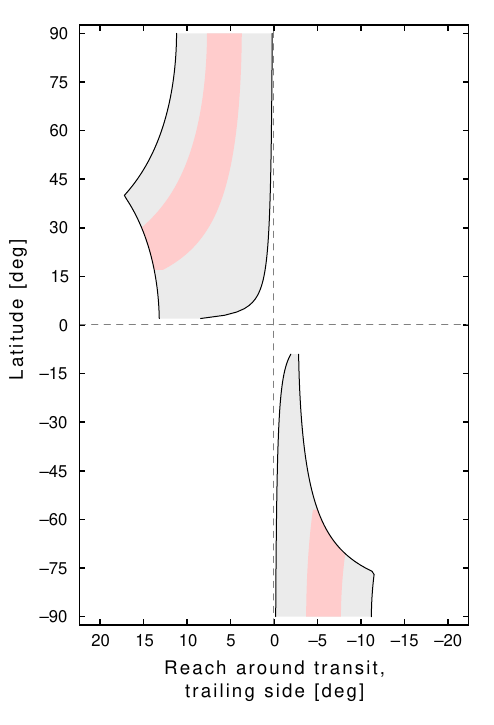}
    \includegraphics[width=0.3\textwidth]{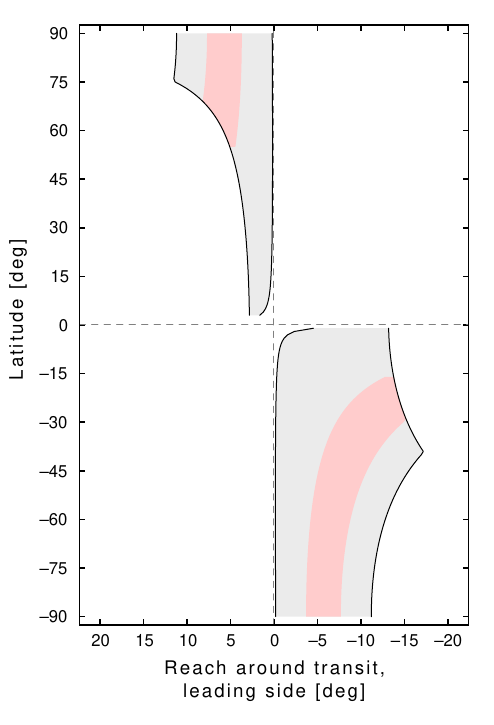} 
    \hfill
    \raisebox{60pt}{
    \begin{minipage}[b]{0.35\textwidth}
        \caption{Visibility windows for trailing ({\it left panel}) and leading ({\it right panel}) observations from any point of the orbit, computed for the SAA and AA allowed ranges and a maximum misalignment of $\ang{3.5}$ between the observed fields and the tessellation. The leading and trailing transit meridians are separated by $\ang{180;;}$.
        The inner pink bands show the visibility windows computed for the same SAA and AA ranges but with no tilt allowed, illustrating the unfeasibility of computing a wide survey in those conditions. 
        }
        \label{fig:viswindow}
    \end{minipage}
    }
\end{figure*}


Figure~\ref{fig:viswindow} shows the 
visibility window, represented as the reach in ecliptic longitude around the transit ecliptic meridian as a function of ecliptic latitude. It is valid for any point in the orbit, with the caveat that the transit meridians vary with orbital movement.
We note that since the AA range no longer includes zero, the transit meridian is barely within reach and observations with ${\rm SAA}=90\,^\circ$ are no longer the norm as they were in the original pre-launch designed survey strategy. 
From any point in the orbit, there are more precisely two visibility windows in the sky:
\begin{itemize}
    \item leading, when observing in the leading direction, i.e., when looking forward in the direction of the orbital movement;
    \item trailing, when looking in the trailing direction, i.e., looking back with respect to the orbital movement to the antipodal side of the sky and with the telescope flipped around its axis. 
\end{itemize} 
Given the asymmetries of the SAA and AA allowed ranges, each of the two visibility windows has two distinct sections, one in the northern and the other in the southern hemisphere. The broader section, indicating longer visibility, is in the north when looking in the trailing direction and in the south if looking in the leading direction. Given the shape of the RoI, only one of the two sections of each visibility window lies on a mainland. From these considerations it follows that if at a period of time the broad section of the leading window lies in a mainland, the broad section of the trailing window necessarily lies in the other mainland. This defines a so-called summer season where the visibility is longer and it is easier to schedule EWS observations. In contrast, there is another period of the year in which the two narrow sections of both visibility windows lie on the mainlands. This epoch is dubbed the winter season, where visibility is simultaneously short on both mainlands and the progress of the EWS is therefore difficult. 

Figure~\ref{fig:viswindow} also shows the 
visibility windows in the case where misalignemnts between FoVs and tiles are not allowed. In this case, the only regions reachable are the ones where it is possible to use the allowed range of AA to counter-rotate a misaligned FoV, such as to align it with the tessellation. As shown in the figure, those regions are extremely narrow. Conversely, allowing for larger misalignements between the observed fields and the tessellation would result in larger visibilities. 
In the extreme case of having no tessellation, i.e., observing with unconstrained position angles, as is the case of most stage~1 observations, visibility is solely determined by the SAA range, broadening towards the poles where visibility is perennial. The drawback of such approach would be to create gaps between observed fields, which would require a procedure to be filled in, either with additional observations or by placing the centres of the tiles closer to each other, in any case leading to a decrease in survey efficiency. The optimal value of maximum misalignment is the one that, even though introducing a highly constrained visibility, it still allows us to observe fields all year round and throughout the RoI, with  minimal field overlap such as to keep the survey efficiency large enough to enable the EWS to reach an area close to the requirement in the nominal mission duration.

Proceeding now with the description of the scheduling process, for the first survey window, and as a function of its duration, \ectile generates a set of patches 
that are possible to observe within the available time, a patch being a connected set of tiles of the tessellation. Usually, it generates at least one patch from each mainland and sometimes also from the islands. One of the patches is subsequently selected, and the others are discarded. The generation and selection of patches are done automatically according to a hard-coded heuristic. 
\begin{itemize}
\item Priority to higher latitudes: promoting a contiguous build-up of the \ac{EWS} layer by layer from the ecliptic poles (where the sky quality is higher) towards the equator, trying to complete two layers (one per hemisphere) longitudinally in roughly 1 year before moving to the next one.
\item Minimising solar angle variations: this conditions the height of a patch, which depends on the speed of the orbit and duration of the ROS.
\item Priority to the mainlands over the islands: promoting larger contiguous coverage.
\item Preference for areas covered by \ac{DES} and \ac{UNIONS} footprints:
DR1 \acp{RSD} prioritise southern versus northern hemisphere due to earlier availability of DES data.
\end{itemize}

Once the patches are selected for a survey window, a search is made to find PSF calibration fields within reach from those patches. If the search is successful, an observation of a PSF field (plus the other calibrations associated in the PSF block) is scheduled within the survey window, with SAA and AA values matching the ones of the contemporary \ac{EWS} patch. The purpose of scheduling the PSF fields in stage~2 and not in stage~1 together with the other calibrations is to ensure matching solar angles, since the PSF varies with the thermal conditions.
SOPS interruptions are also scheduled at this stage. These various placements use some of the time of the survey window, hence leaving less time for the scheduling of the patch and, thus, the patch originally selected is reduced by a few tiles. The final list of tiles within the patch are geometrically connected by the scheduling algorithm, known as the diffusion algorithm \citep{Scaramella-EP1}, and time stamps are assigned to each tile taking into account the starting time and the slew time. The tiles are observed in a sequence that minimises solar angle variations. 
The process continues to the next survey window; a \ac{RSD} contains around 100 survey windows. Given that \Euclid observes roughly 19 ROS tiles per day, a single patch in a typical survey window of 10 days contains up to around 200 tiles.

In addition to the patch generation and selection heuristics, there are other 
parameters, with default values set in {\tt ectile.cfg}, that control the many aspects of scheduling, with the goal of minimising the risk of the diffusion algorithm to fail. These parameters control the following four mechanisms.
\begin{itemize}
\item 
RoI clipping: at the start of the process some areas of the \ac{RoI} are masked with rectangles, effectively reducing its footprint, to help the diffusion algorithm with a smoother  input geometry. As the \ac{EWS} builds, it is sometimes useful to mask out additional areas to steer the patch placement towards certain areas.
\item
Quadrant corners: the diffusion algorithm fails if the shapes of the patches are not approximately rectangles. To enforce this shape, patches near the corners of the \ac{RoI} quadrants are partitioned into segments. The coordinates of the relevant points, both at the bottom and at the sides of the \ac{RoI} quadrants, are stored for this purpose. As the \ac{EWS} builds up and more patches are placed, new sharp corners may appear in the evolving top layer of the footprint. If they are found to cause a scheduling failure, their coordinates can be added to the list of quadrant corners to enforce patch partitioning at the new critical points. 
\item
Patch selection: in the first survey windows, there is no valid reason to choose one patch over the other.  However, later on, the heuristics for choosing the best patch are not so effective, and it is useful to have a mechanism to override the default selection.
\item 
Shrinking margins: to embed some scheduling margin into a generated  patch, its allocated time is decreased by some margin. Default margins are applied to the individual partitions of a patch and also to the total time extent of the patch.
\end{itemize}

At each iteration, the evolving survey map is updated on-screen, adding the resulting scheduled patches of the current iteration. Often, patches show imperfections. Common types of failures are the generation of patches that do not use the full time available in the survey window, creation of patches with irregular shapes that will be difficult to fit with the geometry of neighbouring patches, or the existence of a small set of tiles left unscheduled between two patches. 
The mission planner can  try to fix the issues by changing the default values of the scheduling parameters for the problematic iterations.  The general strategy is to make a full run of the process over the whole survey windows, analyse it to identify iterations where the default choices were not the best, and act at two levels.
\begin{itemize}
\item Customised patch selection. The patch selection of a given survey window can be overridden by selecting a patch from a different quadrant. The new selection is defined in the file {\tt wide-plan.d}. 
\item Customised shrinking margins. It is possible to define alternate specific shrinking margins for selected iterations in {\tt wide-plan.d}. Shrinking margins are specified as a duration, later converted to units of longitude and lead to a modification in the patch width and height, or in a longitudinal shift of its location.
\end{itemize}

Sometimes, these mechanisms are not capable of fixing the issues. In those cases, an alternative is to change the size of a problematic survey window. For this, the mission planner goes back to {\tt plan.d} and shifts a target, either by slightly delaying a calibration block or by a major edit of the calibration plan. This proves to be an effective way to avoid the presence of vertical and narrow gaps of unobserved tiles. Another useful approach is to use the fact that EDF-N has perennial visibility, to place its visits in the observing plan where they conflict the least with the \ac{EWS} survey windows. This strategy was applied in the DR1 \acp{RSD}, but it will not be available in future \acp{RSD} since the EDF-N visits are required to be finished by DR2.
We note that a modification in {\tt plan.d} implies restarting the process from scratch. 

In summary, the computation of the \ac{EWS} is a laborious and time consuming procedure. A single run of the {\tt wide} tool is relatively fast and fully automated. However, a single run is not enough. Most often, the result  contains many shortcomings or low performance. The survey windows can then be optimised using the various mechanisms described above. The optimisation must be done in chronological order, since a change in one iteration may potentially change any of the subsequent windows. A new run must hence be made after each modification, until the final 
survey window is optimised. At that point the \ac{RSD} survey is complete in the form of a binary file ({\tt survey.sur}) and the {\tt surv} tool is applied to it to generate the \ac{RSD} in the official format ({\tt rsd.xml}).

\subsection{\label{sc:surveystrategyreplan} Survey replanning}

A new \ac{RSD} is computed regularly.  
There are various reasons for survey replanning, including: 
\begin{itemize}
    \item to reschedule skipped observations that have been dislodged by other activities (such as ice decontamination campaigns);
    \item to reschedule skipped observations not executed due to instrument outages;
    \item to reschedule lost observations with low quality flags (such as observations affected by solar activity or ice contamination); 
    \item to implement updated calibration specifications;
    \item to implement updated system requirements;
    \item to include additional observations  (such as implementation of the Quick Data Releases programmes);
    \item to implement science optimisation strategies;
    \item to implement recommendations from mission reviews.
\end{itemize}
During the replanning cycle, \ac{SOST} tracks various sources of survey-monitoring information, including the SOC's table of events, and feedback from various interfaces, such as the \ac{SGS} processing, the science coordination group, the \ac{CalF}, the Mission Manager, and the \ac{EST}, and compiles a list of replanning items. 

Survey replanning is the process of computing a new \ac{RSD} from a previous one, by retaining all the observations  up to a restart date, also called the break point,
and computing an observation schedule for the remaining nominal survey time. 
Given the incremental nature of the survey computation, where a plan is computed by iteratively appending observations, a change in one target observation or in a survey window invalidates all the rest of the survey, implying a computation of the full survey after the specified breakpoint. 
Note also that not all aspects of a survey can be subject to replanning. Some parameters were set when building the first operational survey and must not be changed throughout the mission. These are:
\begin{itemize}
    \item the \ac{RoI}, since a modification would imply a new tessellation, invalidating the tiling associated with previous observations;
    \item the tessellation;
    \item the definition of the polar patches, since changing this would impact the shape of all other polar patches, invalidating the already acquired polar patches;
    \item the EDF-S shape.
\end{itemize}

In replanning, the restart date is specified by the index of the first patch of the previous survey to be changed. Everything up to that point is copied from the previous survey, while everything from that point onwards is up for replanning. The restart patch may be an \ac{EWS} patch or any patch from the calibration plan. The replanning process starts with the computation of a new calibration plan, editing the {\tt plan.d} file after the break point to implement the required changes. All the three files {\tt ectile.cfg}, {\tt targets.d}, and {\tt plan.d} may be edited for the purpose of replanning. This means that the Stage~1 replanning may include new targets, rescheduling of fully or partially lost targets, and new
timings and cadences in the calibration plan. Then, the plan is converted to a schedule as usual, with the {\tt plan} tool, generating the intermediate survey.

The computation of a replanned \ac{EWS} requires the definition of new patches, named `infills', which will cover the lost and skipped fields of the previous survey, i.e., the gaps. An infill patch is a sequence of observations (one or more consecutive obsIDs) that has been previously scheduled. The infill patches are treated as new targets. They are registered in a new input file, {\tt backlog.d} (cf. Fig.~\ref{fig:ectile_diagram}), with each infill patch being defined as a specified sequence of obsIDs of the previous survey. An infill patch will behave like any other target and its visibility can be determined with the usual \ectile tools. In some cases, infill patches are inserted in the calibration plan, in their corresponding visibility locations. 
This capability is best suited for small infills and is particularly useful in those cases where the visibility of an infill falls in the middle of a calibration block (and not inside a survey window). However, in most cases, it is more convenient to insert them  into a survey window during the Stage~2 procedure. For this, each infill is registered in {\tt wide-plan.d} as part of a survey window, and is declared as being large or small. In the presence of infills, the processing of a survey window is modified from the standard procedure described in Sect.~\ref{sc:surveystrategy2}. The first operation is now to place the large infill as a new patch, either at its natural placement or shifted according to the margin parameters. 

This placement splits the survey window into two sub-windows. Only after that are the \ac{EWS} patches generated and selected in the usual way. Then, small infills, if any, are placed by carving time in the survey window, just like the PSF and SOPS are inserted at the end of the iteration step. Finally, we note that gap-recovery should be handled as early as possible in the schedule to reduce the chance of gaps piling up and to keep the re-observation dates as close as possible to their original observations. We also stress that when computing the survey statistics, the pre- and post- breakpoint \ac{RSD} information is merged, avoiding multiple counts of the corresponding observations. 

Survey replanning was first used in the computation of \ac{RSD}~2024C from \ac{RSD}~2024B, although no re-observation of \ac{EWS} patches was made at the time, i.e., \ac{RSD}~2024C did not schedule any infills. The re-observations were only introduced in the calibration plan. Later on, the computation of \ac{RSD}~2025A from \ac{RSD}~2024C was the first time that replanning was made with \ac{EWS} gap recovery. 
The first modified \ac{RSD}, \ac{RSD}~2024B, was computed in a different way and is a repairing of the original \ac{RSD}~2024A rather than a fully replanned survey. 

The repairing process computes a new survey from a previous one from a breakpoint onwards, by applying just a small set of changes to calibrations. The calibrations are scheduled in the same order as in the previous survey, but implemented with the updated parameters. The allowed modifications are minor changes of on-target timings, causing a calibration to last slightly more or slightly less than previously scheduled. Other changes are also possible, such as the number of pointings of a calibration, the pattern covering a calibration field, or the grism cycling of certain calibrations.  The \ac{EWS} sequence of observations is left intact, but each tile is observed at a slightly different time, due to the drift in the survey window timings. The \ac{EWS} patches are regenerated with updated orientation and solar angles. Given the very limited visibility of the \ac{EWS} tiles, the repairing procedure is successful only if the drift is kept to a minimum, with a total cumulated drift over the full survey of a few hours at most. PSF calibrations and SOPS are recomputed from scratch and repairing is done one iteration at a time, in chronological order.

\section{\label{sc:surveyresults} RSD DR1 components}

\subsection{\label{sc:DR1definition}  RSD DR1 definition} 

The first operational \ac{RSD}, \ac{RSD}~2024A, started to be executed on February 14 2024. At that time, thin layers of water ice were already forming on optical surfaces due to outgassing, causing a reduction in transmission \citep{Schirmer-EP29}. Selective heating was applied to folding mirror FoM3 and to mirror M3, followed by recalibration activities. This decontamination campaign lasted 9 days, from 12 to 21 March 2024. All observations previously scheduled for this period were calibrations and were not required to be rescheduled. In a first moment, the decontamination restored the transmission to immediate post-launch levels. However, after an initial improvement, \ac{VIS} data started to quickly degrade and on 6 June 2024, a second selective heating of folding mirror FoM3 was performed \citep{DR1-TP026b}. The second decontamination campaign was carried out during 5.5 days ending on 12 June 2024, and this time the scheduled polar cap north-cap-06 and 8 \ac{EWS} tiles were skipped and were rescheduled in a new \ac{RSD}, although not in time for DR1. No other decontamination campaign has been required so far. VIS data observed between the two decontamination campaigns, from 23 March to 7 June 2024 were declared invalid. This period, which became known as the `ice age', was rescheduled in \ac{RSD}~2025A for re-observations executed with the \ac{ROS},  although NISP observations had not been declared invalid. In total, the DR1 \ac{RSD} revisions consist of the following replanned \acp{RSD}:
\begin{itemize}
\item \ac{RSD}~2024B (operating from 15 July 2024), repairing of \ac{RSD}~2024A, with no rescheduled observations;
\item \ac{RSD}~2024C (operating from 16 November 2024), replanning of \ac{RSD}~2024B, rescheduling EDF-N (one visit), CPC-N (one visit), and polar-cap (one patch); 
\item \ac{RSD}~2025A, replanning of \ac{RSD}~2024C (operating from 28 February 2025), rescheduling EDF-N (three visits), CPC-N (four visits), polar cap (three patches), \acp{EAF} (one visit to COSMOS), and 
infills (12 patches, totalling 1119 tiles mostly from the ice age).
\end{itemize}

\begin{figure*}
    \centering
    \includegraphics[width=1.0\linewidth]{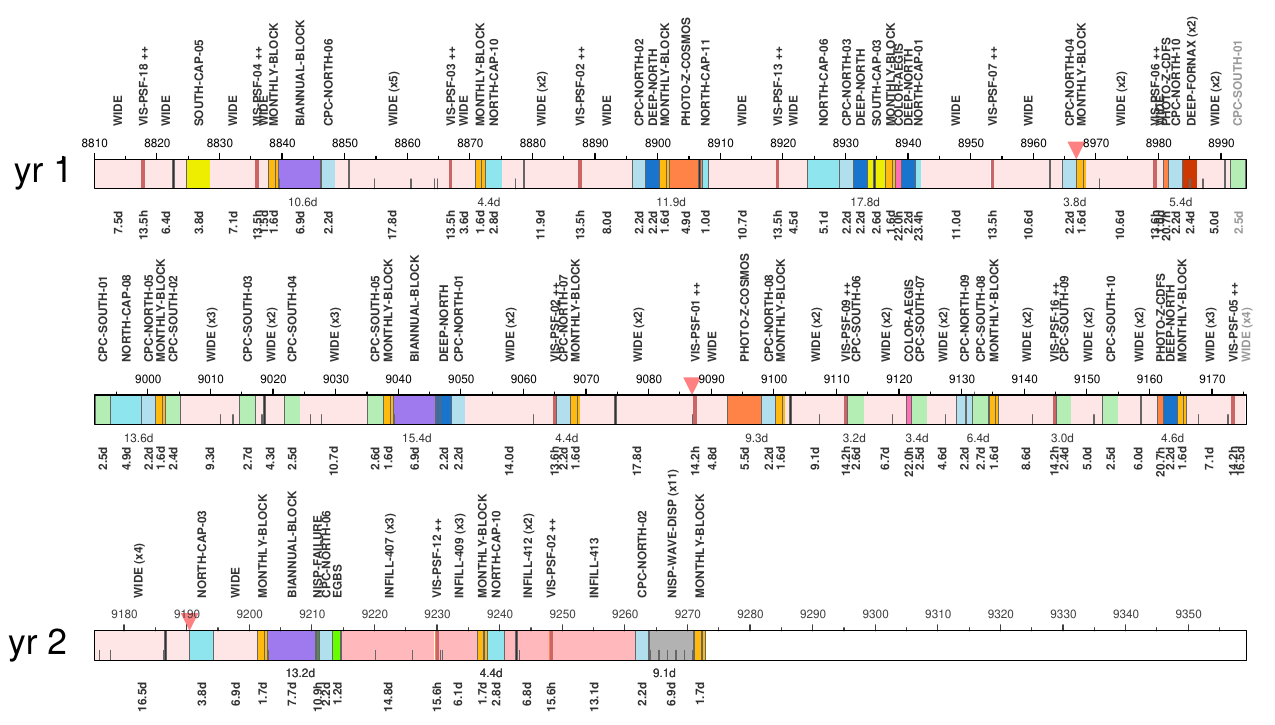}
    \caption{DR1 \ac{RSD} schedule timeline, from 14 February 2024 to 21 May 2025. Each EDF, EAF, and calibration block observation is shown as a box of a unique colour, labelled by the name and duration of the observation. Vertical black lines are the SOP time slots. The running time scale above the boxes is the MJD2000 date. The pink boxes are the survey windows where EWS observations are scheduled, the ones in a darker shade labelled `infill' are re-observations of previous ice-age EWS observations. Colour bars under the EDF and EAF boxes indicate when red or blue grisms are used. Orange triangles mark the starting point of a new \ac{RSD} (\ac{RSD}~2024A, \ac{RSD}~2024B, \ac{RSD}~2024C and \ac{RSD}~2025A), and the blue triangles mark the equinoxes, which is relevant information for scheduling.
    }
    \label{fig:chrono}
\end{figure*}

DR1 includes the various types of observation scheduled in \ac{RSD}~2025A and executed in the 462.8 days that occurred between 14 February 2024 and 21 May 2025. Four of those days were reserved for SOPS and 1.2 days were allocated to the \ac{EGBS}, presented in \citet{Q2-TP002}, whose data were released in Q2.
In the remaining 457.6 days, the \ac{RSD} scheduled visits to \ac{EWS} tiles, \acp{EDF}, \acp{EAF}, and calibration fields. The DR1 schedule is depicted in the timeline shown in Fig.~\ref{fig:chrono}, where each observation type is represented by a unique colour.

Not all \ac{RSD}~2025A observations executed up to 21 May 2025 are part of DR1, because some are not complete or were declared invalid. There are five reasons why some observations are excluded from DR1. These are the following.
\begin{itemize} 
    \item  Ice age: all observations from the ice age period are declared invalid.
    \item Ground-based data: \ac{EWS} observations without external ground-based data are considered incomplete, since ground-based data are necessary to retrieve photometric redshifts within requirements for the weak lensing probe. In the southern sky, \ac{DES} \citep{Abbott:2021} is currently used until deeper data from the Vera C.\ Rubin Observatory Legacy Survey of Space and Time \citep[LSST;][]{Ivezic:2019}
    become available. In the northern sky, the coverage is provided by \ac{UNIONS} \citep {Gwyn:2025}, including a UNIONS extension, except for the most southern part, which is expected to be covered by Rubin’s LSST. These observations will become complete once the ground-based observations will be available.
    \item Instrument outage: observations lacking data from one of the instruments are considered incomplete. There were some instances of the instruments being turned off preemptively, to avoid any instrument damage.
    \item Non-executed observations: there are cases of scheduled observations that were not executed, e.g., due to on-board tests or to give room for decontamination activities. 
    \item ${\rm RA}\,{=}\,0$\,h: observations whose pointings cover the ${\rm RA}\,{=}\,0$\,h line cannot yet be processed by the \ac{SGS} and are temporarily considered invalid. This issue is fixed in the DR2 pipeline.
\end{itemize}

\begin{figure}
    \centering
    \includegraphics[width=1.0\linewidth]{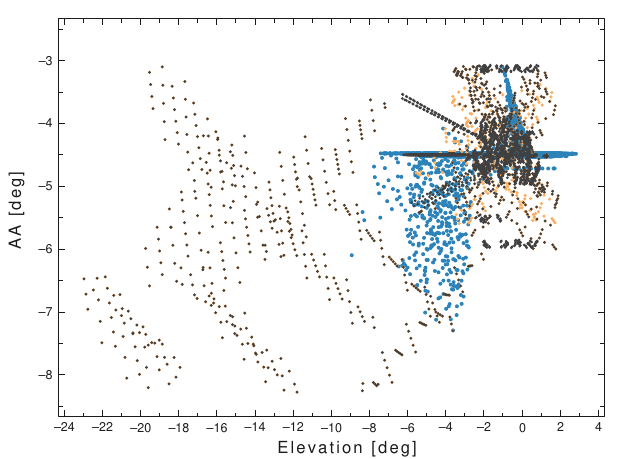}
    \caption{Solar angles map of the DR1 \ac{RSD} pointings. The SAA value is represented by the Elevation, which is defined as $90-$SAA. Each point represents an observation: EWS (orange); EDS and EAF (blue); CPC (brown); and calibrations (black).     
    }
    \label{fig:aasaa}
\end{figure}

During the observing period of 457.6 days, 46.8 days are spent slewing either between dithers or from field to field. In this period, the \ac{RSD} contains a total of \num{11697} observations. 
All observations comply with the solar angle restrictions, with AA kept in the range $\ang{-8.28}$ to $\ang{-3.06}$ and SAA in the range $\ang{87.16}$ to $\ang{113.0}$. The map of (elevation, AA) values for each \ac{RSD} pointing is shown in Fig.~\ref{fig:aasaa}, where elevation is defined as $\ang{90;;} - \mathrm{SAA}$. Most of the EWS pointings are close to zero elevation, indicating that they are observed near transit. This is a feature of the first year of the survey; in later years a longer reach in longitude will have to be applied to cover gaps in the sky that were left unobserved in the first orbits. The full allowed range of SAA is only used for \ac{CPC}-S observations. These are EDF-S observations made with a required cadence, and they often need to be observed far from transit to comply with the cadence. Each visit to the EDF/CPC field is made by keeping their tiles aligned and applying an AA rotation to compensate for a varying SAA, which leads to a strong correlation between the two angles.

\subsection{\label{sc:RSDews} The Euclid Wide Survey}

The \ac{RSD}~2025A \ac{EWS} contains \num{5384} ROS observations, corresponding to \num{21536} pointings. We note that \Euclid observations are identified by a unique number: the `obsID', with each \ac{ROS} observation being assigned a single obsID for the four dithers. 
 The area of the EWS footprint is $2639.1\,{\rm deg}^2$. This area already excludes gaps left around bright stars that amount to $60.5\,{\rm deg}^2$.
A significant fraction of this area is not processed by the \ac{SGS} and is excluded from DR1 for the reasons listed above. The breakdown of the excluded observations is as follows.
\begin{itemize}
\item Ice age: during the ice age, a total of 1166 EWS obsIDs were executed on the \Euclid tiles, consisting of 1092 tiles from the tessellation and 74 from polar patches. All were declared invalid and their observations were rescheduled. 838 of these tiles were reobserved in \ac{RSD}~2025A and their data are part of DR1. Therefore, only the remaining 328 obsIDs, reobserved at a later time, contribute to a decrease in the EWS released area of $147.4\,{\rm deg}^2$, of which $9.7\,{\rm deg}^2$ is in a polar patch.
\item Ground-based data: in the southern hemisphere, when following the strategy of observing in latitude layers from the ecliptic poles to the equator, the telescope passes  areas of the \ac{RoI} that are not covered by \ac{DES}. Some of these areas were nevertheless observed to avoid creating large holes in the contiguous coverage early in the mission. These data need to be excluded from DR1 while waiting for LSST data. The excluded areas due to non-existing ground-based data amount to $302.3\,{\rm deg}^2$ in 633 tiles.
\item Instrument outage: there were five instances of instrument outages, leading to incomplete data. Recovery from an instrument outage requires manual intervention and is only possible once every 24 hours, during the daily telecommunication period. This affected 53 EWS tiles, corresponding to $25.4\,{\rm deg}^2$ of excluded area.
\item Non-executed observations: during this period, a total of 124 EWS scheduled obsIDs were not executed due to various reasons, including a delay in the start of the survey that prevented the execution of obsIDs 1 to 6, on-board testing that replaced ObsIDs 7192 and 7193, and, most importantly, the second decontamination campaign that postponed the observation of the large polar-north-06 patch. The corresponding excluded area is $56.5\,{\rm deg}^2$.
\item ${\rm RA}\,{=}\,0$\,h: there are 39 tiles that cross the meridian and are excluded from DR1. 22 of them are already excluded for having no ground-based coverage and therefore the additional excluded area is only $8.2\,{\rm deg}^2$. 
\end{itemize}

Given these considerations, the resulting DR1 EWS area totals $2099.3\,{\rm deg}^2$, unevenly distributed in the two ecliptic hemispheres: 
\begin{itemize}
\item 
Northern mainland: $560.7\,{\rm deg}^2$,
\item 
Southern mainland: $1538.6\,{\rm deg}^2$,
\end{itemize}
which represents a decrease of $539.8\,{\rm deg}^2$ with respect to the \ac{RSD} EWS scheduled area. The EWS area validated for Level 2 data by the relevant SGS \acp{OU} is further reduced, with \citet{DR1-TP001} quoting a \ac{EWS} DR1 SGS-validated area of $1900\,{\rm deg}^2$.
Part of the additional area reduction is due to radiation impact related to solar activity. There are, in particular, 94 EWS obsIDs with the four dithered frames impacted that are already scheduled for re-observation at a later time. The set of identified obsIDs is 8--14, 2757--2774, 3802--3812, 4861--4875, 5644--5660, 6580--6586, 7040--7052, and 7074--7079. They contribute to a decrease in the effective useful EWS area of $44.5\,{\rm deg}^2$. 

\begin{figure*}
    \centering
    \includegraphics[trim=70 20 65 0,width=1.0\linewidth]{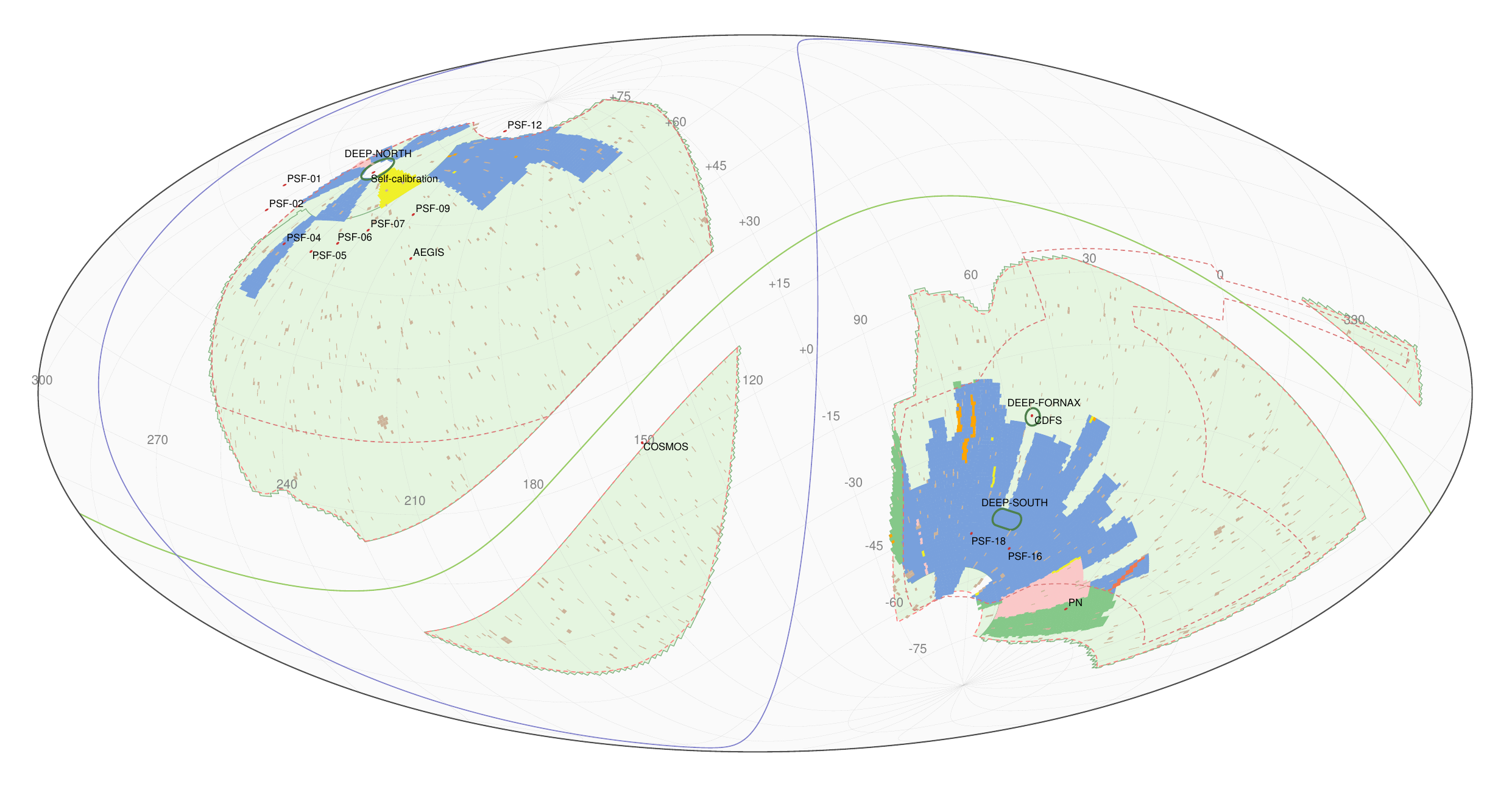}
    \caption{The \Euclid RSD Surveys at DR1, in the equatorial frame rotated around the central point (${\rm RA} = \ang{120;;}, {\rm Dec} = \ang{0;;}$) to improve visibility of the polar regions. The EWS footprint (see Sect.~\ref{sc:RSDews}) is shown in blue. The other colours show the EWS observations that were excluded from DR1: the red stripe of tiles in the south is the ${\rm RA}\,{=}\,0$\,h region, pink patches show invalid ice-age observations not yet re-observed, orange tiles correspond to instrument outages and lack VIS data, yellow regions were scheduled and not executed, and green areas have no ground-based data (see text for details). Brown areas scattered on the map are the bright star exclusion regions. The \ac{RSD} targets (see Sect.~\ref{sc:RSDtargets}) included in DR1, i.e., the observed \acp{EDF}, \acp{EAF}, and all calibration fields (Self-calibration, planetary nebula and PSF fields) are also shown in the map. Dashed red countours mark the UNIONS (in the North) and DES (in the South) footprints where ground-based data are available. The orange line shows the ecliptic and the blue line the Galactic plane. The light green background shows the \Euclid \ac{RoI}, consisting of two mainlands and two islands, where the spacecraft continues its surveying mission.   
    }
    \label{fig:wide}
    \end{figure*}

Figure~\ref{fig:wide} shows the \ac{EWS} DR1 footprint, including the DR1 valid and complete area, and also shows the various types of excluded regions. The non-EWS components of the DR1 \ac{RSD}, i.e., the \ac{EDF}, \ac{EAF} and calibration fields  observed in DR1 (see Sect.~\ref{sc:RSDtargets}), are also shown. 
We note that the remaining fields observed in DR1 corresponding to the non-RSD components of the \Euclid surveys described in Sects.~\ref{sc:pv} and ~\ref{sc:PDC} are included in the full DR1 \Euclid surveys map shown in Sect.~\ref{sc:conclusion}.

\begin{table*}
\centering
\caption{Summary properties of the DR1 Euclid Deep and Auxiliary Fields.
}
\begin{tabular}{llrrccc}
\hline\hline
\noalign{\vskip 2pt}
Type & Field & \multicolumn{1}{c}{RA} & \multicolumn{1}{c}{\ Dec} & Area& No.~Visits & Area \\
&&&&design&  & 1 pass \\
&&&& $[{\rm deg}^2]$&& $[{\rm deg}^2]$\\
\noalign{\vskip 1pt}
\hline
\noalign{\vskip 2pt}
& EDF-N & \ang{269.733} & \ang{66.02}& 20.0 & 11/40 (8R,3B) & 30.6\\
EDS & EDF-S & \ang{61.241} & \ang{-48.42}& 22.7 & 10/45 (10R,0B)&33.0\\
& EDF-F & \ang{52.932} & \ang{-28.09}& 10.0 & 2/52 (1R,1B)& 13.2\\
\hline
\noalign{\vskip 2pt}
& COSMOS & \ang{150.120} & \ang{2.21}& 2.0 & 27/75 & 2.6\\
EAF & CDFS & \ang{53.118} & \ang{-27.81}& 0.5 & 34/34 & 0.8 \\
& AEGIS & \ang{214.827} & \ang{52.82}& 1.0 & 18/18 & 1.4\\
\hline
\end{tabular}
\label{tab:targets}
 \footnotesize
 \tablefoot{Columns give the coordinates of the centres, the designed areas where the full depth will eventually be reached, the number of visits in DR1 as a fraction of the final number of visits planned (also showing the breakdown in ROS and ROS-Blue visits for the \acp{EDF}), and 
the area covered by at least one dither in DR1.}
\end{table*}

\begin{table*}
\centering
\caption{Summary properties of the DR1 \ac{RSD} calibration fields, grouped in three types. 
}
\begin{tabular}{llrrccccc}
\hline\hline
\noalign{\vskip 2pt}
Type& Field & \multicolumn{1}{c}{RA} & \multicolumn{1}{c}{\ Dec} & Area& CALBLOCK & No.~ObsIDs & Area& Sequence \\
&&&&design&  && 1 pass& \\
&&&& $[{\rm deg}^2]$&&& $[{\rm deg}^2]$ &\\
\noalign{\vskip 1pt}
\hline
\noalign{\vskip 2pt}
&SelfCal-centre & \ang{268.813}& \ang{65.29}& 2.0 & F-001& 11 &2.3 & ROS-D1\\
&&&&& F-009 & 2 &&NISP (special)\\
&&&&& F-010 & 5 &&NISP (special)\\
EUDF &&&&& F-011 & 1 &&NISP (special)\\
&&&&& F-016 & 2 &&VIS (special)\\
&&&&& F-018 & 1 &&NISP (special)\\
&SelfCal-faint & \ang{268.000}& \ang{65.00}&& F-007 & 10 &&VIS (special)\\
\hline
\noalign{\vskip 2pt}
PN &SMC-SMP-20 & \ang{14.035}& \ang{-70.30}& 0.5& F-010 & 5 & 2.2 &NISP (special)\\
&&&&& F-010 & 1 &&ROS \\
\hline
\noalign{\vskip 2pt}
&PSF-01 & \ang{292.48}& \ang{53.53}& 0.5& F-008 & 1 & 0.7 & VIS (special)\\
& & & & & F-013 & 1 & & ROS\\
& & & & & F-015 & 1 & & VIS (special)\\
&PSF-02 & \ang{285.03}& \ang{49.83}& 0.5& F-008 & 2 & & VIS (special)\\
&PSF-04 & \ang{265.72}& \ang{48.09}& 0.5& F-008 & 1 & & VIS (special)\\
PSF &PSF-05 & \ang{254.20}& \ang{49.37}& 0.5& F-008 & 1 & & VIS (special)\\
&PSF-06 & \ang{247.27}& \ang{52.69}& 0.5& F-008 & 1 & & VIS (special)\\
&PSF-07 & \ang{239.80}& \ang{56.89}& 0.5& F-008 & 1 & & VIS (special)\\
&PSF-09 & \ang{225.35}& \ang{61.91}& 0.5& F-008 & 1 & & VIS (special)\\
&PSF-012 & \ang{218.16}& \ang{82.06}& 0.5& F-008 & 1 & & VIS (special)\\
&PSF-016 & \ang{59.25}& \ang{-55.16}& 0.5& F-008 & 1 & & VIS (special)\\
&PSF-018 & \ang{73.52}& \ang{-49.10}& 0.5& F-008 & 1 & & VIS (special)\\
\hline
\end{tabular}
\label{tab:calibs}
 \footnotesize
 \tablefoot{Columns give the coordinates of the centres, the designed areas, the larger area covered by at least one dither in DR1, the CALBLOCKS that contribute to each field, and the number of obsIDs and sequences executed. The Self-calibration field (in its two offset centres) gets most of the visits. The PSF fields are observed in three CALBLOCKS, as indicated for PSF-01.}
\end{table*}

\subsection{\label{sc:RSDtargets} The Euclid RSD targets}

The remaining 6313 observations correspond to the \acp{EDF}, \acp{EAF}, and calibration fields mentioned in Sect.~\ref{sc:surveystrategy1}, here grouped under the title `RSD targets'. EDF and EAF observations are made with the ROS, with one observation (with 4 dithers) corresponding to one obsID. Calibration fields are observed with non-ROS special observing sequences consisting on many pointings (typically much more than 4), but that are assigned a single obsID for the full sequence (with a few exceptions). For calibration observations, the quoted number of observations corresponds to individual pointings, therefore, unlike for the ROS observations reported in Sect.~\ref{sc:RSDews}, the number of corresponding obsIDs is much lower than the quoted 6313 observations. 

Table~\ref{tab:targets} summarises the \acp{EDF} and \acp{EAF} observed for DR1, showing the coordinates of their centres, areas, and number of visits. Since each visit is made from a different position in the orbit, each visit has a different position angle on the sky. This implies that the overlap is maximum at the centre of the field, decreasing towards the edge, and thus the area with at least one visit is much larger than the central `design' area, which is the one that will reach the full depth at the end of the mission. The difference between the 1-pass and the full-depth areas is smaller in the case of EDF-F, due to its lower ecliptic latitude, and therefore shorter visibility, which strongly constrains the range of allowed observation positions in the orbit. 
We note that one visit represents one pass over the full field, thus containing a large number of obsIDs, depending on the size of the field.
The total number of planned visits is also given, providing an indication of the current progress towards full depth. The total number of visits to reach full depth is a function of ecliptic latitude, to compensate for the higher sky background at lower latitudes. We also note that during one of the EDF-N visits included in DR1 (patchID 44), there was a VIS instrument outage and half of the VIS data from that visit was lost. There were a few other visits lost (incidentally always in EDF-N), but those were fully discarded. 

The \ac{EDS} is executed on the three \acp{EDF}, and corresponds to 1271 obsIDs. The \acp{EDF} are observed either with the standard ROS that uses the RGS grisms or with the ROS-Blue variant, where all the four dithered frames acquire the spectroscopic data in the same direction using the BGS grism. The number of visits in Table~\ref{tab:targets} discriminates between `red' (denoted R) and `blue' (denoted B) visits. The goal at the end of the mission is to achieve a 5:3 blue-to-red ratio. The ratio at DR1 is very far from the goal, with a significantly large number of red visits. This is due to the need of performing the first ten EDF-N visits, as well as the first ten EDF-S visits, with the RGS in order to obtain as soon as possible a \ac{CPC} sample compatible with the \ac{EWS} observing mode.

Additional data on the EDF-N and EDF-S fields were taken in the \ac{PV} phase and are also included in DR1 (see Sect.~\ref{sc:pvfields}).
We also note that a fraction of the \ac{EDF} data was already released in Q1 \citep{Q1-TP001}. Indeed, Q1 consisted of one visit to each of the three \acp{EDF}, totalling 63.1\,${\rm deg}^2$, plus data from the LDN 1641 field (see Sect.~\ref{sc:pvfields}), with dozens of papers already published using that data set. The location and shape of the three \acp{EDF} are included in the all-sky map of Fig.~\ref{fig:wide}.

\Euclid makes repeated observations of six \acp{EAF} throughout the mission \citep{EuclidSkyOverview}, eventually reaching a depth of 5 times that of the EWS average. The data are used to calibrate photometric redshifts and to quantify the impact of colour gradients on shape measurement in the presence of a chromatic PSF. The observing strategy is to make repeated passes to the full area of these fields, increasing depth (each pass applies small shifts in the pointings). The passes alternate between the ROS and ROS-Blue sequences. DR1 includes data from three of these fields, as shown in Table~\ref{tab:targets}. Again, the number of visits reported is the number of passes over the full fields, each corresponding to multiple obsIDs. 

Only three of the \acp{EAF} are included in DR1, as shown in the all-sky map of Fig.~\ref{fig:wide}.
The visits to CDFS and AEGIS are already completed in DR1. Each of these two fields was observed in two periods, separated by 6 months, implying that half of the passes share a common direction that is $\ang{180.0}$-rotated with respect to the position angle of the second half of the visits. The third field, COSMOS, is the largest one and is $36\,\%$ complete in depth. Additional data on COSMOS and CDFS were taken in the \ac{PV} phase and are also included in DR1 (see Sect.~\ref{sc:pvfields}). 
  
Observations to acquire data for calibrations  define another set of fields in DR1, observed with a variety of observing sequences, as shown in Table~\ref{tab:calibs}. We point out that the various special sequences listed are different from each other, with a broad range of exposure times, and some including off-sky observations.

Several CALBLOCKS are executed in the Self-calibration field, with a diversity of observing sequences, most notably the 11 ROS-D1 observations made as part of the monthly F-001 self-calibration. The monthly F-007 VIS nonlinearity calibration is also executed in the same field, slightly offset to a region with a lower density of stars. Moreover, the Self-calibration field is contained in EDF-N and it was also a frequent target used in the \ac{PV} phase (see Sect.~\ref{sc:pvfields}). All these observations will contribute to form the deepest \Euclid field, named the \ac{EUDF}, as mentioned in \citet{EuclidSkyOverview}.

The single visit to the \ac{PN} in the \ac{SMC} has a complex sequence, with a total of 456 pointings, most only in NISP, also including four ROS observations with the PN placed near the four corners of the FoV, which is responsible for the larger one-pass area reported in Table~\ref{tab:calibs}. The \ac{RSD} visit is its second one, having also been observed during the \ac{PV} phase (see Sect.~\ref{sc:pvfields}).

The remaining calibration fields included in DR1 are 10 of the set of 24 fields defined for PSF calibration observations, with one of the fields, PSF-02, being used twice. Visits to PSF fields include the main CALBLOCK-F-008 VIS sequence, with 100 pointings, and also the CALBLOCK-F-013 single ROS observation as well as a short CALBLOCK-F-015 sequence. The observations are preceded by the CALBLOCK-F-012 closed shutter sequence. The sequence of three CALBLOCKS is executed for all these fields, although Table~\ref{tab:calibs} only shows it for one field for the sake of brevity. We note that PSF-16 also contains data from the \ac{PDC} campaign (see Sect.~\ref{sc:PDCcf}).

The DR1 \ac{RSD} calibration fields, i.e., Self-calibration, SMC-SMP-20 (PN), and the 10 PSF fields, are included in the all-sky map of Fig.~\ref{fig:wide}. PSF-01 and PSF-02 are outside the \ac{RoI}, and hence are not covered by a ROS EWS tile. A dedicated ROS observation was scheduled on those fields, but not on time for DR1. 
PSF-12, at a high declination, is outside the reach of ground-based telescopes and was later removed from the set of PSF fields. All the DR1 RSD target fields are also represented in the full DR1 \Euclid surveys map shown in Sect.~\ref{sc:conclusion}.

\begin{figure*}
    \centering
    \includegraphics[width=0.34\linewidth]{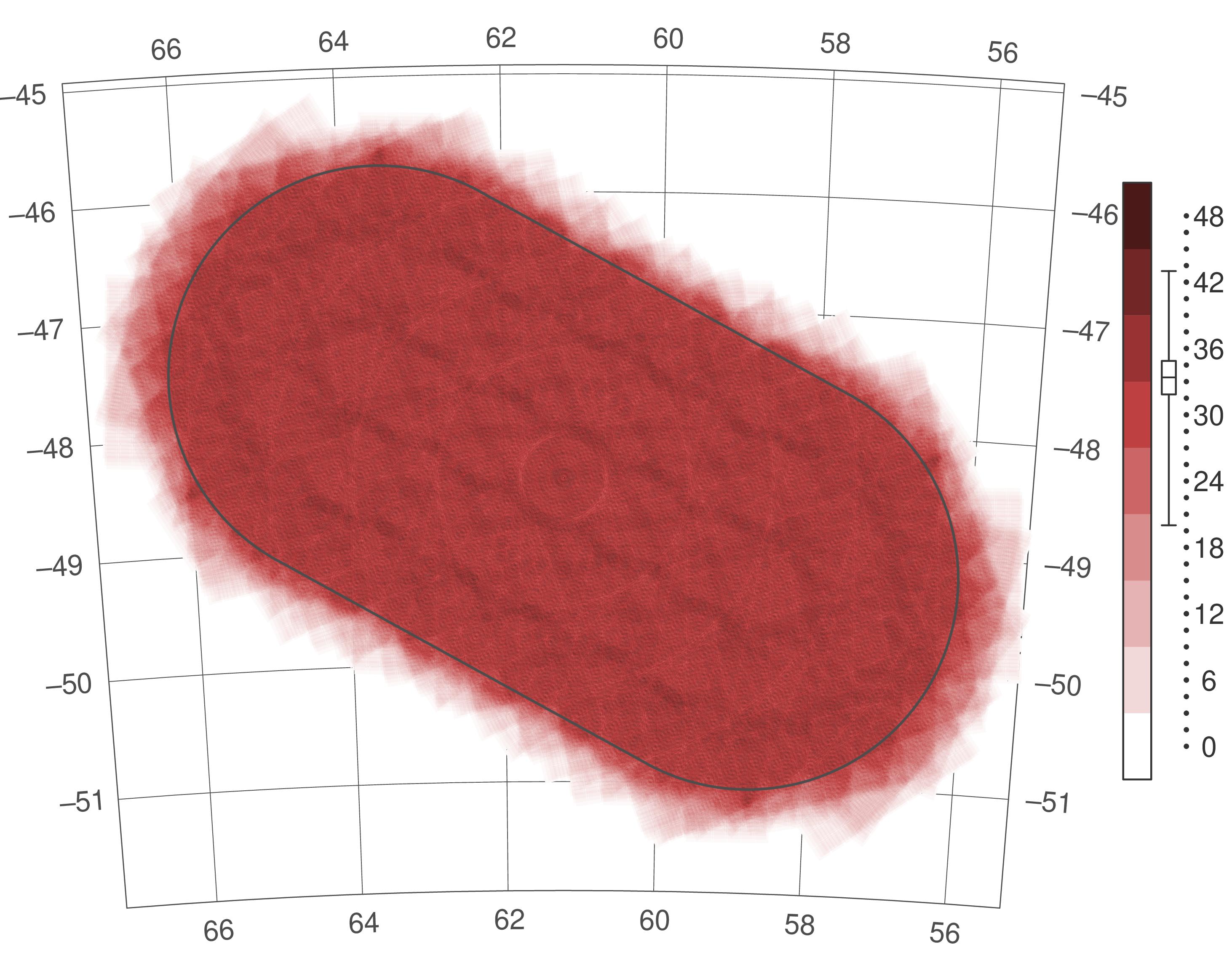} 
    \hfill
    \includegraphics[width=0.32\linewidth]{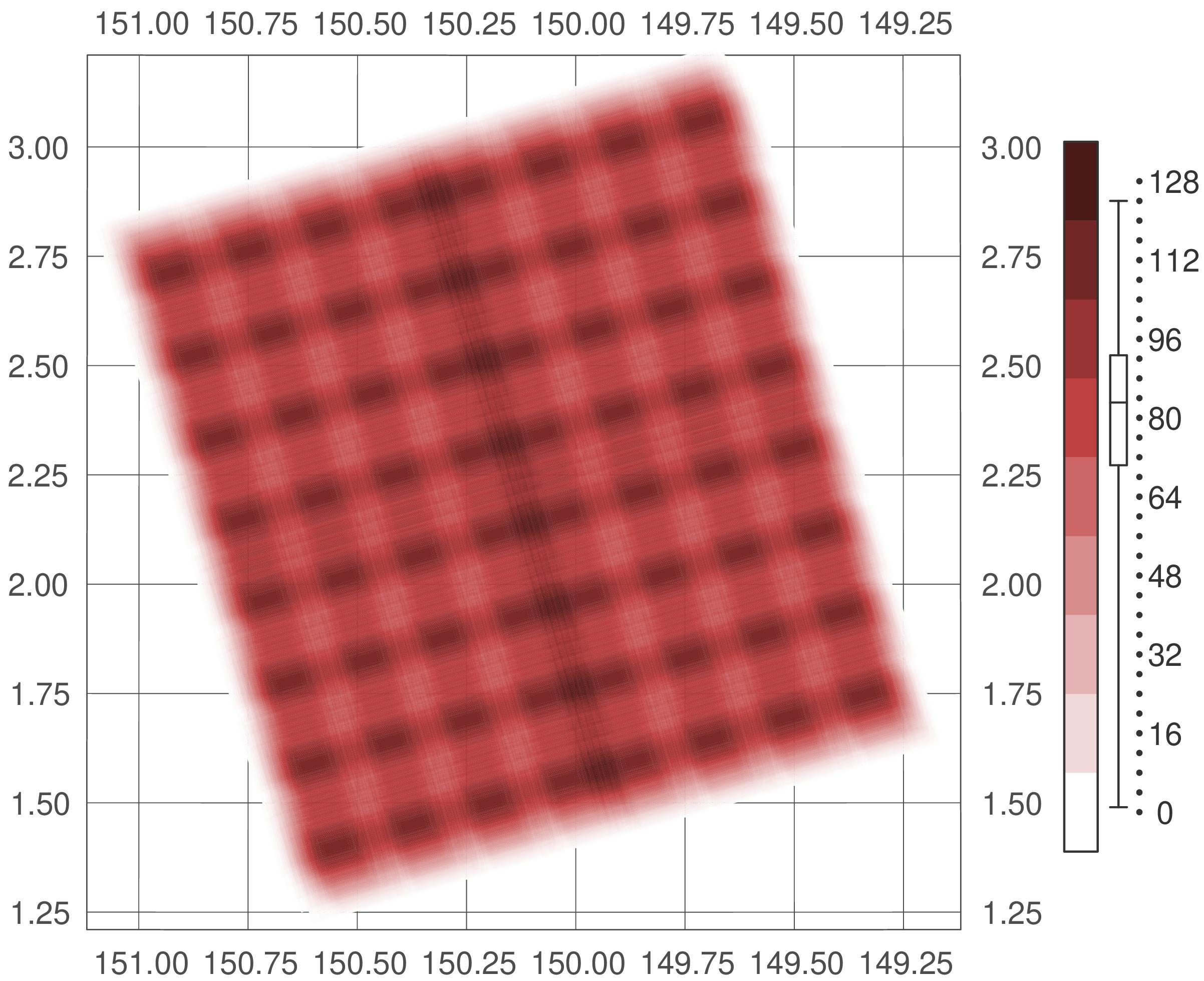} 
    \hfill
    \includegraphics[width=0.28\linewidth]{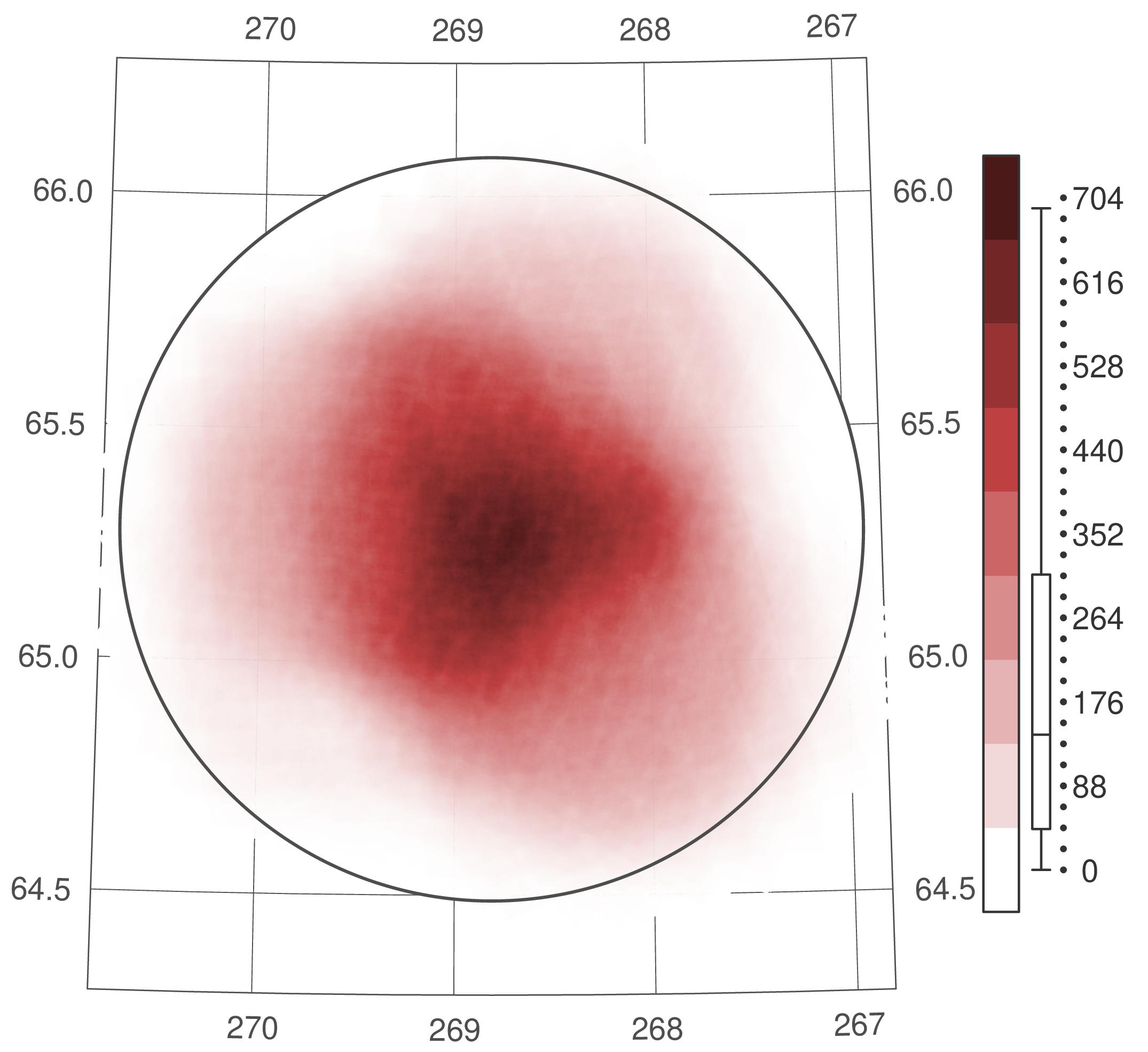}   
    \caption{Exposure maps showing the pixel coverage of EDF-S (in VIS; {\it left panel}), COSMOS (in NISP; {\it central panel}), and the Self-calibration field (in VIS; {\it right panel}). The scale includes a quartile bar, with the lower tick showing the first quartile (the median of the lower half of the exposure values), the middle tick showing the second quartile (the median of the full set of values), and the upper tick showing the third quartile (the median of the upper half of the exposure values). The box within the bar shows the inter-quartile range, i.e., the spread of the middle half of the points. 
   }
    \label{fig:exposure}
\end{figure*}

Figure~\ref{fig:exposure} shows three exposure maps containing the number of single-exposures at pixel level, resulting from the DR1 \ac{RSD} observations. The figures highlight features of the \acp{EDF}, \acp{EAF}, and Self-calibration field, and provide a comparison between VIS and NISP exposures. The maps shown are the VIS EDF-S, the NISP COSMOS, and the VIS Self-calibration field.

The exposure map of EDF-S contains 10 ROS visits (see Table~\ref{tab:targets}), hence each pointing is covered by 40 dithers. The number of exposures goes up to 48 in some pixels due to tile overlap, with a median of 33. The coverage is very homogeneous, with half of the pixels having between 31.5 and 34.5 exposures, and none with less than 20 exposures. Due to the execution of observations from various directions, circular patterns form in the map. 

The exposure map of COSMOS corresponds to 27 ROS visits (see Table~\ref{tab:targets}), hence to 108 exposures. The number of exposures goes up to 122, with
a median of 85. The coverage is less homogeneous than in the EDF-S case, with a much larger spread in values. This happens because all visits are made in a single direction (and in the opposite $180^\circ$ rotated direction), with small offsets applied to the pointings. The distribution of counts is not symmetric around the peak, with an excess of pixels on the lower side. The pattern formed by the contrast between high and low counts has an  orthogonal shape, following the layout of the detector gaps, and is not circularised by multiple directions as in the EDF-S case. In general, the patterns are more noticeable in NISP than in VIS maps due to their different layout of gaps. 

The exposure map of the Self-Calibration field corresponds to the 11 F-001 visits (see Table~\ref{tab:calibs}), with each visit containing 76 ROS-D1 pointings, there is a total of 836 exposures. The visits are made at a roughly monthly cadence, thus with different directions resulting in a roughly circular footprint of 2.0\,${\rm deg}^2$, deeper in the centre. The applied offsets ensure a more homogeneous coverage of the central region without reaching the maximum possible number of exposures at the central point. We note that the Self-calibration field does not include the other calibration observations made in the same field with special observing sequences, nor the ones centred in SelfCal-faint nor the overlapping EDF-N visits.

Appendix \ref{app:expmaps} provides the complete set of VIS and NISP exposure maps for all \acp{EDF}, \acp{EAF}, and Self-calibration (only CALBLOCK-F-001) DR1 observations. 
It also contains the exposure maps of the \ac{PN} field observations, and the VIS exposure map of the PSF-18 observations, which is representative of all  other PSF fields.

\section{\label{sc:socsurvey} Operational survey}

The production of the \Euclid survey does not end with the computation of the \ac{RSD} and related files and documentation. Upon receiving an \ac{RSD}, the \Euclid \ac{SOC} verifies its scientific validation and reports its conclusions in the \ac{RIAR}, proceeding then with its operational validation using \ac{ESS} tools. Afterwards, the \ac{SOC} processes the \ac{RSD}  for operational use generating the \ac{OSS}. This procedure involves a second input besides the RSD, namely the \acp{ICR} provided by the \ac{IOT}. Finally, the \ac{SOC} generates the operational products to be transmitted to the \ac{MOC} that operates the spacecraft. 

A brief mention of the two \ac{ESS} tools is appropriate. The Euclid Survey System Planning Tool \citep[\esspt ;][]{Gomez-Alvarez18}, based on the \textit{Herschel} mission planning infrastructure, significantly extended and adapted for the \Euclid mission, is a standalone Java tool. 
It loads an \ac{RSD} and populates tables listing all patches (ID, observing dates, coordinates, and solar angles), and pointings. It contains a sky panel, showing the time-indexed survey patches in various coordinate frames and sky projections, with the possibility of drawing contours and overlaying maps and catalogues. It also produces survey statistics and verifies the survey compliance with the system requirements. 

\esspt could also be used to generate the full \ac{EWS}, as an alternative to \ectile. For this, the user manually draws a patch on the sky panel, and the tool automatically schedules the patch with ROS pointings. The patch is then filled by tiles, starting from the bottom-right corner and progressing up and down in a zigzag way, with
user inserting the time of the first pointing or, alternatively, its \ac{SAA}. However, the manual patch placement procedure and the lack of integration with the other aspects of the complex scheduling (such as the implementation of the calibration plan), makes it unpractical to use \esspt to produce the \ac{RSD}, and the tool is mainly used for survey visualisation and characterisation. 

The second \ac{ESS} tool, the Euclid Survey System Operational Tool (\essot), is the one used by the SOC to process the \ac{RSD}. 
It comprises all the functionalities of \esspt plus the functionalities needed to generate the \ac{OSS}, and the functionalities required to maintain the survey implementation throughout the mission. These include the following.
\begin{itemize}
\item \ac{OSS} generation (Sect.~\ref{sc:socOSS}).
\item Operational products generation (Sect.~\ref{sc:socOP}).
\item Operational reporting: automated report generation of the status of the survey; \ac{QLA} ingestion to visualise the temporal and spatial evolution of selected values; and data volume and consumables estimation.
\item Mission replanning (Sect.~\ref{sc:socreplan}): the planning can be modified by removing current observations or importing new ones.
\end{itemize}


\subsection{\label{sc:socOSS} OSS generation}

While the \ac{RSD} provides a high-level description of the observations to be performed, the \ac{OSS} contains the detailed implementation required for their execution. The \ac{OSS} therefore represents the operational implementation of the survey. 

The \ac{OSS} is generated by the \ac{SOC} using the \essot. Following validation of the \ac{RSD}, the tool processes the \ac{RSD} schedule and generates the corresponding \ac{OSS}. This is done with two mechanisms applicable according to the type of observation sequence. In the case of ROS observations scheduled in the \ac{RSD}, the detailed implementation is derived  through the ROS processing performed by \essot, which populates the OSS \ac{XML} file with the relevant ROS information, such as the grism and filter to be actuated in each frame.

In the case of calibration activities, the detailed implementation is made by inserting the references of the corresponding \acp{ICR} in the OSS. An \ac{ICR} is the mechanism by which the VIS and NISP \acp{IOT} specify sequences of instrument commands not covered by the \ac{RSD}. This mechanism is mainly used to specify the detailed calibration activities and consists of an \ac{XML} file generated from two inputs produced by the \ac{IOT}. The first input is an instruction file containing lists of target and dithering instructions; incidentally this information can be automatically retrieved from the \ac{CalF}. The second input are CSV configuration files containing instrument operations instructions, structured as a script with the possibility to loop over blocks of instructions. There is one configuration file per CALBLOCK, containing the telecommand (as defined by MOC, based on the true instrument commands), the parameters defining the details of the telecommand, and the execution timing of the telecommand per pointing. Each pointing has a maximum window of 3000\,s, which is the maximum duration the spacecraft can hold telecommunication contact with the ground station while holding a fixed pointing.    

The OSS file incorporates the activities generated by the \essot for ROS observations, together with the applicable \acp{ICR}. Each new release of an \ac{OSS} includes all the activities already executed and the future planned ones up to the nominal end of the mission.

\subsection{\label{sc:socOP} Operational products generation} 

The SOC produces two operational products that are delivered to \ac{MOC} at each \ac{PC}. The operational planning products 
are generated for a specific \ac{PC}. Each \ac{PC} corresponds to the interval between two consecutive SOPS windows. The corresponding planning products are delivered by the \ac{SOC} to the \ac{MOC} at least 2 weeks before their on-board execution period.

The first operational product is the set of \acp{SSR} to be delivered to \ac{MOC} flight dynamics (in the form of the SSR\_FD file). The SSR\_FD file is generated by the \essot, derived from the \ac{OSS}, and contains the slew timeline together with the pointing information defining the spacecraft attitude for a given \ac{PC}. An SSR\_DEBUG file containing additional diagnostic information is also generated for internal \ac{SOC} use to support validation of the SSR\_FD generation process.

\essot also generates the ESS\_EVENT file, which is intended for internal use within the \ac{SOC}. This file contains the detailed implementation of ROS observations translated into SOC events, and also includes the implementation of calibration activities defined through \acp{ICR}. We note that, depending on the complexity of the activity, the calibrations may be represented either as SOC events or directly as telecommand sequences. 

The ESS\_EVENT file, together with the \acp{ICR} and the Mission Information Base is used as input to another set of tools, the \ac{SCS} which is the system in charge of centralising and generating the commanding requests from the SOC, merging \ac{ESS} and \ac{IOT} requests. 
The \ac{SCS} interprets the events and converts them into telecommand sequences according to a set of SOC-defined rules based on the  
Mission Information Base definitions, incorporating ICR-specific information when required. The \ac{SCS} also assigns the corresponding parameter values to the generated telecommand sequences. The resulting output is encoded in the Payload Operation Request (POR) \ac{XML} format, a file used to schedule and automate telecommands. This output file is packaged together with the \ac{PC} delivery manifest, forming the so-called Command Request File Group (CRFG) product, which is the second operational product delivered to the \ac{MOC} at each \ac{PC}.

\subsection{\label{sc:socreplan} SOC survey replanning}

Section~\ref{sc:surveystrategyreplan} described strategic survey replanning implemented at the \ac{SOST} level. However, sometimes there is the need for a quick localised modification of the survey plan while in operations, such as a one-off change that needs to be performed quickly, sacrificing temporarily a small part of the scheduled survey. These requests are studied and approved by the \ac{CCB}
and a corresponding \ac{ICR} is issued. 
Such modifications do not require a full-scale strategic replanning and are implemented by the SOC with one of two procedures:

\begin{itemize}
\item
fast replanning, known as the palaeolithic approach;
\item local optimised replanning, known as the neolithic approach.
\end{itemize}

In the palaeolithic approach, 
the change request is directly implemented. 
For example a new list of  calibration pointings is requested to be observed at a given point in time, replacing the scheduled observations. This results
in  spatial and time gaps in the survey and no attempt is made to fill them. 
In cases where the ICR specifies earliest and latest start times (i.e., it includes a temporal margin), some level of performance optimisation is applied by introducing the modification in a least-damaging period.
For this purpose, the \essot provides tools to find the worst \ac{FoM} within the ICR time window to guide the user selection. 
The spatial gaps introduced in this approach should be recovered at a later time, through a strategic replanning of the survey.
This was the procedure followed to schedule the two decontamination campaigns. In the first campaign, the \acp{ICR} had a time margin that allowed us to implement the decontamination activities in replacement of calibrations that did not need to be rescheduled. In the second decontamination the \ac{ICR} timings were stricter and the removed observations were later scheduled in a new \ac{RSD}.

In contrast, the neolithic approach performs a local optimised replanning of the replaced observations. It applies only to fixed-time \acp{ICR}. The \ac{FoM} of the replaced observations is compared to their neighbours and
worse-quality neighbours are replaced by the removed observations that are thus reinserted with shifted
starting times.

\section{\label{sc:pv} PV phase}

We turn now to the observations performed in the epoch before the start of the \ac{RSD} execution, which were implemented by the \ac{SOC}. A large part of these observations contribute to the DR1 data set, thus also being part of the \Euclid surveys. The first period after launch was the spacecraft commissioning. Most commissioning activities were not scheduled through standard SOC tools and lack metadata. They are therefore considered incomplete and are not included in DR1. Following spacecraft commissioning, in early August 2023, the \ac{PV} activities of technical characterisation and calibrations started, with a two-month plan of observations.
Due to issues with the fine-guidance sensor
\citep[FGS;][]{Laureijs2024}, the \ac{PV} phase was paused at the end of August 2023. It resumed a few days later, with tasks that did not require high pointing stability. At the end of September 2023, the FGS was recommissioned and PV activities restarted for an additional 2-month period until early December 2023. The full \ac{PV} schedule is given in \cite{DR1-TP026}. It consists of 31 blocks of activities referenced as CALBLOCK-PV-XXX, and one labelled CALBLOCK-F-014, of which 11 fulfil VIS and NISP calibration products (PV-001 through PV-009, PV-035 and F-014), eight are tuned for NISP calibration products (PV-010 through PV-019), and 12 for VIS (PV-021 through PV-034). Three of the blocks are observed with ROS (PV-005, PV-007, F-014), six with ROS-D1 (PV-001, 003, 004, 006, 023, and 035), and the remaining implement special observing sequences using only one or both instruments. 

\begin{table*}
\centering
\caption{Summary properties of the PV fields, grouped by type.}
\begin{tabular}{llrrl@{\hspace{0pt}}c}
\hline\hline
\noalign{\vskip 2pt}
Type & Field & RA & Dec & CALBLOCKS & No.~ObsIDs\\
\hline
\noalign{\vskip 2pt}
ERO & 17 fields & & & PV-007 & 25\\
\hline
\noalign{\vskip 2pt}
& CDFS & \ang{53.118} & \ang{-27.81}& PV-005 & 1\\
& COSMOS & \ang{150.120} & \ang{2.21}& PV-005 & 1\\
& EDF-S & \ang{61.241} & \ang{-48.42}& PV-005 & 1\\
Characterisation & EDF-N & \ang{269.733} & \ang{66.02}& PV-005 & 2\\
 & Scan-90 (z1--z22) & \ang{270.000} to \ang{310.159}& \ang{66.56} to \ang{-23.51}& PV-006 & 1\\
  & Scan-95 (z23--z35) & \ang{244.546} to \ang{317.293}& \ang{80.00} to \ang{-26.85}& PV-006 & 1\\
& Scan-100 (z35--z48) & \ang{279.438} to \ang{322.841}& \ang{81.38} to \ang{-25.22}& PV-006 & 1\\
& Scan-Gal (s1--s6) & \ang{257.104} to \ang{98.221}& \ang{75.71} to \ang{-57.32}& PV-006 & 1\\
\hline
\noalign{\vskip 2pt}
& WDJ175318+644502 & \ang{268.345} & \ang{64.76}& PV-002 & 3\\
Calibrator & GRW+70 & \ang{204.650}& \ang{70.28}& PV-003 & 5\\
& VIS-AST-1 & \ang{90.000}& \ang{-66.50}& PV-023 & 2\\
& SMC-SMP-20 & \ang{14.035}& \ang{-70.30}& PV-009 & 5\\
\hline
\noalign{\vskip 2pt}
& $\alpha$Lyr & \ang{279.858} & \ang{39.50}& PV-004 & 1\\
 & HD \num{165906} & \ang{271.109} & \ang{64.83}& PV-004 & 1\\
Stray light & SL1 & \ang{279.390} & \ang{64.82}& PV-035 & 1\\
 & SL2 & \ang{268.810} to \ang{314.540}& \ang{65.29} to \ang{48.23}& PV-035 & 7\\
& SL3 & \ang{267.977} to \ang{9.276}& \ang{66.61} to \ang{-78.25}& PV-035 & 7\\
& SL4 & (\ang{268.445}) & (\ang{66.10})& PV-035 & 1\\
& SL5 & (\ang{314.539}) & (\ang{73.42})& PV-035 & 1\\
  \hline
  \noalign{\vskip 2pt}
  & Selfcal-centre& \ang{268.813} & \ang{65.29}& PV-001, 009, 016, 017, \\
& & & & 019, 027, 028, 034, F-014 & 26\\
EUDF & Selfcal-faint& \ang{268.000} & \ang{65.00}& PV-011, 012, 014, 018, \\
& & & & 021, 022, 024, 026, 027, \\
& & & & 028, 029, 030, 031, 034  & 60\\
& NEP& \ang{270.000} & \ang{66.56}& PV-008, 018, F-014 & 22\\
\hline
\noalign{\vskip 2pt}
& EBF-06 & \ang{257.790} & \ang{19.47}& PV-028, 034 & 7\\
EBF & EBF-08 & \ang{43.945}& \ang{-17.63}& PV-022, 027, 028, 029, 034 & 10\\
& EBF-10 & \ang{350.268}& \ang{1.32}& PV-022 & 1\\
& EBF-12 & \ang{0.251}& \ang{-75.21}& PV-022, 029, 030, 031, 033 & 12\\
\hline
\noalign{\vskip 2pt}
FGS & LDN 1641 & \ang{85.754} & \ang{-8.39}& OTHER & 4\\
\hline
\end{tabular}
\label{tab:pv}
 \footnotesize
 \tablefoot{Each field may be observed for multiple purposes in different CALBLOCKS. The number of obsIDs is also given. One obsID may consist of observations of one field or a sequence of fields, in which case the coordinate range of the sequence is given, or in some cases the median point is given (in parenthesis). Most of the observations are done with special observing sequences, while some use the \ac{ROS}, or the single-dither ROS-D1. See text for details.}
\end{table*}

\subsection{\label{sc:pvimp} PV phase implementation}

Long before launch, a PV plan started to be prepared in the form of a document associated with a timeline built using the \texttt{Excel} software. This plan was used in feasibility studies concerned with 
the total PV duration and the daily downlink volume, among other parameters. Already before launch, the procedure was changed to a fully electronic system that used the actual SOC planning products in a feedback loop to the PV planning process. 

The mechanism to construct the \ac{PV} phase timeline is based on \acp{ICR}. Each calibration block is defined by the on-sky pointings and instrument commands, which constitute the required inputs to the SOC planning tools. The pointings are generated at the SOC, while the instrument commands are provided by the two instrument teams through a dedicated interface based on CSV files. For simplicity, most calibration blocks are devoted to a single instrument, with a few exceptions that observe with the ROS.
From these inputs, the SOC generates the \acp{ICR} with a tool written using the \texttt{LabVIEW} programming environment. One \ac{ICR}, or group of \acp{ICR}, is generated for each of the PV execution blocks.

The next step is to build the PV plan, i.e., to order the calibration blocks on a common timeline. This is done by the \texttt{PV scheduler}, a tool also written using the \texttt{LabVIEW}  environment that in addition also permits the exchange of observational conditions and targets. Building the PV plan is a highly constrained problem. The most stringent constraint is target visibility in cases where specific targets are required and limited visibility windows exist. We also needed to place fundamental calibrations early in the schedule, since  their calibration products were required for the data processing of higher-level observations (wavelength calibrations are a notable example).
An additional constraint comes from the interruptions every 28 days through the SOPS windows that were also needed during PV. Not only do they prevent the placement of long calibration blocks that should not be interrupted, 
but they create a thermal impact. In addition, there was the need for rescheduling either because of failed observations, missed targets, deficient data quality, or additional engineering observations.
To ease their sequencing, calibration blocks were broken up into smaller functional and self-consistent subsets called `variants'. 114 variants were created out of the 31 executed PV CALBLOCKS. These helped in dealing with rescheduling, multiple executions, interruptions by SOPS windows or by other calibration blocks. The \texttt{PV scheduler} could then combine variants to again form higher-level indivisible execution blocks. 
The final PV timeline contains 132 execution blocks. 

The PV schedule is then fed into \essot that performs a time consistency and constraints check. After removal of inconsistencies and fine-tuning, the plan is fed into a \texttt{LabVIEW} visualisation tool. The complex procedure of PV scheduling requires a very flexible and accurate \texttt{PV scheduler}, and the recurrent preparation of a visual plan that allows communication of the updated PV scheduling status to all stakeholders. During the four months of PV, a total of ten sets of SOC planning products were produced in the form of SSR\_FD files, operational products delivered to \ac{MOC} flight dynamics for execution. In the end, the PV schedule as executed made the most efficient use of \Euclid’s in-orbit time, with almost no unused time gaps.

\subsection{\label{sc:pvfields} PV fields}

The \ac{PV} phase includes a very diverse set of observations. Table~\ref{tab:pv} provides the list of fields that resulted from the valid executions of on-sky \ac{PV} CALBLOCKS. 
In the table, the fields are grouped by types. The first type consists of the \ac{ERO}. \ac{ERO}, or CALBLOCK-PV-007, contains six science programmes, executed in 17 fields, prepared to demonstrate \Euclid’s capabilities in advance of the main mission, and constitute a first set of \ac{PV} science fields. These data, acquired in 24 hours using the \ac{ROS}, led to the public release of the first \Euclid images in November 2023, and to a series of papers. The details of the six programmes and their 17 fields can be found in \cite{EROData} and references therein.

\begin{figure*}
    \centering
    \includegraphics[width=1.0\linewidth]{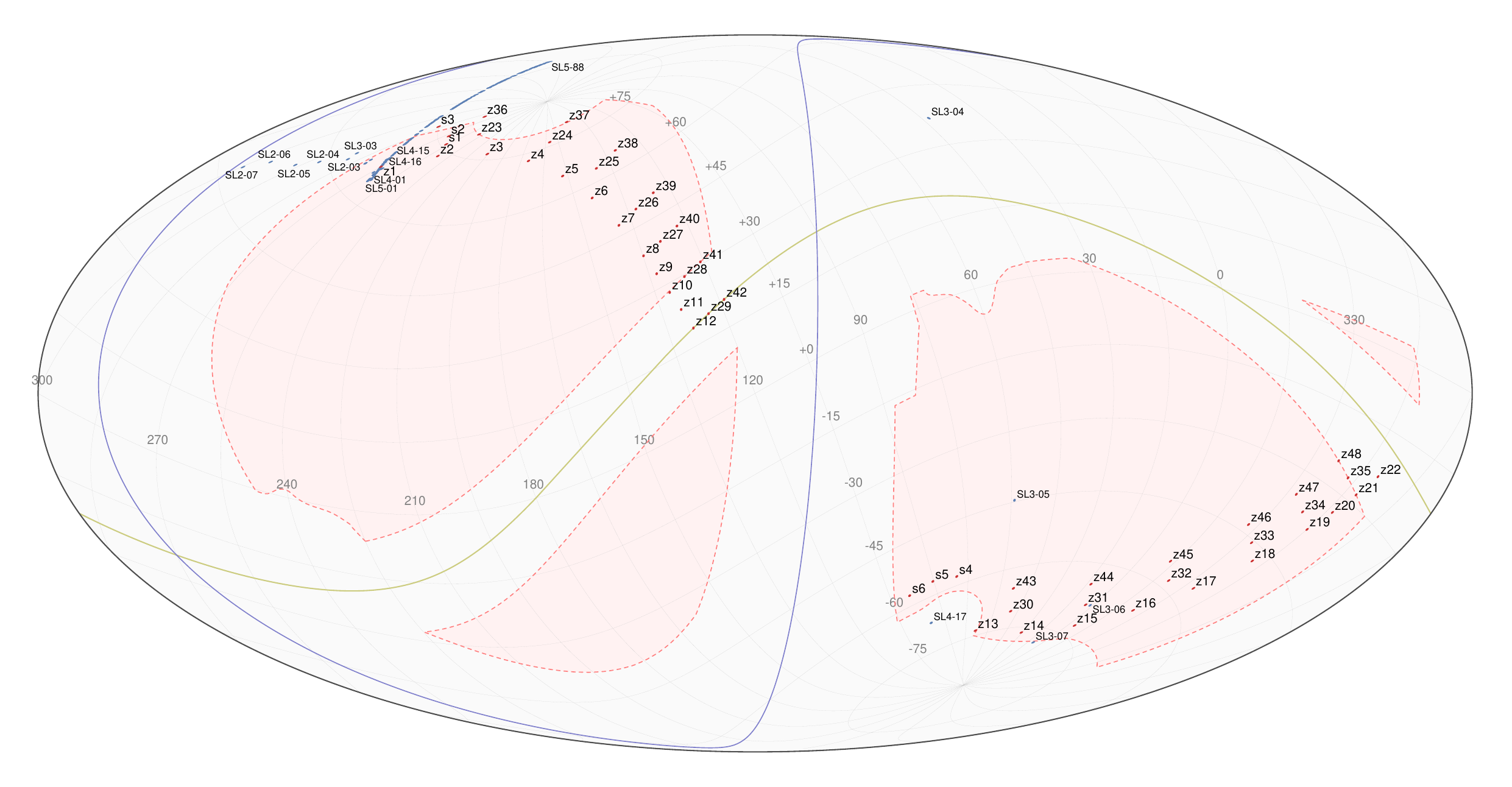}
    \caption{The 174 fields observed in the ten PV-006 and PV-035 sequences,  in a full-sky map in the equatorial frame rotated around the central point (${\rm RA} = \ang{120;;}, {\rm Dec} = \ang{0;;}$) to improve the visibility of the polar regions. The orange line shows the ecliptic and the blue line the Galactic plane. The pink background shows the \Euclid \ac{RoI}. The 48 `z' pointings from PV-006 scan the ecliptic latitude in three sequences each with a constant SAA, while the six `s' pointings, also from PV-006, scan the Galactic latitude. The fields from the five PV-035 sequences are labelled SL1 through SL5. Most of them lie on the blue strip on the north, and are too crowded to label individually. 
    }
    \label{fig:zodi}
\end{figure*}

The second type includes the ECSURV-defined observations for survey validation (CALBLOCK-PV-005) and background model characterisation (CALBLOCK-PV-006). The goal of the former is to ensure early on that \Euclid performs as expected, that the deep field locations are without problems, and that survey-like data volume can be processed on the ground. They consist of one visit, executed with the ROS, to some of the \acp{EAF} and \acp{EDF} prior to their visits scheduled in the \ac{RSD}. The fields selected for this programme were CDFS, COSMOS, EDF-S, 
and two observations of EDF-N (one pass to the full field and one pass to the central 3$\times$3 tiles). The observations for characterisation of the zodiacal background models provide data used to validate the zodiacal model that guided the original choice of the \Euclid \ac{RoI}  \citep[][see also Euclid Collaboration: Trombetti et al in prep.] {Maris2019}. The observations consist of three sets of pointings scanning the ecliptic latitude at fixed values of \ac{SAA}, executed with ROS-D1. We note that each scan, consisting of several pointings, has a single observation ID. In particular, there is a set of 22 pointings made with ${\rm SAA}=\ang{90.0}$, 13 pointings made with ${\rm SAA}=\ang{95.0}$, and 13 pointings made with ${\rm SAA}=\ang{100.0}$. An additional set of 6 pointings, across the Galactic plane at constant ecliptic latitude, was performed to estimate diffuse star light contribution. The location of the 54 pointings of the five sequences is shown in Fig.~\ref{fig:zodi}, and each one constitutes a \Euclid PV field. 

The third type of observations consists of four calibrator fields observations that contain the VIS/NISP-P (CALBLOCK-PV-002) and NISP-S (CALBLOCK-PV-003) absolute photometric standards, a \ac{LMC} high-density stellar field for the VIS astrometric solution (CALBLOCK-PV-023) and a compact \ac{PN} with strong emission lines for the NISP wavelength dispersion (CALBLOCK-PV-009) that will be re-observed in the \ac{RSD}. Most of these observations are ROS-like.

The fourth type contains the fields defined to defined to characterize the various stray light (expected and parasitic). The internal stray light check observes bright stars with an ROS-D1 sequence (CALBLOCK-PV-004). The external stray light check (CALBLOCK-PV-035) was a later addition to the PV plan, following the discovery of strong stray light from the Sun when observing at the pre-originally planned solar attitude  \citep{EuclidSkyOverview}. this was aimed at sampling the AA/SAA parameter space to identify a survey operational range where the stray light is suppressed to a minimum. Five sequences of pointings were executed: SL1, with 13 observations in the same position with different \ac{AA} values; SL2 and SL3, each consisting of a sequence of seven pointings with different \acp{SAA} each executed with seven \ac{AA} values; SL4, a sequence of 17 pointings each executed with two \ac{AA} values; and SL5, a sequence of 88 single pointings. PV-035 defined 120 fields in total, which are shown in the full sky map of Fig.~\ref{fig:zodi}.

The fifth type are observations made close to the \ac{NEP} in the Self-calibration field, and in some cases pointing off-centre  to an area within the field with lower density of bright stars (SelfCal-faint). Some of the observations are done off-sky, some on-sky and some have a sequence of both. These repeated visits to the \ac{NEP} area, together with the recurrent visits to the Self-calibration field in the \ac{RSD}, will result in the \ac{EUDF}. The \ac{EUDF} is currently being built with a dedicated data processing pipeline with data from PV-001 and F-014 PV observations, and from F-001 RSD observations (McCracken et al in prep.).

From \Euclid's orbit, the \ac{NEP} region can only be observed with SAA close to \ang{90;;}. The set of \acp{EBF} was defined by looking for alternative regions with low density of bright stars. They were selected to have a minimal density of stars brighter than 12\,mag, and located at lower ecliptic latitudes, hence observable with different values of SAA.  Some instrument calibrations, including additional observations of CALBLOCKS also executed in Selfcal-faint, were performed in four of these fields. This set of fields, defining the sixth type of PV fields, was no longer used in the \ac{RSD} due to FGS guiding 
problems in the sparse fields and was  replaced by Selfcal-faint.

 Finally, there are the observations introduced for specific purposes outside the PV scope. In order to distinguish these from the regular PV observations they were given a dedicated ID, labelled `OTHER', already at the moment of scheduling. Most of these special engineering observations are not included in DR1, with the exception of the dark cloud LDN1641, an extended dust-obscured part of the Orion A Cloud. This observation was made as part of the recommissioning of the FGS and ended up forming another PV science field. This field was already released as part of Q1 \citep{Q1-TP001}.

All these fields are part of DR1 as LE1 data. 
However, PV LE2 data cannot be obtained with the standard \ac{SGS} processing. For this reason, the only PV fields present in LE2 data are the Self-calibration (PV-001 and F-014 observations only) and COSMOS (PV-005). 

\section{\label{sc:PDC} PDC campaign}

 The \ac{PV} phase was followed by the \ac{PDC} campaign. The rationale for this specialised calibration campaign is that the VIS PSF depends on the telescope optical path errors that vary with the thermo-elastic state of the spacecraft. In each pointing of the survey, the Sun illuminates the sunshield from a slightly different direction, primarily associated with variations in SAA and AA solar angles, producing a thermo-elastic variation of the spacecraft elements, including the position and shape of the telescope optical surfaces.
In addition to the attitude of the spacecraft, the internal dissipation caused by the instrument cadence also greatly affects the thermo-elastic status of the telescope \citep{EU-Anselmi}. These time-varying modulations result in a slightly different set of telescopes, where deviations from the reference fiducial telescope model translate into PSF variations. There is thus the need to perform observations with the aim of obtaining an accurate model of the in-orbit optical wavefront model of the telescope and its variation with the thermal state of the payload module. The procedure is performed by applying the so-called phase diversity technique, described in \cite{DR1-TP026}. The resulting wavefront error \citep{Neumann26} can be used to construct a basis set of models that may then be fitted to other  data (Euclid Collaboration: Miller et al. in prep.).

\subsection{\label{sc:PDCimp} PDC campaign implementation}

The \ac{PDC} campaign observation programme requires sampling of the space of spacecraft attitudes, and needs much more observations than those possible in the time allocated for the PV phase (where a first CALBLOCK-PV-008 phase-diversity test was performed). Indeed, the observing campaign lasted for about 70 days, until the start of the \ac{RSD} execution on 14 February 2024.

\subsubsection{\label{sc:PDCcf} PDC calibration fields}

The observation plan of the \ac{PDC} campaign 
consists of seven runs of observations, labelled PDC-2 through PDC-8, with a first run, PDC-1, previously executed during the PV phase on the NEP. Each run is performed in a different and fixed thermal state and focus conditions and with the instruments operated at a constant cadence.  The requirement on thermal stability was defined as a less than 10-mK variation during out-of-focus observing sequences, and less than 60-mK variation for in-focus observations. In terms of solar attitude it was also verified that SAA variations have a greater thermal impact than AA variations, in agreement with \cite{EU-Anselmi}. 

\begin{table}
\centering
\caption{Summary properties of the PDC calibration fields.}
\begin{tabular}{lrrl@{\hspace{6pt}}rl}
\hline\hline
\noalign{\vskip 2pt}
Field & \multicolumn{1}{c}{RA} & \multicolumn{1}{c}{\ Dec} & \multicolumn{1}{c}{Run} & Duration & Focus\\
&&&& [days]\\
\noalign{\vskip 1pt}
\hline
\noalign{\vskip 2pt}
NEP & $\ang{270.00}$& $\ang{66.56}$& PDC-2 & 3.2 & out\\
 & & & PDC-3 & 3.2 & out\\
PSF-24 & $\ang{265.07}$& $\ang{71.35}$& PDC-4 & 1.25 &in\\
 & & & PDC-5 & 3.2 & out\\
Polarised & $\ang{328.00}$& $\ang{70.76}$& PDC-5 & 3.2 & out\\
PSF-16 & $\ang{59.25}$& $\ang{-55.16}$& PDC-6 & 3.2 & out\\
PSF-10 & $\ang{221.47}$& $\ang{67.43}$& PDC-7 & 1.25 & in\\
PSF-23 & $\ang{87.60}$& $\ang{-47.50}$& PDC-8 & 1.25 & in\\
\hline
\end{tabular}
\label{tab:pdccalib}
 \tablefoot{Run PDC-5 was the longest one, since it included observations of two low- and high-polarisation fields. The last column indicated if the observations were taken in-focus or out-of-focus.}
\end{table}

Within the stringent requirements of maximum SAA drift, four runs were implemented with sequences of out-of-focus VIS observations taking 3.2 days each, with the exception of PDC-5 that contained two sets of observations, one in a low-polarisation and the other in a high-polarisation field. The remaining three runs implemented sequences of in-focus VIS observations of 1.25 days duration. The implementation was made by the \ac{SOC} through the \ac{ICR} mechanism. 
Whenever possible, the observations were made in the set of predefined PSF VIS fields used in the regular VIS PSF calibration observations during the EWS. An additional `polarised' was defined, as well as a new PSF field at high ecliptic latitude (PSF-24) to ensure visibility to perform the tandem-sequences of PDC-5.
NEP was used in PDC-2 and PDC-3 when no PSF field was visible. The PDC calibration fields used in this campaign are listed in Table~\ref{tab:pdccalib}.

\begin{table*}
\caption{Summary properties of the seven PDC science fields.}
\begin{tabular}{llrrcrrc@{\hspace{6pt}}rrrl}
\hline \hline
\noalign{\vskip 2pt}
Survey & Field & \multicolumn{1}{c}{RA} & \multicolumn{1}{c}{\ Dec} & Run & SAA & AA & No.~ObsIDs & Depth & Area & Area & Seq. \\
&&&&&&&&median&1 pass&4 passes&\\
&&&&&&&&&$[{\rm deg}^2]$&$[{\rm deg}^2]$&\\
\noalign{\vskip 1pt}
\hline
\noalign{\vskip 2pt}
ILS & ILS-WAVES-S & $\ang{355.7}$ & $\ang{-31.2}$ & PDC-2 & $\ang{89.7}$ & $\ang{-4.5}$ & 135 &  1.5 & 26.5 & 0.0 & ROS \\
\hline
\noalign{\vskip 2pt}
 & EMDS-WAVES-S & $\ang{8.2}$ & $\ang{-30.7}$ & PDC-3 & $\ang{90.2}$ & $\ang{-6.0}$ & 132 &  4.3 & 14.0 & 9.8 & VIS \\
 & EMDS-HSC-HL-1 & $\ang{208.9}$ & $\ang{43.5}$ & PDC-5 & $\ang{97.3}$ & $\ang{-3.5}$ & 134 &  4.3 & 14.1 & 10.0 & VIS \\
EMDS  & EMDS-HSC-HL-2 & $\ang{237.2}$ & $\ang{44.2}$ & PDC-6 & $\ang{87.3}$ & $\ang{-3.5}$ & 143 &  4.7 & 12.3 & 9.2 & VIS \\
& EMDS-HSC-HL-3 & $\ang{214.7}$ & $\ang{44.3}$ & PDC-7 & $\ang{107.9}$ & $\ang{-4.5}$ & 140 &  2.8 & 13.4 & 3.6 & ROS \\
& EMDS-HSC-HL-4 & $\ang{230.2}$ & $\ang{43.6}$ & PDC-8 & $\ang{101.2}$ & $\ang{-4.5}$ & 137 &  3.0 & 10.9 & 3.8 & ROS \\
\hline
\noalign{\vskip 2pt}
EES & SSO-ECLIPTIC & $\ang{9.8}$ & $\ang{4.2}$ & PDC-4 & $\ang{95.9}$ & $\ang{-8.0}$ & 149 &  3.5 & 16.8 & 4.5 & ROS \\
\hline
\end{tabular}
\label{tab:pdc}
 \tablefoot{Columns include the coordinates of the centres of the fields and the values of the solar angles fixed in the run. One pass is a 4-dithered sequence (either ROS or VIS-only, depending on the field), and corresponds to 1 obsID. The median number of passes across the field in the field's observing sequence, the area covered by 1 pass and the area covered by at least 4 passes, in the field's observing sequence, is also given, as well as the observing sequence.}
\end{table*}

\begin{figure}
    \centering
    \includegraphics[width=1.0\linewidth]{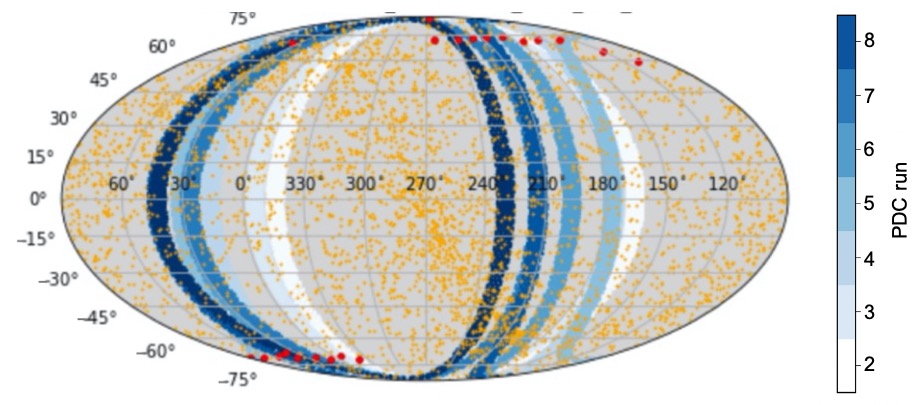}
    \caption{Visibility windows for the seven runs of PDC-2 through PDC-8, on a full-sky map in the ecliptic frame. Each visibility window is shaped as a blue-shaded stripe $7^\circ$ wide at the ecliptic and is computed for a specific value of the relevant input parameters (SAA, duration, and starting date). The set of PSF fields, potential targets for the calibration part of the run, are shown as red dots. Orange dots, which are distinctively denser in the Galactic plane, represent stars. The runs are designed to avoid pointing to bright stars in the stripes.
    }
    \label{fig:thermstab}
\end{figure}

\subsubsection{\label{sc:PDCsf} PDC science fields}

The stability in the thermal conditions required in the PDC campaign is much stricter than that used in standard \ac{EWS} observations. It is thus necessary to allocate thermal stabilisation periods before each calibration run, where the spacecraft keeps its  solar attitude fixed at the target SAA/AA value used in the subsequent calibration sequence. The first runs empirically confirmed the optimal duration of the thermal stabilisation period to be 7 days. The total thermalisation period is thus 50 days, accounting for $70\,\%$ of the duration of the PDC campaign. 

The requirement of observing with a fixed attitude during 7 days with an imposed starting date, translates into the visibility windows shown in Fig.~\ref{fig:thermstab}, one for each run. Those windows indicate that viable observing strategies must point along circles at a constant angle to the Sun. With the sky drifting roughly $1^\circ$ per day, an ROS taking 1/20th of a day, and \Euclid's FoV being $\theta$ degrees wide depending on its orientation (with $\theta$ between $\ang{0.7}$ and $\ang{1.0}$), a contiguous survey can be obtained by returning to the same ecliptic latitude $b$ after $20\,\theta\,{/}\,\cos b$ ROSs (with each new ROS stepping a full field along the meridian, to arrive at nominal \ac{EWS} depth), or after $20\,\theta/(N\cos b)$ ROSs for $N$ times the depth (by stepping $1/N$th of a field). In other words, the shorter the height of a patch of FoVs the larger the overlaps.

These constraints make it impractical to use the thermal stabilisation periods to start building the EWS, since the observations would be made in isolated patches difficult to connect later on with the EWS RSD progression. Moreover, the observations would be less efficient, with larger overlaps and heterogeneous depth. Alternative proposals that would benefit from the opportunity of taking multiple exposures in a confined region, adjusted to the natural drifting of the sky in the telescope frame, had to be considered.

The observations of the \ac{PDC} science programmes were implemented by the \ac{SOC} using a purpose-modified version of the \essot scheduling tool. 
In particular, a new patch-scheduling strategy was developed to automatically generate sequences of FoVs with large overlaps in user-defined patches. The FoVs are observed with the ROS and follow a time sequence that keeps SAA and AA constant during the 7-day run, defining a patch with height strongly correlated with its width and depth. Moreover, for the same height, patches at higher ecliptic latitudes have larger FoV overlaps. As mentioned before, shorter patches have larger FoV overlap, but this results in a heterogeneous depth across the patch, with varying coverage from the borders to the inner regions. This led to the introduction of a further constraint when defining the survey strategy, namely ensuring the coverage variation across a patch to remain below approximately 30
In addition to developing this new patch-scheduling strategy, \essot also implemented a mechanism to avoid bright stars, and created a new feature of tilted patch geometry. 

When executing the first run, PDC-2, a sudden drop of around 60\,mK occurred in the base-plate temperature when starting the VIS calibration sequence. This was found to be caused by turning off NISP, that had been active in the thermalisation period for ROS observations of the first PDC science field. In the second run, PDC-3, ROS was again used in the first pointings of the thermalisation period, but was soon turned off in advance of the next calibration sequence. NISP was kept off 
during the thermalisation periods of all the subsequent runs that included out-of-focus observations. In result of this, the ROS was only used during the thermalisation periods that preceded the in-focus-only sequences, where the thermal constraints are weaker (runs PDC-4, 7, and 8). This measure had a significant impact on \ac{EMDS} where only two of the five fields have NISP data. 

The selection process of the PDC science fields is described in \cref{sc:emds,sc:ils,sc:sso}.
Three scientific programmes were defined and executed: the \ac{EMDS}; the \ac{ILS} field; and the \ac{EES}. The three programmes provided seven \Euclid PDC science fields, 
with the \ac{EMDS} comprising five of the fields. 
Table~\ref{tab:pdc} summarizes the seven PDC science fields and their properties computed from the DR1 LE2 data.  One pass is defined as one 4-dithered sequence, either ROS or VIS depending on the field, 1 ObsID corresponds to a 4-dithered sequence, and the depth is given in number of passes. The total area covered by at least 1 pass in VIS is $119\,{\rm deg}^2$, while the total area covered by at least 1 pass in NISP is a significantly lower $71\,{\rm deg}^2$, leading to a total area of $68\,{\rm deg}^2$ covered by at least 1 pass in ROS. 

\subsection{\label{sc:emds} The Euclid Medium-Deep Survey}

\begin{figure*}
    \centering
    \includegraphics[width=1.0\linewidth]{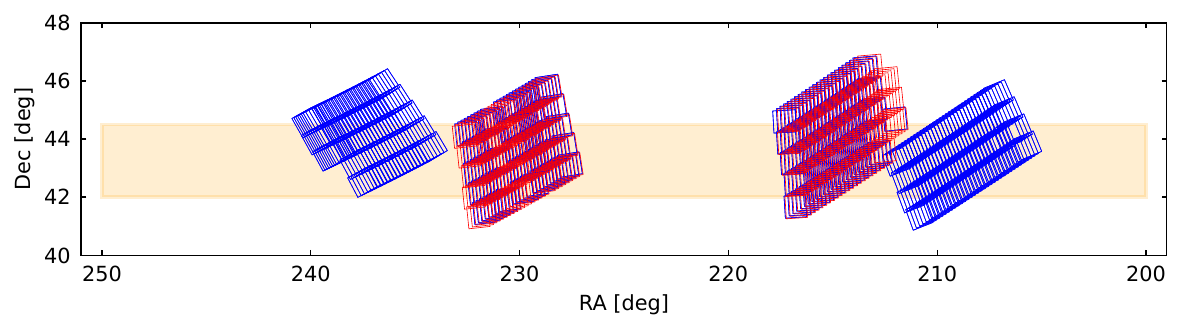}
    \caption{Layout of the four \ac{EMDS} fields in the northern sky. From left to right: EMDS-HSC-HIGHLAT-2; EMDS-HSC-HIGHLAT-4; EMDS-HSC-HIGHLAT-3; and EMDS-HSC-HIGHLAT-1.
    VIS exposures are plotted in blue and NIR exposures in red, with a small offset for clarity. The area with available HSC-SSP data is indicated in orange. Note that unlike for the \ac{EWS}, there is considerable overlap between exposures, resulting in the over 4 times deeper exposure time in VIS. 
    }
    \label{fig:mdsn}
\end{figure*}

The Euclid Consortium's survey scientist at the time had since long advocated the usefulness of \Euclid having an intermediate survey, 1 magnitude deeper and 1 order of magnitude smaller than the \ac{EWS}. This idea from the start dubbed a `medium deep survey' would fill the gap between the \ac{EDS} and the \ac{EWS}, since the former is 2 magnitudes deeper and over two orders of magnitude smaller than the latter.
Such a survey would be achieved with six \ac{ROS} visits (five additional visits to the one executed by the \ac{EWS}) over an area of at least $500 \,{\rm deg}^2$. 

The PDC thermalisation periods offered a unique opportunity to make repeated visits of the same area, and implement this long-standing proposal, in spite of not being sufficiently long to enable the coverage of the large area originally intended. 
It was thus a question of finding the right latitude and patch height to achieve the desired depth. On the other hand, the possible longitudes are constrained by the time of the observing runs, i.e., by the visibility windows of Fig.~\ref{fig:thermstab}, which also implies that the survey cannot be contiguous but will need to be spread over separate fields.

An additional input to the problem of finding the optimal location for the \ac{EMDS} fields is the fact that,  as in the \ac{EWS}, \Euclid data need to be complemented with ground-based photometric data to be maximally useful. An extra requirement on the \ac{EMDS} is therefore the presence of suitably deep multi-band surveys. This was largely achieved by leveraging the Hyper Supreme-Cam Subaru Strategic Programme (HSC-SSP) in the North \citep{HSCSSP:2017}, and the Kilo-Degree Survey (KiDS-VIKING) in the South \citep{KiDS:2013}. Both surveys add depth to the ground-based external data gathered for the EWS, and also form the basis for major spectroscopic surveys, with the Subaru Prime Focus Spectrograph (PFS) in the north \citep{PFS:2014} and the Wide Area VISTA Extragalactic Survey (WAVES) on the 4MOST spectroscopic facility in the south \citep{Driver:2019}. Fortunately, these two surveys consist of long constant-declination strips on the sky, providing available fields at a wide range of ecliptic longitudes. 

\begin{figure}
    \centering
    \includegraphics[width=1.0\linewidth]{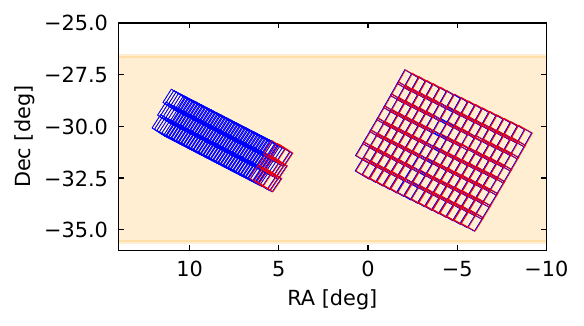}
    \caption{Layout of EMDS-WAVES-S (left) and ILS-WAVES-S  (right) in the southern sky at similar ecliptic latitude. The area with KiDS-VIKING data is indicated in orange. Note that the \ac{ILS} field is a taller patch with larger offsets (50\,\%\ overlaps between tiles), resulting in an exposure time that is roughly double than that of the \ac{EWS}, while the larger overlaps of the \ac{EMDS} field result in a larger exposure time that is over 4 times that of the \ac{EWS}.}
    \label{fig:mdss}
\end{figure}

The EMDS programme was allocated in five of the runs, defining the five separate fields listed in Table~\ref{tab:pdc}, four overlapping the HSC-SSP high latitude area and one in the KiDS-VIKING region. Observations started in the south during the PDC-3 run, from 2023-12-13T10:19:22 to 2023-12-20T06:31:46, and then moved to the northern strip during runs PDC-5 to PCD-8, in the following observing epochs: 2024-01-02T00:18:02 to 2024-01-08T22:57:48; 2024-01-15T13:31:43 to 2024-01-22T06:02:13; 2024-01-26T00:45:01 to 2024-02-02T00:34:55; and 2024-02-03T11:52:19 to 2024-02-10T11:34:15.

The five \ac{EMDS} fields are shown in Figs.~\ref{fig:mdsn} and \ref{fig:mdss}, overlaid on  the ground-based target areas. Their footprints show the individual VIS and NISP pointings computed with DR LE1 data. This choice highlights the difference between the LE1 and LE2 contents. For instance, EMDS-WAVES-S, the left field in Fig.~\ref{fig:mdss}, shows a number of NISP-only exposures on the edge of the field, that are not present in LE2 that only contains the ROS pointings for this field. These correspond to the first 13 pointings of the PDC-3 run that were still executed with NISP before being turned off.

The four EMDS-HSC fields, shown in Fig. ~\ref{fig:mdsn} were named following the observing order, which does not follow an ordered sequence of longitudes due to the non-monotonic sequence of SAA values used in the runs. The fields observed in the two final runs, EMDS-HSC-HIGHLAT-3 and 4, are the only ones where NISP was turned on. Unfortunately, their 4-pass areas are significantly smaller than the 4-pass areas of the other EMDS fields. This happens  for three reasons. On the one hand, the total overlap in exposures of two instruments is naturally smaller than the overlap of exposures of a single instrument. On the other hand, both runs experienced instrument outages that led to the loss of VIS data in around 25 obsIDs and NISP data in around 28 obsIDs in each run. In addition, these two fields are also the tallest ones, i.e, the most extended ones in ecliptic latitude. This implies that the resulting overlap between FoVs is smaller than in the shorter fields. This effect is striking in Fig.~\ref{fig:mdss}, which shows the two \Euclid PDC WAVES-S fields, the EMDS and the ILS (see Sect.~\ref{sc:ils}). The ILS field is a much taller patch, resulting in larger offsets and in a much smaller multiple exposed area. 

In summary, the area with at least 4 times the depth of the EWS  observed in both VIS and NISP in the EMDS only amounts to $7.3\,{\rm deg}^2$ (see 
Table~\ref{tab:pdc}). In contrast, the total area of the EMDS covered with at least 4 passes in VIS is $43\,{\rm deg}^2$, while the total area of the EMDS covered with at least 1 pass in VIS is $67\,{\rm deg}^2$. Both depth and area are far from those originally proposed for an EMDS. At the moment, there are no defined plans for the continuation of this survey, except the fact that all five fields will be observed one more time as part of the \ac{EWS}, some for DR2 and others for DR3, depending on future \ac{RSD} replanning. Exposure maps of the seven PDC science fields, produced with DR1 LE2 data, are shown in Appendix \ref{app:expmaps}.

\begin{figure}
    \centering
    \includegraphics[width=0.8\linewidth]{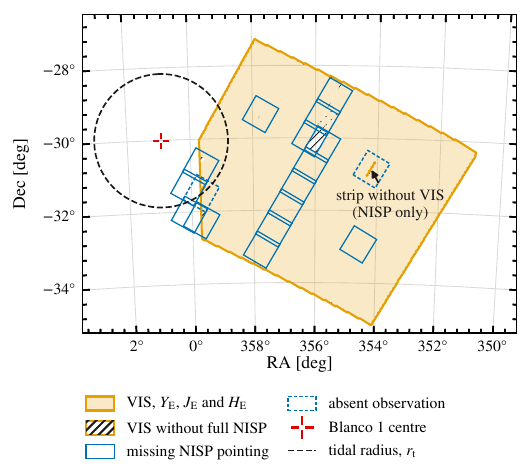}
    \caption{ILS field covered in all \Euclid bands except in the labelled areas. The field intersects the Blanco 1 open cluster tidal radius.}
    \label{fig:ils}
\end{figure}

\subsection{\label{sc:ils} The ILS field}

The two ILS scientists were granted a slot in thermalisation blocks 2 or 3 to propose a joint science programme. 
The field selection was made by consensus between the ILS scientists. Its properties were defined so that the field is part of the 4MOST/WAVES Wide South spectroscopic survey 
\citep{Driver:2019}, which has previous $u$-, $g$-, $r$-, $i$-, $Z$-, $Y$-, $J$-, $H$-, and $K_{\rm s}$-band imaging coverage by the VIKING \citep{Edge:2013} and KiDS surveys 
\citep{KiDS:2013, Wright:2024}
on VISTA and VST respectively, 
and that it should include members of a nearby young open cluster.  
The pre-existing, moderate-depth $Y$, $J$, and $H$ imaging can be used for proper motions and a high density of redshifts will be obtained by the WAVES survey. 
The ILS programme aligned with the WAVES survey is primarily to study dwarf galaxies in the local cosmological volume using Euclid's imaging for detection and characterization of 
low-surface brightness galaxies. 
This is complemented by the WAVES Wide survey prioritization for obtaining spectra of galaxies at redshifts $<0.2$. 
A second \Euclid epoch is scheduled for DR3, when the field will be observed as part of the EWS.
 
The observations were performed from 
2023-12-03T02:04:48 to 2023-12-10T02:22:58
in the PDC-2 run. The area covered by any number of exposures is $35.6\,{\rm deg}^2$, the area covered by at least 1 ROS is $26.5\,{\rm deg}^2$ (see Table~\ref{tab:pdc}), and the depth is 1--2 $\times$ ROS depth (1.5 for ROS and 1.75 for VIS-only exposures). The sequence of observations included 135 obsIDs, with two pointings missing in VIS, 16 in \YE, 16 in \JE, and 15 in \HE, as shown in Fig.~\ref{fig:ils}.

This programme managed to cover part of the nearby (distance 234\,pc) Blanco 1 open cluster's corona and tidal tails and nothing of its core. Ten new comoving substellar-mass ultracool cluster member candidates have been identified using the \Euclid photometry and spectra, and proper motions determined by comparison with ground-based imaging. These results extend the cluster sequence to cooler temperatures and fainter luminosities, build up on previous identifications of ultracool dwarfs based on ERO \citep{Martin:2025} and Q1 \citep{Zerjal:2026} data, and demonstrate the potential of \Euclid to cover large area with deep observations that are crucial to reveal the very low mass population of nearby open clusters, particularly in the tidal tails. More observations of this cluster are needed for full coverage and proper motion studies of substellar candidate members.

\subsection{\label{sc:sso} The Euclid Ecliptic Survey}

The remaining thermal stabilisation period was allocated to perform a survey on the ecliptic plane. The on-sky distribution of \acp{SSO} is heavily skewed towards the ecliptic plane. Based on the expected \Euclid footprint, it has been estimated that about 150\,000 \acp{SSO} would be observed during the EWS \citep{Carry2018}. The estimate is, however, dependent on the actual size-frequency distributions of SSO populations and ecliptic latitude excursion of the survey. It was nevertheless clear that observing closer to the ecliptic plane would significantly increase the number of SSOs detected in each field. The \Euclid Science Working Group SSOs therefore proposed a PDC survey focusing on the ecliptic plane.

\begin{figure}
    \centering
    \includegraphics[width=1.0\linewidth]{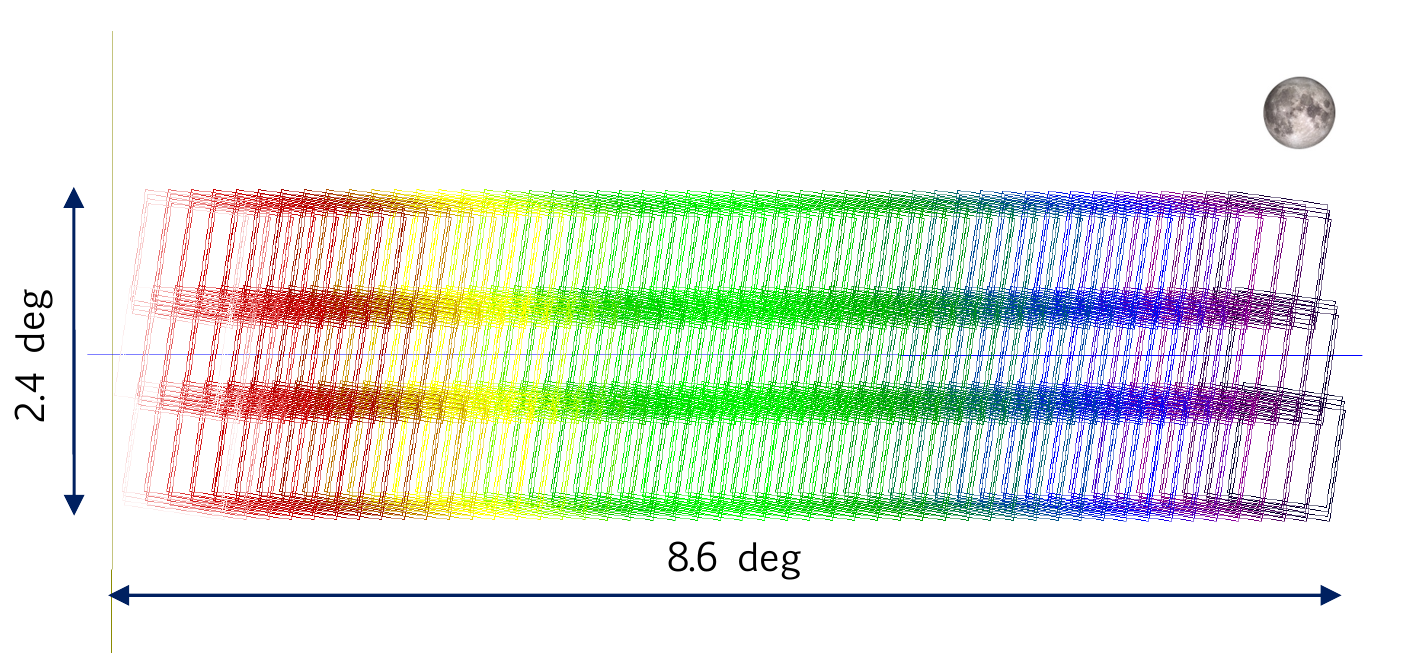}
    \caption{Observing strategy of the EES, which starts on the rightmost squares on 23 December 2023 (black) and ends on the leftmost squares on 31 December 2023 (pink). The approximate centre point is on the ecliptic (horizontal line) at the ecliptic longitude of $\ang{10.7}$. Each square corresponds to a single VIS exposure. Ecliptic longitude increases from right to left, and the ecliptic latitude increases from bottom to top (ecliptic north is up). The Moon is displayed for scale.}
    \label{fig:ees}
\end{figure}

The Euclid Ecliptic Survey, the PDC survey for \acp{SSO}, was designed to maximise the number of observations of \acp{SSO}, both in number of objects and in number of detections per object. During the campaign, from 
2023-12-23T20:36:57 to 2023-12-31T14:18:20 in run PDC-4, the telescope was pointed at, or very close to, the ecliptic plane and approximately followed the average motion of objects in the main asteroid belt, thanks to the high overlap observing strategy implemented, which ensures that the asteroids are seen in as many exposures as possible (Fig.~\ref{fig:ees}). This strategy allowed us to detect more \acp{SSO} and extract more astrometry over a longer timespan, as well as denser, longer rotational lightcurves, than would be possible in the EWS, which avoids the ecliptic plane and typically produces four detections of each moving target in its ROS observations. 
The total area covered by any exposure is about $19\,{\rm deg}^2$, while roughly $17\,{\rm deg}^2$ are covered by at least 1 ROS (see Table~\ref{tab:pdc}).

The outcome of the \ac{EES} is presented in two papers. P\"ontinen et al. (in prep.) presents the survey and the extraction of astrometry and photometry of about 4000 trailed moving objects (virtually all interior to Jupiter's orbit) from VIS exposures with tools developed for that particular task \citep{Pontinen:2020} as well as NISP photometry for the same objects. The data allow us to test current knowledge on the orbit, size, and VIS/NISP taxonomic distributions beyond the completeness level reached by the other contemporary surveys. The other paper \citep{Irureta-Goyena:2026} describes extracted lightcurves from the streaks of 2321 known asteroids detected by P\"ontinen et al. (in prep.), 93\,\% of which lack information about their rotation periods. Comparing to asteroids with rotation periods reported in the literature, they found that 98\,\% of the rotation periods extracted from EES agree to within 15\,\% of the rotation periods found in the literature. While the large majority of the asteroids analysed have rotation rates in agreement with the typical ranges for their physical size, there are also 16 superfast rotators with orbital periods shorter than the canonical 2.2 hour spin barrier. 

\begin{table*}
\centering
\caption{The Euclid Surveys at DR1.}
\begin{tabular}{llllll}
\hline\hline
\noalign{\vskip 2pt}
Survey component & Epoch & Contents & Procedure & \multicolumn{1}{c}{Scheduled} & \multicolumn{1}{c}{Release} \\
&&&& \multicolumn{1}{c}{notes} & \multicolumn{1}{c}{notes}\\
\noalign{\vskip 1pt}
\hline
\noalign{\vskip 2pt}
Wide survey & RSD &  EWS tiles & \ectile & $2639\,{\rm deg}^2$ & $2099\,{\rm deg}^2$ (LE1),  $1900\,{\rm deg}^2$ (LE2)\\
\hline
  \noalign{\vskip 2pt}
Deep survey   &  RSD & EDF-N, EDF-S, EDF-F & \ectile & 7 EDF-N & $28\%$, $22\%$, $4\%$ of final depth  \\
&&&&visits lost&  \\
   &  PV & EDF-N, EDF-S & ICR & &not included in LE2\\
\hline
  \noalign{\vskip 2pt}
Auxiliary fields & RSD & COSMOS, CDFS, AEGIS & \ectile &  &$36\%$, $100\%$, $100\%$ of final depth\\
 & PV & COSMOS, CDFS & ICR &  &  only COSMOS in LE2\\
\hline
  \noalign{\vskip 2pt}
Calibration fields &  RSD & Self-calibration, PN, PSFs & \ectile & & \\
&   PV & Self-calibration, EBFs, & ICR &  &   only PV-001 and F-014 in LE2\\
&& calibrators, LDN1641&&& LDN also in Q1 \\
&  PDC & PSFs & ICR & defocus &  \\
\hline
  \noalign{\vskip 2pt}
PV science fields & PV & PV-006, PV-007, SLs& ICR & &  not in LE2, PV-007 also in ERO  \\
\hline
  \noalign{\vskip 2pt}
PDC surveys & PDC & EMDS & \essot &  &  $64.7\,{\rm deg}^2$, of which $36.4\,{\rm deg}^2$ with 4 passes\\
& & ILS field, EES & \essot &&$26.5\,{\rm deg}^2$, $16.8\,{\rm deg}^2$\\
\hline
\end{tabular}
\label{tab:summarytable}
 \tablefoot{For each survey component, the table indicates the observation epochs, its fields, and the scheduling procedure applied. Information is given on the EWS area variation between RSD-scheduled, LE1, and LE2, as well as on the DR1 depths of the EDFs and EAFs. The ROS-observed areas of the PDC surveys are given and some release notes on the PV fields are included.}
\end{table*}

\begin{figure*}
    \centering
    \rotatebox{90}{
        \begin{minipage}{0.992\textheight}
            \centering
        \includegraphics[width=\linewidth]{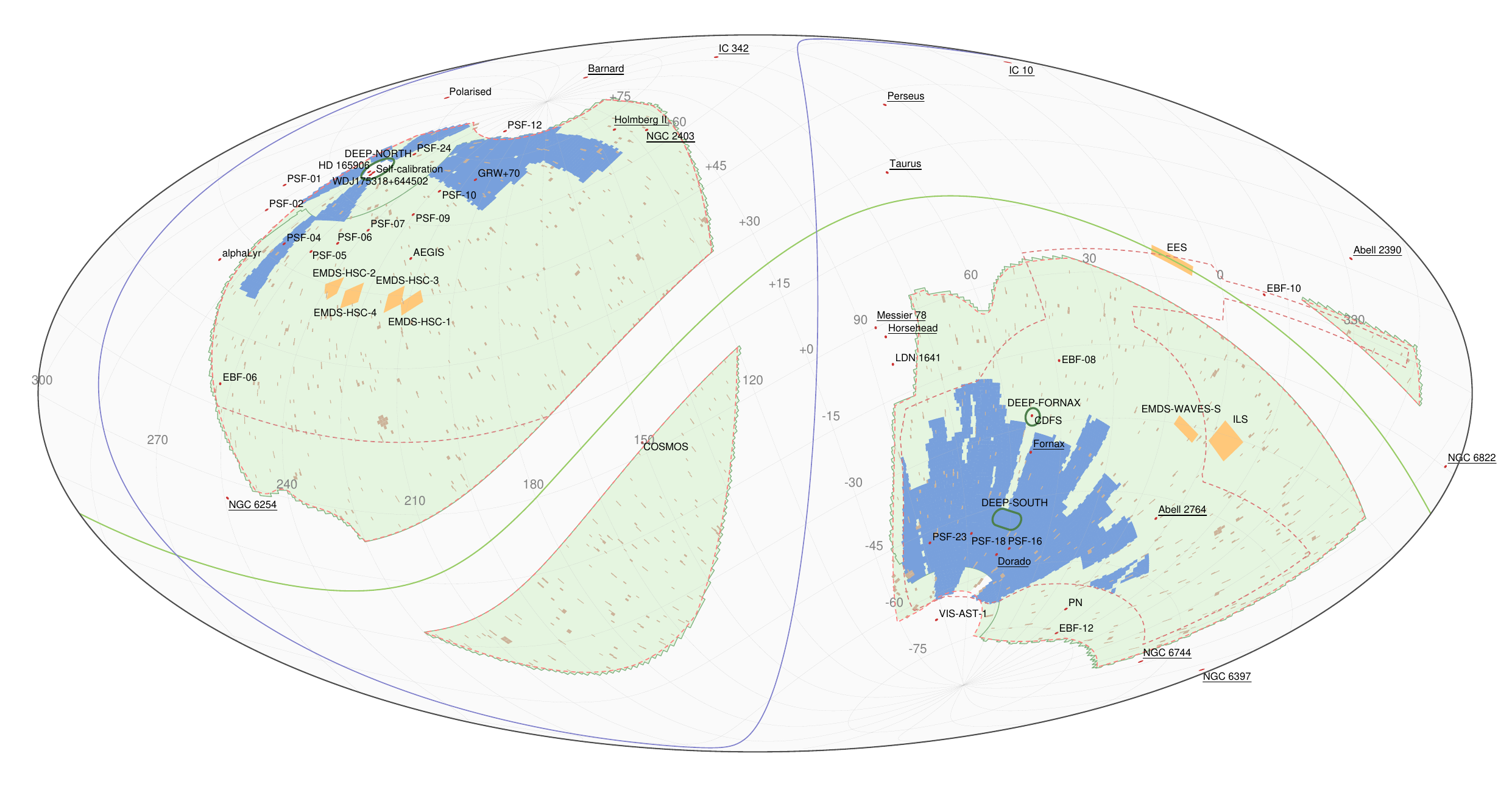}
            \caption{The complete set of \Euclid Surveys at DR1, with the exception of the 174 PV-006 and PV-035 fields to avoid cluttering the map (see Fig.~\ref{fig:zodi}), are shown in the equatorial frame rotated around the central point (${\rm RA} = \ang{120;;}, {\rm Dec} = \ang{0;;}$) to improve visibility of the polar regions. The EWS footprint (Sect.~\ref{sc:RSDews}) is shown in blue. The \ac{RSD} targets (Sect.~\ref{sc:RSDtargets}) included in DR1, i.e., the observed \acp{EDF}, \acp{EAF}, and calibration fields (Self-calibration, planetary nebula and PSF fields) are also represented in the map, as are the PV fields,  including the 17 PV-007 ERO fields with  underlined names (Sect.~\ref{sc:pvfields}), 
            the PDC calibration fields (Sect.~\ref{sc:PDCcf}) and the PDC surveys shown in orange (Sect.~\ref{sc:PDCsf}). Brown areas scattered on the map are the bright star exclusion regions. Dashed red countours mark the UNIONS (in the North) and DES (in the South) footprints where ground-based data are available. The orange line shows the ecliptic, and the blue line the Galactic plane. The light green background is the \Euclid \ac{RoI}, consisting of two mainlands and two islands, where the EWS continues its progress beyond DR1.   
    \label{fig:summaryfig} 
    }
        \end{minipage}
    }
\end{figure*}

\section{\label{sc:conclusion} Conclusions}

We have presented the status of the \Euclid surveys in the DR1 data release of 12 November 2026 \citep{DR1-TP001}. DR1 is the first major \Euclid data release, following the Early Release Observation of 23 May 2024 \citep{EROData}, consisting of PV-007 data, the Quick Data Release Q1 of 19 March 2025 \citep{Q1-TP001}, consisting of data from one visit to each of the \acp{EDF} and the LDN 1641 dark cloud, and the Quick Data Release Q2 of 24 June 2026 \citep{Q2-TP002}, consisting of the Euclid Galactic Bulge Survey. 

\Euclid surveys refer to the various components of the overall programme that are executed during the three main phases of the mission, namely, the PV phase, the PDC campaign performed in the early science operations phase, and the much longer science operations phase, dubbed in this paper the RSD phase. The focus of this paper is on the survey implementation procedures applied in the three phases by both the Euclid Consortium and the SOC, as well as on the scheduling properties of the \Euclid fields defined by the various survey components. The \Euclid surveys are summarised in Fig.~\ref{fig:summaryfig} and in Table~\ref{tab:summarytable}. Figure~\ref{fig:summaryfig} shows the location and extent of all \Euclid fields included in the DR1 data release, while Table~\ref{tab:summarytable} provides a summary of the contents, epoch of observation, and scheduling procedure, of the \Euclid surveys,

%
%

\begin{acknowledgements} 
\AckEC  
\AckDRone
\end{acknowledgements}

%
%

\bibliography{Euclid, Survey} 

@ARTICLE{DR1-TP001,
        author ={{Euclid Collaboration: Aussel}, H. and others},
        title = "{Euclid Data Release 1 (DR1). Data release overview}",
        journal = {arXiv, submitted},
        year = 2026
        }

@ARTICLE{DR1-TP002,
        author ={{Euclid Collaboration: McCracken}, H.J. and {Benson}, K. and {Cropper}, M. and others},
        title = "{Euclid Data Release 1 (DR1). VIS processing and data products}",
        journal = {A\&A, submitted},
        year = 2026
        }

@ARTICLE{DR1-TP003,
        author ={{Euclid Collaboration: Polenta}, G. and {Frailis}, M. and {Alavi}, A. and others},
        title = "{Euclid Data Release 1 (DR1). Near-infrared processing and data products}",
        journal = {A\&A, submitted},
        year = 2026
        }

@ARTICLE{DR1-TP006,
        author ={{Euclid Collaboration: Corcho-Caballero}, P. and {Balbinot}, E. and {De Jong}, J. and others},
        title = "{Euclid Data Release 1 (DR1). Colouring Euclid with external ground-based imaging data}",
        journal = {A\&A, submitted},
        year = 2026
        }

@ARTICLE{DR1-TP009,
        author ={{Euclid Collaboration: K\"ummel}, M. and {Caro}, F. and {Dimauro}, P. and others},
        title = "{Euclid Data Release (DR1). The MERge processing}",
        journal = {A\&A, submitted},
        year = 2026
        }

@ARTICLE{DR1-TP011,
        author ={{Euclid Collaboration: Hartley}, W. and {Paltani}, S. and {Dubath}, F. and others},
        title = "{Euclid Data Release 1 (DR1). Photometric redshift pipeline overview and validation}",
        journal = {A\&A, submitted},
        year = 2026
        }

@ARTICLE{DR1-TP023,
        author ={{Euclid Collaboration: Scodeggio}, M. and {Copin}, Y. and others},
        title = "{Euclid Data Release 1 (DR1). From spectrograms to spectra: The SIR spectroscopic processing function}",
        journal = {in preparation},
        year = 2026
        }

@ARTICLE{DR1-TP025,
        author ={{Euclid Collaboration: Le Brun}, V. and {Bethermin}, M. and {Moresco}, M. and others},
        title = "{Euclid Data Release 1 (DR1). Performance of the redshift measurement and spectroscopic analysis}",
        journal = {A\&A, submitted},
        year = 2026
        }

@ARTICLE{DR1-TP026,
        author ={{Euclid Collaboration: Schirmer}, M. and {Paterson}, K. and {Alavi}, A. and others},
        title = "{Euclid Data Release (DR1). Calibration strategy and observations}",
        journal = {A\&A, submitted},
        year = 2026
        }

@ARTICLE{DR1-TP026b,
        author ={{Euclid Collaboration: Schirmer}, M. and {Paterson}, K. and {Altieri}, B.  and others},
        title = "{Euclid Data Release (DR1). Spacecraft performance in the L2 environment}",
        journal = {A\&A, submitted},
        year = 2026
        }

@ARTICLE{DR1-TP033,
        author ={{Euclid Collaboration: Ruffenach}, M. and {Frailis}, M. and {Zacchei}, A. and others},
        title = "{Euclid Data Release 1 (DR1). An overview of the Science Ground Segment architecture and operations}",
        journal = {A\&A, submitted},
        year = 2026
        }

@ARTICLE{Q2-TP002,
       author = {{Beaulieu}, J.-P. and {Bachelet}, E. and {Gilles}, M. and others},
       title = "{Euclid Quick Data Release (Q2) -- The Euclid Galactic Bulge Survey}",
      journal = {A\&A, accepted},
         year = 2026,
        month = jun,
          eid = {arXiv:2606.25883},
        pages = {arXiv:2606.25883},
archivePrefix = {arXiv},
       eprint = {2606.25883},
 primaryClass = {astro-ph.GA},
       adsurl = {https://ui.adsabs.harvard.edu/abs/2026arXiv260625883B}
}

@ARTICLE{Q1-TP001,
       author = {{Euclid Collaboration: Aussel}, H. and {Tereno}, I. and {Schirmer}, M. and others},
       title = "{Euclid Quick Data Release (Q1). I. Data release overview}",
      journal = {\aap},
         year = 2026,
        month = jun,
       volume = {711},
          eid = {A1},
        pages = {A1},
          doi = {10.1051/0004-6361/202554610},
       adsurl = {https://ui.adsabs.harvard.edu/abs/2026A&A...711A...1E}
}

@ARTICLE{EuclidSkyOverview,
author = {{Euclid Collaboration: Mellier}, Y. and {Abdurro'uf} and {Acevedo~Barroso}, J.A. and others},
	title = {Euclid - I. Overview of the Euclid mission},
	DOI= "10.1051/0004-6361/202450810",
	url= "https://doi.org/10.1051/0004-6361/202450810",
	journal = {A\&A},
	year = 2025,
	volume = 697,
	pages = "A1",
}

@ARTICLE{EuclidSkyVIS,
author = {{Euclid Collaboration: Cropper}, M. and {Al-Bahlawan}, A. and {Amiaux}, J. and others},
	title = {Euclid - II. The VIS instrument},
	DOI= "10.1051/0004-6361/202450996",
	url= "https://doi.org/10.1051/0004-6361/202450996",
	journal = {A\&A},
	year = 2025,
	volume = 697,
	pages = "A2",
}

@ARTICLE{EuclidSkyNISP,
author = {{Euclid Collaboration: Jahnke}, K. and {Gillard}, W. and {Schirmer}, M. and others},
	title = {Euclid - III. The NISP Instrument},
	DOI= "10.1051/0004-6361/202450786",
	url= "https://doi.org/10.1051/0004-6361/202450786",
	journal = {A\&A},
	year = 2025,
	volume = 697,
	pages = "A3",
}

@ARTICLE{EROData,
author = {{Cuillandre}, J.-C. and {Bertin}, E. and {Bolzonella}, M. and others},
	title = {Euclid: Early Release Observations – Programme overview and pipeline for compact- and diffuse-emission photometry},
	DOI= "10.1051/0004-6361/202450803",
	url= "https://doi.org/10.1051/0004-6361/202450803",
	journal = {A\&A},
	year = 2025,
	volume = 697,
	pages = "A6",
}

@misc{DR1cite,
author = "{Euclid Data Release 1}",
howpublished = "\url{https://doi.org/10.57780/esa-d079c51}",
year = 2026
}

@ARTICLE{EU-Anselmi,
       author = {{Euclid Collaboration: Anselmi}, A. and {Laureijs}, R. and {Racca}, G.~D. and others},
        title = "{Euclid: VII. Structural-thermal-optical performance}",
      journal = {\aap},
         year = 2026,
        month = may,
       volume = {709},
          eid = {A15},
        pages = {A15},
          doi = {10.1051/0004-6361/202557524},
archivePrefix = {arXiv},
       eprint = {2512.01075},
 primaryClass = {astro-ph.IM},
       adsurl = {https://ui.adsabs.harvard.edu/abs/2026A&A...709A..15E}
}

@ARTICLE{Schirmer-EP29,
       author = {{Euclid Collaboration: Schirmer}, M. and {Th{\"u}rmer}, K. and {Bras}, B. and others},
        title = "{Euclid preparation. XXIX. Water ice in spacecraft Part I: The physics of ice formation and contamination}",
      journal = {\aap},
         year = 2023,
        month = jul,
       volume = {675},
          eid = {A142},
        pages = {A142},
          doi = {10.1051/0004-6361/202346635},
archivePrefix = {arXiv},
       eprint = {2305.10107},
 primaryClass = {astro-ph.IM},
       adsurl = {https://ui.adsabs.harvard.edu/abs/2023A&A...675A.142E}
}

@ARTICLE{Scaramella-EP1,
       author = {{Euclid Collaboration: Scaramella}, R. and {Amiaux}, J. and {Mellier}, Y. and others},
        title = "{Euclid preparation. I. The Euclid Wide Survey}",
      journal = {\aap},
         year = 2022,
        month = jun,
       volume = {662},
          eid = {A112},
        pages = {A112},
          doi = {10.1051/0004-6361/202141938},
archivePrefix = {arXiv},
       eprint = {2108.01201},
 primaryClass = {astro-ph.CO},
       adsurl = {https://ui.adsabs.harvard.edu/abs/2022A&A...662A.112E}
}

@ARTICLE{Neumann26,
       author = {{Neumann}, D. and {Miller}, L. and {Hoekstra}, H. and others},
       title = "{Euclid: Precise inference of defocus wavefront error from image diffraction spike measurements}",
      journal = {A\&A, accepted},
         year = 2026,
        month = jun,
          eid = {arXiv:2606.25625},
        pages = {arXiv:2606.25625},
archivePrefix = {arXiv},
       eprint = {2606.25625},
 primaryClass = {astro-ph.CO},
       adsurl = {https://ui.adsabs.harvard.edu/abs/2026arXiv260625625N}
}

@INPROCEEDINGS{Gomez-Alvarez18,
       author = {{G{\'o}mez-Alvarez}, Pedro and {Dupac}, Xavier and {Buenadicha}, Guillermo and {Vavrek}, Roland and {Hoar}, John and {Laureijs}, Ren{\'e} and {Scaramella}, Roberto and {Cuillandre}, J.~C. and {Dinis}, Joao and {Amiaux}, Jer{\^o}me and {Tereno}, Ismael},
        title = "{The Euclid survey planning system}",
    booktitle = {Software and Cyberinfrastructure for Astronomy V},
         year = 2018,
       editor = {{Guzman}, Juan C. and {Ibsen}, Jorge},
       series = {SPIE Conf. Ser.},
       volume = {10707},
        month = jul,
          eid = {1070712},
        pages = {1070712},
          doi = {10.1117/12.2314262},
       adsurl = {https://ui.adsabs.harvard.edu/abs/2018SPIE10707E..12G}
}

@article{Abbott:2021,
   title={The Dark Energy Survey Data Release 2},
   volume={255},
   ISSN={1538-4365},
   url={http://dx.doi.org/10.3847/1538-4365/ac00b3},
   DOI={10.3847/1538-4365/ac00b3},
   number={2},
   journal={ApJS},
   publisher={American Astronomical Society},
   author={Abbott, T. M. C. and Adamów, M. and Aguena, M. and Allam, S. and Amon, A. and Annis, J. and Avila, S. and Bacon, D. and Banerji, M. and Bechtol, K. and Becker, M. R. and Bernstein, G. M. and Bertin, E. and Bhargava, S. and Bridle, S. L. and Brooks, D. and Burke, D. L. and Carnero Rosell, A. and Carrasco Kind, M. and Carretero, J. and Castander, F. J. and Cawthon, R. and Chang, C. and Choi, A. and Conselice, C. and Costanzi, M. and Crocce, M. and da Costa, L. N. and Davis, T. M. and De Vicente, J. and DeRose, J. and Desai, S. and Diehl, H. T. and Dietrich, J. P. and Drlica-Wagner, A. and Eckert, K. and Elvin-Poole, J. and Everett, S. and Evrard, A. E. and Ferrero, I. and Ferté, A. and Flaugher, B. and Fosalba, P. and Friedel, D. and Frieman, J. and García-Bellido, J. and Gaztanaga, E. and Gelman, L. and Gerdes, D. W. and Giannantonio, T. and Gill, M. S. S. and Gruen, D. and Gruendl, R. A. and Gschwend, J. and Gutierrez, G. and Hartley, W. G. and Hinton, S. R. and Hollowood, D. L. and Honscheid, K. and Huterer, D. and James, D. J. and Jeltema, T. and Johnson, M. D. and Kent, S. and Kron, R. and Kuehn, K. and Kuropatkin, N. and Lahav, O. and Li, T. S. and Lidman, C. and Lin, H. and MacCrann, N. and Maia, M. A. G. and Manning, T. A. and Maloney, J. D. and March, M. and Marshall, J. L. and Martini, P. and Melchior, P. and Menanteau, F. and Miquel, R. and Morgan, R. and Myles, J. and Neilsen, E. and Ogando, R. L. C. and Palmese, A. and Paz-Chinchón, F. and Petravick, D. and Pieres, A. and Plazas, A. A. and Pond, C. and Rodriguez-Monroy, M. and Romer, A. K. and Roodman, A. and Rykoff, E. S. and Sako, M. and Sanchez, E. and Santiago, B. and Scarpine, V. and Serrano, S. and Sevilla-Noarbe, I. and Smith, J. Allyn and Smith, M. and Soares-Santos, M. and Suchyta, E. and Swanson, M. E. C. and Tarle, G. and Thomas, D. and To, C. and Tremblay, P. E. and Troxel, M. A. and Tucker, D. L. and Turner, D. J. and Varga, T. N. and Walker, A. R. and Wechsler, R. H. and Weller, J. and Wester, W. and Wilkinson, R. D. and Yanny, B. and Zhang, Y. and Nikutta, R. and Fitzpatrick, M. and Jacques, A. and Scott, A. and Olsen, K. and Huang, L. and Herrera, D. and Juneau, S. and Nidever, D. and Weaver, B. A. and Adean, C. and Correia, V. and de Freitas, M. and Freitas, F. N. and Singulani, C. and Vila-Verde, G. and },
   year={2021},
   pages={20} }

@article{Gwyn:2025,
title = "UNIONS: The Ultraviolet Near-infrared Optical Northern Survey",
author = "Stephen Gwyn and McConnachie, \{Alan W.\} and Jean-Charles Cuillandre and Chambers, \{Kenneth C.\} and Magnier, \{Eugene A.\} and \{de Boer\}, Thomas and Hudson, \{Michael J.\} and Masamune Oguri and Hisanori Furusawa and Hendrik Hildebrandt and Raymond Carlberg and Ellison, \{Sara L.\} and Junko Furusawa and Rapha{\"e}l Gavazzi and Rodrigo Ibata and Yannick Mellier and Ken Osato and H. Aussel and Lucie Baumont and Manuel Bayer and Olivier Boulade and Patrick C{\^o}t{\'e} and David Chemaly and Cail Daley and Pierre-Alain Duc and Florence Durret and A. Ellien and S{\'e}bastien Fabbro and Leonardo Ferreira and Fitriana, \{Itsna K.\} and \{Le Floc{\textquoteright}h\}, Emeric and Yoshinobu Fudamoto and Hua Gao and Goh, \{L. W. K.\} and Tomotsugu Goto and Sacha Guerrini and Axel Guinot and Vincent H{\'e}nault-Brunet and Francois Hammer and Yuichi Harikane and Kohei Hayashi and Nick Heesters and Kohei Ichikawa and Martin Kilbinger and Kuzma, \{P. B.\} and Qinxun Li and Liaudat, \{Tob{\'i}as I.\} and Chien-Cheng Lin and Oliver M{\"u}ller and Martin, \{Nicolas F.\} and Yoshiki Matsuoka and Medina, \{Gustavo E.\} and Hironao Miyatake and Satoshi Miyazaki and Mpetha, \{Charlie T.\} and Tohru Nagao and Navarro, \{Julio F.\} and Masafumi Niwano and Itsuki Ogami and Nobuhiro Okabe and Masafusa Onoue and Paek, \{Gregory S. H.\} and Parker, \{Laura C.\} and Patton, \{David R.\} and Peters, \{Fabian Hervas\} and Simon Prunet and Rub{\'e}n S{\'a}nchez-Janssen and M. Schultheis and Federico Sestito and Smith, \{Simon E. T.\} and J.-L. Starck and Else Starkenburg and Connor Stone and Christopher Storfer and Yoshihisa Suzuki and T. Erben and Salvatore Taibi and Thomas, \{G. F.\} and Yoshiki Toba and Hisakazu Uchiyama and David Valls-Gabaud and Venn, \{Kim A.\} and \{Van Waerbeke\}, Ludovic and Wainscoat, \{Richard J.\} and Scott Wilkinson and Anna Wittje and Taketo Yoshida and TianFang Zhang and Yuxing Zhong",
year = "2025",
month = dec,
day = "31",
doi = "10.3847/1538-3881/ae03ab",
language = "English",
volume = "170",
 pages = {324},
journal = "AJ",
issn = "0004-6256",
publisher = "American Astronomical Society",
number = "6",
}

@ARTICLE{Martin:2025,
       author = {{Mart{\'\i}n}, E.~L. and {{\v{Z}}erjal}, M. and {Bouy}, H. and {Martin-Gonzalez}, D. and {Mu{\~n}oz Torres}, S. and {Barrado}, D. and {Olivares}, J. and {P{\'e}rez-Garrido}, A. and {Mas-Buitrago}, P. and {Cruz}, P. and {Solano}, E. and {Zapatero Osorio}, M.~R. and {Lodieu}, N. and {B{\'e}jar}, V.~J.~S. and {Zhang}, J.-Y. and {del Burgo}, C. and {Hu{\'e}lamo}, N. and {Laureijs}, R. and {Mora}, A. and {Saifollahi}, T. and {Cuillandre}, J.-C. and {Schirmer}, M. and {Tata}, R. and {Points}, S. and {Phan-Bao}, N. and {Goldman}, B. and {Casewell}, S.~L. and {Reyl{\'e}}, C. and {Smart}, R.~L. and {Dominguez-Tagle}, C. and {Escobar}, A. and {Sedighi}, N. and {Tsilia}, S. and {Vitas}, N. and {Ayadi}, A. and {Aghanim}, N. and {Altieri}, B. and {Andreon}, S. and {Auricchio}, N. and {Baldi}, M. and {Balestra}, A. and {Bardelli}, S. and {Basset}, A. and {Bender}, R. and {Bonino}, D. and {Branchini}, E. and {Brescia}, M. and {Brinchmann}, J. and {Camera}, S. and {Capobianco}, V. and {Carbone}, C. and {Carretero}, J. and {Casas}, S. and {Castellano}, M. and {Cavuoti}, S. and {Cimatti}, A. and {Congedo}, G. and {Conselice}, C.~J. and {Conversi}, L. and {Copin}, Y. and {Corcione}, L. and {Courbin}, F. and {Courtois}, H.~M. and {Cropper}, M. and {Da Silva}, A. and {Degaudenzi}, H. and {Di Giorgio}, A.~M. and {Dinis}, J. and {Dubath}, F. and {Dupac}, X. and {Dusini}, S. and {Ealet}, A. and {Farina}, M. and {Farrens}, S. and {Ferriol}, S. and {Fosalba}, P. and {Frailis}, M. and {Franceschi}, E. and {Fumana}, M. and {Galeotta}, S. and {Garilli}, B. and {Gillard}, W. and {Gillis}, B. and {Giocoli}, C. and {G{\'o}mez-Alvarez}, P. and {Grazian}, A. and {Grupp}, F. and {Guzzo}, L. and {Haugan}, S.~V.~H. and {Hoar}, J. and {Hoekstra}, H. and {Holmes}, W. and {Hook}, I. and {Hormuth}, F. and {Hornstrup}, A. and {Hu}, D. and {Hudelot}, P. and {Jahnke}, K. and {Jhabvala}, M. and {Keih{\"a}nen}, E. and {Kermiche}, S. and {Kiessling}, A. and {Kilbinger}, M. and {Kitching}, T. and {Kohley}, R. and {Kubik}, B. and {K{\"u}mmel}, M. and {Kunz}, M. and {Kurki-Suonio}, H. and {Le Mignant}, D. and {Ligori}, S. and {Lilje}, P.~B. and {Lindholm}, V. and {Lloro}, I. and {Maino}, D. and {Maiorano}, E. and {Mansutti}, O. and {Marggraf}, O. and {Martinet}, N. and {Marulli}, F. and {Massey}, R. and {Medinaceli}, E. and {Mei}, S. and {Melchior}, M. and {Mellier}, Y. and {Meneghetti}, M. and {Meylan}, G. and {Mohr}, J.~J. and {Moresco}, M. and {Moscardini}, L. and {Niemi}, S.-M. and {Padilla}, C. and {Paltani}, S. and {Pasian}, F. and {Pedersen}, K. and {Percival}, W.~J. and {Pettorino}, V. and {Pires}, S. and {Polenta}, G. and {Poncet}, M. and {Popa}, L.~A. and {Pozzetti}, L. and {Racca}, G.~D. and {Raison}, F. and {Rebolo}, R. and {Renzi}, A. and {Rhodes}, J. and {Riccio}, G. and {Rix}, Hans-Walter and {Romelli}, E. and {Roncarelli}, M. and {Rossetti}, E. and {Saglia}, R. and {Sapone}, D. and {Sartoris}, B. and {Sauvage}, M. and {Scaramella}, R. and {Schneider}, P. and {Secroun}, A. and {Seidel}, G. and {Seiffert}, M. and {Serrano}, S. and {Sirignano}, C. and {Sirri}, G. and {Stanco}, L. and {Tallada-Cresp{\'\i}}, P. and {Taylor}, A.~N. and {Teplitz}, H.~I. and {Tereno}, I. and {Toledo-Moreo}, R. and {Tsyganov}, A. and {Tutusaus}, I. and {Valenziano}, L. and {Vassallo}, T. and {Verdoes Kleijn}, G. and {Wang}, Y. and {Weller}, J. and {Williams}, O.~R. and {Zucca}, E. and {Baccigalupi}, C. and {Willis}, G. and {Simon}, P. and {Mart{\'\i}n-Fleitas}, J. and {Scott}, D.},
        title = "{Euclid: Early Release Observations ─ A glance at free-floating newborn planets in the {\ensuremath{\sigma}} Orionis cluster}",
      journal = {\aap},
         year = 2025,
        month = may,
       volume = {697},
          eid = {A7},
        pages = {A7},
          doi = {10.1051/0004-6361/202450793},
archivePrefix = {arXiv},
       eprint = {2405.13497},
 primaryClass = {astro-ph.EP},
       adsurl = {https://ui.adsabs.harvard.edu/abs/2025A&A...697A...7M}
}

@ARTICLE{Zerjal:2026,
       author = {{Zerjal}, M. and {Dominguez-Tagle}, C. and {Vitas}, N. and {Sedighi}, N. and {Mart{\'\i}n}, E.~L. and {Zapatero Osorio}, M.~R. and {Olivares}, J. and {Mu{\~n}oz Torres}, S. and {Tsilia}, S. and {Zhang}, J.-Y. and {Barrado}, D. and {B{\'e}jar}, V.~J.~S. and {Bouy}, H. and {Burgasser}, A. and {Cruz}, P. and {Lodieu}, N. and {Mas Buitrago}, P. and {Phan-Bao}, N. and {Solano}, E. and {Tata}, R. and {Goldman}, B. and {Mohandasan}, A. and {Reyl{\'e}}, C. and {Smart}, R.~L. and {Aghanim}, N. and {Altieri}, B. and {Amara}, A. and {Andreon}, S. and {Auricchio}, N. and {Baccigalupi}, C. and {Baldi}, M. and {Balestra}, A. and {Bardelli}, S. and {Battaglia}, P. and {Biviano}, A. and {Bonchi}, A. and {Branchini}, E. and {Brescia}, M. and {Brinchmann}, J. and {Camera}, S. and {Ca{\~n}as-Herrera}, G. and {Capobianco}, V. and {Carbone}, C. and {Carretero}, J. and {Casas}, S. and {Castellano}, M. and {Castignani}, G. and {Cavuoti}, S. and {Chambers}, K.~C. and {Cimatti}, A. and {Colodro-Conde}, C. and {Congedo}, G. and {Conselice}, C.~J. and {Conversi}, L. and {Copin}, Y. and {Courbin}, F. and {Courtois}, H.~M. and {Cropper}, M. and {Cuby}, J.-G. and {Da Silva}, A. and {Degaudenzi}, H. and {De Lucia}, G. and {Dolding}, C. and {Dole}, H. and {Douspis}, M. and {Dubath}, F. and {Dupac}, X. and {Dusini}, S. and {Escoffier}, S. and {Farina}, M. and {Faustini}, F. and {Ferriol}, S. and {Fotopoulou}, S. and {Frailis}, M. and {Franceschi}, E. and {Galeotta}, S. and {George}, K. and {Gillis}, B. and {Giocoli}, C. and {G{\'o}mez-Alvarez}, P. and {Gracia-Carpio}, J. and {Granett}, B.~R. and {Grazian}, A. and {Grupp}, F. and {Haugan}, S.~V.~H. and {Hoar}, J. and {Holmes}, W. and {Hormuth}, F. and {Hornstrup}, A. and {Jahnke}, K. and {Jhabvala}, M. and {Keih{\"a}nen}, E. and {Kermiche}, S. and {Kiessling}, A. and {Kubik}, B. and {Kuijken}, K. and {K{\"u}mmel}, M. and {Kunz}, M. and {Kurki-Suonio}, H. and {Le Boulc'h}, Q. and {Le Brun}, A.~M.~C. and {Ligori}, S. and {Lilje}, P.~B. and {Lindholm}, V. and {Lloro}, I. and {Mainetti}, G. and {Maino}, D. and {Maiorano}, E. and {Mansutti}, O. and {Marggraf}, O. and {Martinelli}, M. and {Martinet}, N. and {Marulli}, F. and {Massey}, R. and {Medinaceli}, E. and {Mei}, S. and {Mellier}, Y. and {Meneghetti}, M. and {Merlin}, E. and {Meylan}, G. and {Mora}, A. and {Moresco}, M. and {Moscardini}, L. and {Nakajima}, R. and {Neissner}, C. and {Niemi}, S.-M. and {Padilla}, C. and {Paltani}, S. and {Pasian}, F. and {Pedersen}, K. and {Percival}, W.~J. and {Pettorino}, V. and {Pires}, S. and {Polenta}, G. and {Poncet}, M. and {Popa}, L.~A. and {Pozzetti}, L. and {Raison}, F. and {Rebolo}, R. and {Renzi}, A. and {Rhodes}, J. and {Riccio}, G. and {Romelli}, E. and {Roncarelli}, M. and {Saglia}, R. and {Sakr}, Z. and {Sapone}, D. and {Sartoris}, B. and {Schewtschenko}, J.~A. and {Schirmer}, M. and {Schneider}, P. and {Secroun}, A. and {Seidel}, G. and {Seiffert}, M. and {Serrano}, S. and {Simon}, P. and {Sirignano}, C. and {Sirri}, G. and {Stanco}, L. and {Steinwagner}, J. and {Tallada-Cresp{\'\i}}, P. and {Taylor}, A.~N. and {Tereno}, I. and {Toft}, S. and {Toledo-Moreo}, R. and {Torradeflot}, F. and {Tsyganov}, A. and {Tutusaus}, I. and {Valenziano}, L. and {Valiviita}, J. and {Vassallo}, T. and {Verdoes Kleijn}, G. and {Veropalumbo}, A. and {Wang}, Y. and {Weller}, J. and {Zacchei}, A. and {Zamorani}, G. and {Zerbi}, F.~M. and {Zucca}, E. and {Mart{\'\i}n-Fleitas}, J. and {Scottez}, V.},
        title = "{Euclid: Quick Data Release (Q1) ─ A photometric search for ultracool dwarfs in the Euclid Deep Fields}",
      journal = {\aap},
         year = 2026,
        month = jun,
       volume = {711},
          eid = {A37},
        pages = {A37},
          doi = {10.1051/0004-6361/202556980},
archivePrefix = {arXiv},
       eprint = {2503.22497},
 primaryClass = {astro-ph.SR},
       adsurl = {https://ui.adsabs.harvard.edu/abs/2026A&A...711A..37Z}
}

@ARTICLE{Ivezic:2019,
       author = {{Ivezi{\'c}}, {\v{Z}}eljko and {Kahn}, Steven M. and {Tyson}, J. Anthony and {Abel}, Bob and {Acosta}, Emily and {Allsman}, Robyn and {Alonso}, David and {AlSayyad}, Yusra and {Anderson}, Scott F. and {Andrew}, John and {Angel}, James Roger P. and {Angeli}, George Z. and {Ansari}, Reza and {Antilogus}, Pierre and {Araujo}, Constanza and {Armstrong}, Robert and {Arndt}, Kirk T. and {Astier}, Pierre and {Aubourg}, {\'E}ric and {Auza}, Nicole and {Axelrod}, Tim S. and {Bard}, Deborah J. and {Barr}, Jeff D. and {Barrau}, Aurelian and {Bartlett}, James G. and {Bauer}, Amanda E. and {Bauman}, Brian J. and {Baumont}, Sylvain and {Bechtol}, Ellen and {Bechtol}, Keith and {Becker}, Andrew C. and {Becla}, Jacek and {Beldica}, Cristina and {Bellavia}, Steve and {Bianco}, Federica B. and {Biswas}, Rahul and {Blanc}, Guillaume and {Blazek}, Jonathan and {Blandford}, Roger D. and {Bloom}, Josh S. and {Bogart}, Joanne and {Bond}, Tim W. and {Booth}, Michael T. and {Borgland}, Anders W. and {Borne}, Kirk and {Bosch}, James F. and {Boutigny}, Dominique and {Brackett}, Craig A. and {Bradshaw}, Andrew and {Brandt}, William Nielsen and {Brown}, Michael E. and {Bullock}, James S. and {Burchat}, Patricia and {Burke}, David L. and {Cagnoli}, Gianpietro and {Calabrese}, Daniel and {Callahan}, Shawn and {Callen}, Alice L. and {Carlin}, Jeffrey L. and {Carlson}, Erin L. and {Chandrasekharan}, Srinivasan and {Charles-Emerson}, Glenaver and {Chesley}, Steve and {Cheu}, Elliott C. and {Chiang}, Hsin-Fang and {Chiang}, James and {Chirino}, Carol and {Chow}, Derek and {Ciardi}, David R. and {Claver}, Charles F. and {Cohen-Tanugi}, Johann and {Cockrum}, Joseph J. and {Coles}, Rebecca and {Connolly}, Andrew J. and {Cook}, Kem H. and {Cooray}, Asantha and {Covey}, Kevin R. and {Cribbs}, Chris and {Cui}, Wei and {Cutri}, Roc and {Daly}, Philip N. and {Daniel}, Scott F. and {Daruich}, Felipe and {Daubard}, Guillaume and {Daues}, Greg and {Dawson}, William and {Delgado}, Francisco and {Dellapenna}, Alfred and {de Peyster}, Robert and {de Val-Borro}, Miguel and {Digel}, Seth W. and {Doherty}, Peter and {Dubois}, Richard and {Dubois-Felsmann}, Gregory P. and {Durech}, Josef and {Economou}, Frossie and {Eifler}, Tim and {Eracleous}, Michael and {Emmons}, Benjamin L. and {Fausti Neto}, Angelo and {Ferguson}, Henry and {Figueroa}, Enrique and {Fisher-Levine}, Merlin and {Focke}, Warren and {Foss}, Michael D. and {Frank}, James and {Freemon}, Michael D. and {Gangler}, Emmanuel and {Gawiser}, Eric and {Geary}, John C. and {Gee}, Perry and {Geha}, Marla and {Gessner}, Charles J.~B. and {Gibson}, Robert R. and {Gilmore}, D. Kirk and {Glanzman}, Thomas and {Glick}, William and {Goldina}, Tatiana and {Goldstein}, Daniel A. and {Goodenow}, Iain and {Graham}, Melissa L. and {Gressler}, William J. and {Gris}, Philippe and {Guy}, Leanne P. and {Guyonnet}, Augustin and {Haller}, Gunther and {Harris}, Ron and {Hascall}, Patrick A. and {Haupt}, Justine and {Hernandez}, Fabio and {Herrmann}, Sven and {Hileman}, Edward and {Hoblitt}, Joshua and {Hodgson}, John A. and {Hogan}, Craig and {Howard}, James D. and {Huang}, Dajun and {Huffer}, Michael E. and {Ingraham}, Patrick and {Innes}, Walter R. and {Jacoby}, Suzanne H. and {Jain}, Bhuvnesh and {Jammes}, Fabrice and {Jee}, M. James and {Jenness}, Tim and {Jernigan}, Garrett and {Jevremovi{\'c}}, Darko and {Johns}, Kenneth and {Johnson}, Anthony S. and {Johnson}, Margaret W.~G. and {Jones}, R. Lynne and {Juramy-Gilles}, Claire and {Juri{\'c}}, Mario and {Kalirai}, Jason S. and {Kallivayalil}, Nitya J. and {Kalmbach}, Bryce and {Kantor}, Jeffrey P. and {Karst}, Pierre and {Kasliwal}, Mansi M. and {Kelly}, Heather and {Kessler}, Richard and {Kinnison}, Veronica and {Kirkby}, David and {Knox}, Lloyd and {Kotov}, Ivan V. and {Krabbendam}, Victor L. and {Krughoff}, K. Simon and {Kub{\'a}nek}, Petr and {Kuczewski}, John and {Kulkarni}, Shri and {Ku}, John and {Kurita}, Nadine R. and {Lage}, Craig S. and {Lambert}, Ron and {Lange}, Travis and {Langton}, J. Brian and {Le Guillou}, Laurent and {Levine}, Deborah and {Liang}, Ming and {Lim}, Kian-Tat and {Lintott}, Chris J. and {Long}, Kevin E. and {Lopez}, Margaux and {Lotz}, Paul J. and {Lupton}, Robert H. and {Lust}, Nate B. and {MacArthur}, Lauren A. and {Mahabal}, Ashish and {Mandelbaum}, Rachel and {Markiewicz}, Thomas W. and {Marsh}, Darren S. and {Marshall}, Philip J. and {Marshall}, Stuart and {May}, Morgan and {McKercher}, Robert and {McQueen}, Michelle and {Meyers}, Joshua and {Migliore}, Myriam and {Miller}, Michelle and {Mills}, David J.},
        title = "{LSST: From Science Drivers to Reference Design and Anticipated Data Products}",
      journal = {\apj},
         year = 2019,
        month = mar,
       volume = {873},
       number = {2},
          eid = {111},
        pages = {111},
          doi = {10.3847/1538-4357/ab042c},
archivePrefix = {arXiv},
       eprint = {0805.2366},
 primaryClass = {astro-ph},
       adsurl = {https://ui.adsabs.harvard.edu/abs/2019ApJ...873..111I}
}

@ARTICLE{Driver:2019,
       author = {{Driver}, S.~P. and {Liske}, J. and {Davies}, L.~J.~M. and {Robotham}, A.~S.~G. and {Baldry}, I.~K. and {Brown}, M.~J.~I. and {Cluver}, M. and {Kuijken}, K. and {Loveday}, J. and {McMahon}, R. and {Meyer}, M.~J. and {Norberg}, P. and {Owers}, M. and {Power}, C. and {Taylor}, E.~N. and {WAVES Team}},
        title = "{4MOST Consortium Survey 7: Wide-Area VISTA Extragalactic Survey (WAVES)}",
      journal = {The Messenger},
         year = 2019,
        month = mar,
       volume = {175},
        pages = {46-49},
          doi = {10.18727/0722-6691/5126},
archivePrefix = {arXiv},
       eprint = {1903.02473},
 primaryClass = {astro-ph.GA},
       adsurl = {https://ui.adsabs.harvard.edu/abs/2019Msngr.175...46D}
}

@ARTICLE{Edge:2013,
       author = {{Edge}, A. and {Sutherland}, W. and {Kuijken}, K. and {Driver}, S. and {McMahon}, R. and {Eales}, S. and {Emerson}, J.~P.},
        title = "{The VISTA Kilo-degree Infrared Galaxy (VIKING) Survey: Bridging the Gap between Low and High Redshift}",
      journal = {The Messenger},
         year = 2013,
        month = dec,
       volume = {154},
        pages = {32-34},
       adsurl = {https://ui.adsabs.harvard.edu/abs/2013Msngr.154...32E}
}

@ARTICLE{KiDS:2013,
       author = {{de Jong}, Jelte T.~A. and {Verdoes Kleijn}, Gijs A. and {Kuijken}, Konrad H. and {Valentijn}, Edwin A.},
        title = "{The Kilo-Degree Survey}",
      journal = {Exp. Astron.},
         year = 2013,
        month = jan,
       volume = {35},
       number = {1-2},
        pages = {25-44},
          doi = {10.1007/s10686-012-9306-1},
archivePrefix = {arXiv},
       eprint = {1206.1254},
 primaryClass = {astro-ph.CO},
       adsurl = {https://ui.adsabs.harvard.edu/abs/2013ExA....35...25D}
}

@ARTICLE{Wright:2024,
       author = {{Wright}, Angus H. and {Kuijken}, Konrad and {Hildebrandt}, Hendrik and {Radovich}, Mario and {Bilicki}, Maciej and {Dvornik}, Andrej and {Getman}, Fedor and {Heymans}, Catherine and {Hoekstra}, Henk and {Li}, Shun-Sheng and {Miller}, Lance and {Napolitano}, Nicola R. and {Xia}, Qianli and {Asgari}, Marika and {Brescia}, Massimo and {Buddelmeijer}, Hugo and {Burger}, Pierre and {Castignani}, Gianluca and {Cavuoti}, Stefano and {de Jong}, Jelte and {Edge}, Alastair and {Giblin}, Benjamin and {Giocoli}, Carlo and {Harnois-D{\'e}raps}, Joachim and {Jalan}, Priyanka and {Joachimi}, Benjamin and {John William}, Anjitha and {Joudaki}, Shahab and {Kannawadi}, Arun and {Kaur}, Gursharanjit and {La Barbera}, Francesco and {Linke}, Laila and {Mahony}, Constance and {Maturi}, Matteo and {Moscardini}, Lauro and {Nakoneczny}, Szymon J. and {Paolillo}, Maurizio and {Porth}, Lucas and {Puddu}, Emanuella and {Reischke}, Robert and {Schneider}, Peter and {Sereno}, Mauro and {Shan}, HuanYuan and {Sif{\'o}n}, Crist{\'o}bal and {St{\"o}lzner}, Benjamin and {Tr{\"o}ster}, Tilman and {Valentijn}, Edwin and {van den Busch}, Jan Luca and {Verdoes Kleijn}, Gijs and {Wittje}, Anna and {Yan}, Ziang and {Yao}, Ji and {Yoon}, Mijin and {Zhang}, Yun-Hao},
        title = "{The fifth data release of the Kilo Degree Survey: Multi-epoch optical/NIR imaging covering wide and legacy-calibration fields}",
      journal = {\aap},
         year = 2024,
        month = jun,
       volume = {686},
          eid = {A170},
        pages = {A170},
          doi = {10.1051/0004-6361/202346730},
archivePrefix = {arXiv},
       eprint = {2503.19439},
 primaryClass = {astro-ph.GA},
       adsurl = {https://ui.adsabs.harvard.edu/abs/2024A&A...686A.170W}
}

@article{HSCSSP:2017,
       author = {{Aihara}, Hiroaki and {Arimoto}, Nobuo and {Armstrong}, Robert and {Arnouts}, St{\'e}phane and {Bahcall}, Neta A. and {Bickerton}, Steven and {Bosch}, James and {Bundy}, Kevin and {Capak}, Peter L. and {Chan}, James H.~H. and {Chiba}, Masashi and {Coupon}, Jean and {Egami}, Eiichi and {Enoki}, Motohiro and {Finet}, Francois and {Fujimori}, Hiroki and {Fujimoto}, Seiji and {Furusawa}, Hisanori and {Furusawa}, Junko and {Goto}, Tomotsugu and {Goulding}, Andy and {Greco}, Johnny P. and {Greene}, Jenny E. and {Gunn}, James E. and {Hamana}, Takashi and {Harikane}, Yuichi and {Hashimoto}, Yasuhiro and {Hattori}, Takashi and {Hayashi}, Masao and {Hayashi}, Yusuke and {He{\l}miniak}, Krzysztof G. and {Higuchi}, Ryo and {Hikage}, Chiaki and {Ho}, Paul T.~P. and {Hsieh}, Bau-Ching and {Huang}, Kuiyun and {Huang}, Song and {Ikeda}, Hiroyuki and {Imanishi}, Masatoshi and {Inoue}, Akio K. and {Iwasawa}, Kazushi and {Iwata}, Ikuru and {Jaelani}, Anton T. and {Jian}, Hung-Yu and {Kamata}, Yukiko and {Karoji}, Hiroshi and {Kashikawa}, Nobunari and {Katayama}, Nobuhiko and {Kawanomoto}, Satoshi and {Kayo}, Issha and {Koda}, Jin and {Koike}, Michitaro and {Kojima}, Takashi and {Komiyama}, Yutaka and {Konno}, Akira and {Koshida}, Shintaro and {Koyama}, Yusei and {Kusakabe}, Haruka and {Leauthaud}, Alexie and {Lee}, Chien-Hsiu and {Lin}, Lihwai and {Lin}, Yen-Ting and {Lupton}, Robert H. and {Mandelbaum}, Rachel and {Matsuoka}, Yoshiki and {Medezinski}, Elinor and {Mineo}, Sogo and {Miyama}, Shoken and {Miyatake}, Hironao and {Miyazaki}, Satoshi and {Momose}, Rieko and {More}, Anupreeta and {More}, Surhud and {Moritani}, Yuki and {Moriya}, Takashi J. and {Morokuma}, Tomoki and {Mukae}, Shiro and {Murata}, Ryoma and {Murayama}, Hitoshi and {Nagao}, Tohru and {Nakata}, Fumiaki and {Niida}, Mana and {Niikura}, Hiroko and {Nishizawa}, Atsushi J. and {Obuchi}, Yoshiyuki and {Oguri}, Masamune and {Oishi}, Yukie and {Okabe}, Nobuhiro and {Okamoto}, Sakurako and {Okura}, Yuki and {Ono}, Yoshiaki and {Onodera}, Masato and {Onoue}, Masafusa and {Osato}, Ken and {Ouchi}, Masami and {Price}, Paul A. and {Pyo}, Tae-Soo and {Sako}, Masao and {Sawicki}, Marcin and {Shibuya}, Takatoshi and {Shimasaku}, Kazuhiro and {Shimono}, Atsushi and {Shirasaki}, Masato and {Silverman}, John D. and {Simet}, Melanie and {Speagle}, Joshua and {Spergel}, David N. and {Strauss}, Michael A. and {Sugahara}, Yuma and {Sugiyama}, Naoshi and {Suto}, Yasushi and {Suyu}, Sherry H. and {Suzuki}, Nao and {Tait}, Philip J. and {Takada}, Masahiro and {Takata}, Tadafumi and {Tamura}, Naoyuki and {Tanaka}, Manobu M. and {Tanaka}, Masaomi and {Tanaka}, Masayuki and {Tanaka}, Yoko and {Terai}, Tsuyoshi and {Terashima}, Yuichi and {Toba}, Yoshiki and {Tominaga}, Nozomu and {Toshikawa}, Jun and {Turner}, Edwin L. and {Uchida}, Tomohisa and {Uchiyama}, Hisakazu and {Umetsu}, Keiichi and {Uraguchi}, Fumihiro and {Urata}, Yuji and {Usuda}, Tomonori and {Utsumi}, Yousuke and {Wang}, Shiang-Yu and {Wang}, Wei-Hao and {Wong}, Kenneth C. and {Yabe}, Kiyoto and {Yamada}, Yoshihiko and {Yamanoi}, Hitomi and {Yasuda}, Naoki and {Yeh}, Sherry and {Yonehara}, Atsunori and {Yuma}, Suraphong},
        title = "{The Hyper Suprime-Cam SSP Survey: Overview and survey design}",
      journal = {\pasj},
         year = 2018,
        month = jan,
       volume = {70},
          eid = {S4},
        pages = {S4},
          doi = {10.1093/pasj/psx066},
archivePrefix = {arXiv},
       eprint = {1704.05858},
 primaryClass = {astro-ph.IM},
       adsurl = {https://ui.adsabs.harvard.edu/abs/2018PASJ...70S...4A}
}

@ARTICLE{PFS:2014,
       author = {{Takada}, Masahiro and {Ellis}, Richard S. and {Chiba}, Masashi and {Greene}, Jenny E. and {Aihara}, Hiroaki and {Arimoto}, Nobuo and {Bundy}, Kevin and {Cohen}, Judith and {Dor{\'e}}, Olivier and {Graves}, Genevieve and {Gunn}, James E. and {Heckman}, Timothy and {Hirata}, Christopher M. and {Ho}, Paul and {Kneib}, Jean-Paul and {Le F{\`e}vre}, Olivier and {Lin}, Lihwai and {More}, Surhud and {Murayama}, Hitoshi and {Nagao}, Tohru and {Ouchi}, Masami and {Seiffert}, Michael and {Silverman}, John D. and {Sodr{\'e}}, Laerte and {Spergel}, David N. and {Strauss}, Michael A. and {Sugai}, Hajime and {Suto}, Yasushi and {Takami}, Hideki and {Wyse}, Rosemary},
        title = "{Extragalactic science, cosmology, and Galactic archaeology with the Subaru Prime Focus Spectrograph}",
      journal = {\pasj},
         year = 2014,
        month = feb,
       volume = {66},
       number = {1},
          eid = {R1},
        pages = {R1},
          doi = {10.1093/pasj/pst019},
archivePrefix = {arXiv},
       eprint = {1206.0737},
 primaryClass = {astro-ph.CO},
       adsurl = {https://ui.adsabs.harvard.edu/abs/2014PASJ...66R...1T}
      }

@ARTICLE{Irureta-Goyena:2026,
       author = {{Irureta-Goyena}, B.~Y. and {Altieri}, B. and {Kneib}, J.-P. and {P{\"o}ntinen}, M. and {Hainaut}, O.~R. and {Alarcon}, M.~R. and {Granvik}, M. and {Nucita}, A.~A. and {Carry}, B. and {Devogele}, M. and {Mahlke}, M. and {Vavrek}, R. and {M{\"u}ller}, T. and {Vilenius}, E. and {Snodgrass}, C. and {Kohley}, R. and {Lemon}, C. and {G{\'o}mez-Alvarez}, P. and {Verdoes Kleijn}, G. and {Licandro}, J. and {Kruk}, S. and {Conversi}, L. and {Franco}, A. and {Buenadicha}, G. and {Mas-Buitrago}, P. and {Kuijken}, K. and {Andreon}, S. and {Baccigalupi}, C. and {Baldi}, M. and {Balestra}, A. and {Battaglia}, P. and {Biviano}, A. and {Branchini}, E. and {Brescia}, M. and {Camera}, S. and {Capobianco}, V. and {Carbone}, C. and {Carretero}, J. and {Casas}, R. and {Castellano}, M. and {Castignani}, G. and {Cavuoti}, S. and {Chambers}, K.~C. and {Cimatti}, A. and {Colodro-Conde}, C. and {Congedo}, G. and {Conselice}, C.~J. and {Copin}, Y. and {Courbin}, F. and {Courtois}, H.~M. and {Cropper}, M. and {Degaudenzi}, H. and {De Lucia}, G. and {Dolding}, C. and {Dole}, H. and {Dubath}, F. and {Dupac}, X. and {Farina}, M. and {Farinelli}, R. and {Ferriol}, S. and {Frailis}, M. and {Fumana}, M. and {Galeotta}, S. and {George}, K. and {Gillis}, B. and {Giocoli}, C. and {Gracia-Carpio}, J. and {Grazian}, A. and {Grupp}, F. and {Haugan}, S.~V.~H. and {Hoekstra}, H. and {Holmes}, W. and {Hook}, I.~M. and {Hormuth}, F. and {Hornstrup}, A. and {Jahnke}, K. and {Jhabvala}, M. and {Kiessling}, A. and {Kubik}, B. and {K{\"u}mmel}, M. and {Kunz}, M. and {Kurki-Suonio}, H. and {Le Brun}, A.~M.~C. and {Ligori}, S. and {Lilje}, P.~B. and {Lindholm}, V. and {Lloro}, I. and {Mainetti}, G. and {Mansutti}, O. and {Marggraf}, O. and {Martinelli}, M. and {Martinet}, N. and {Marulli}, F. and {Massey}, R.~J. and {Medinaceli}, E. and {Mei}, S. and {Merlin}, E. and {Meylan}, G. and {Mora}, A. and {Moscardini}, L. and {Nakajima}, R. and {Neissner}, C. and {Niemi}, S.-M. and {Padilla}, C. and {Paltani}, S. and {Pasian}, F. and {Pedersen}, K. and {Percival}, W.~J. and {Pettorino}, V. and {Polenta}, G. and {Popa}, L.~A. and {Raison}, F. and {Rebolo}, R. and {Renzi}, A. and {Rhodes}, J. and {Riccio}, G. and {Romelli}, E. and {Roncarelli}, M. and {Saglia}, R. and {Sakr}, Z. and {Sapone}, D. and {Schirmer}, M. and {Schneider}, P. and {Secroun}, A. and {Sihvola}, E. and {Simon}, P. and {Sirignano}, C. and {Sirri}, G. and {Stanco}, L. and {Tallada-Cresp{\'\i}}, P. and {Tereno}, I. and {Toft}, S. and {Toledo-Moreo}, R. and {Torradeflot}, F. and {Tutusaus}, I. and {Valiviita}, J. and {Vassallo}, T. and {Wang}, Y. and {Weller}, J. and {Zerbi}, F.~M. and {Garc{\'\i}a-Bellido}, J. and {Mart{\'\i}n-Fleitas}, J. and {Scottez}, V. and {Helou}, G. and {Scott}, D.},
        title = "{Euclid: Asteroid rotation periods from the Euclid Ecliptic Survey}",
      journal = {\aap},
         year = 2026,
        month = jun,
       volume = {710},
          eid = {A27},
        pages = {A27},
          doi = {10.1051/0004-6361/202659419},
archivePrefix = {arXiv},
       eprint = {2604.25652},
 primaryClass = {astro-ph.EP},
       adsurl = {https://ui.adsabs.harvard.edu/abs/2026A&A...710A..27I}
}

@ARTICLE{Pontinen:2020,
       author = {{P{\"o}ntinen}, M. and {Granvik}, M. and {Nucita}, A.~A. and {Conversi}, L. and {Altieri}, B. and {Auricchio}, N. and {Bodendorf}, C. and {Bonino}, D. and {Brescia}, M. and {Capobianco}, V. and {Carretero}, J. and {Carry}, B. and {Castellano}, M. and {Cledassou}, R. and {Congedo}, G. and {Corcione}, L. and {Cropper}, M. and {Dusini}, S. and {Frailis}, M. and {Franceschi}, E. and {Fumana}, M. and {Garilli}, B. and {Grupp}, F. and {Hormuth}, F. and {Israel}, H. and {Jahnke}, K. and {Kermiche}, S. and {Kitching}, T. and {Kohley}, R. and {Kubik}, B. and {Kunz}, M. and {Laureijs}, R. and {Lilje}, P.~B. and {Lloro}, I. and {Maiorano}, E. and {Marggraf}, O. and {Massey}, R. and {Meneghetti}, M. and {Meylan}, G. and {Moscardini}, L. and {Padilla}, C. and {Paltani}, S. and {Pasian}, F. and {Pires}, S. and {Polenta}, G. and {Raison}, F. and {Roncarelli}, M. and {Rossetti}, E. and {Saglia}, R. and {Schneider}, P. and {Secroun}, A. and {Serrano}, S. and {Sirri}, G. and {Taylor}, A.~N. and {Tereno}, I. and {Toledo-Moreo}, R. and {Valenziano}, L. and {Wang}, Y. and {Wetzstein}, M. and {Zoubian}, J.},
        title = "{Euclid: Identification of asteroid streaks in simulated images using StreakDet software}",
      journal = {\aap},
         year = 2020,
        month = dec,
       volume = {644},
          eid = {A35},
        pages = {A35},
          doi = {10.1051/0004-6361/202037765},
archivePrefix = {arXiv},
       eprint = {2011.05297},
 primaryClass = {astro-ph.EP},
       adsurl = {https://ui.adsabs.harvard.edu/abs/2020A&A...644A..35P}
}

@ARTICLE{Carry2018,
       author = {{Carry}, B.},
        title = "{Solar system science with ESA Euclid}",
      journal = {\aap},
         year = 2018,
        month = jan,
       volume = {609},
          eid = {A113},
        pages = {A113},
          doi = {10.1051/0004-6361/201730386},
archivePrefix = {arXiv},
       eprint = {1711.01342},
 primaryClass = {astro-ph.EP},
       adsurl = {https://ui.adsabs.harvard.edu/abs/2018A&A...609A.113C}
}

@INPROCEEDINGS{Laureijs2024,
       author = {{Laureijs}, R. and {Vavrek}, R. and {Racca}, G.~D. and {Kohley}, R. and {Ferruit}, P. and {Pettorino}, V. and {B{\"o}nke}, T. and {Calvi}, A. and {Gaspar Venancio}, L. and {Campos}, L. and {Maiorano}, E. and {Piersanti}, O. and {Prezelus}, S. and {Ragnit}, U. and {Rosato}, P. and {Rosso}, C. and {Rozemeijer}, H. and {Short}, A. and {Strada}, P. and {Stramaccioni}, D. and {Szafraniec}, M. and {Altieri}, B. and {Buenadicha}, G. and {Dupac}, X. and {G{\'o}mez Cambronero}, P. and {Henares Vilaboa}, K. and {Hernandez de la Torre}, C. and {Hoar}, J. and {Lopez-Caniego Alcarria}, M. and {Marcos Arenal}, P. and {Martin Fleitas}, J. and {Miluzio}, M. and {Mora}, A. and {Nieto}, S. and {Perez Bonilla}, R. and {Teodoro Idiago}, P. and {Cordero}, F. and {Mendes}, J. and {Renk}, F. and {Rudolph}, A. and {Schmidt}, M. and {Schwartz}, J. and {Mellier}, Y. and {Aussel}, H. and {Berth{\'e}}, M. and {Casenove}, P. and {Cropper}, M. and {Cuillandre}, J.~C. and {Dinis}, J. and {Gregorio}, A. and {Kuijken}, K. and {Maciaszek}, T. and {Miller}, L. and {Scaramella}, R. and {Schirmer}, M. and {Tereno}, I. and {Zacchei}, A. and {Awan}, S. and {Candini}, G.~P. and {Liebing}, P. and {Nakajima}, R. and {Dusini}, S. and {Battaglia}, P. and {Medinaceli}, E. and {Sirignano}, C. and {Baldry}, I. and {Baugh}, C. and {Bernardeau}, F. and {Castander}, F. and {Cimatti}, A. and {Gillard}, W. and {Guzzo}, L. and {Hoekstra}, H. and {Jahnke}, K. and {Kitching}, T. and {Martin}, E. and {Mohr}, J. and {Percival}, W. and {Rhodes}, J.},
        title = "{The Euclid mission: status after launch and early operations}",
    booktitle = {Space Telescopes and Instrumentation 2024: Optical, Infrared, and Millimeter Wave},
         year = 2024,
       editor = {{Coyle}, Laura E. and {Matsuura}, Shuji and {Perrin}, Marshall D.},
       series = {SPIE Conf. Ser.},
       volume = {13092},
        month = aug,
          eid = {130920M},
        pages = {130920M},
          doi = {10.1117/12.3020343},
       adsurl = {https://ui.adsabs.harvard.edu/abs/2024SPIE13092E..0ML}
}

@INPROCEEDINGS{Maris2019,
       author = {{Maris}, Michele and {Scaramella}, Roberto and {Burigana}, Carlo and {Romelli}, Erik and {Amiaux}, Jerome and {Carvalho}, Carla Sofia and {Cuillandre}, Jean-Charles and {Cunha Da Silva}, Antonio Jose and {De Rosa}, Adriano and {Dinis}, Joao and {Hudelot}, Patrick and {Maiorano}, Ettore and {Tereno}, Ismael and {Trombetti}, Tiziana},
        title = "{Modelling of Zodiacal Light Emission for Space Missions}",
    booktitle = {Astronomical Data Analysis Software and Systems XXVI},
         year = 2019,
       editor = {{Molinaro}, Marco and {Shortridge}, Keith and {Pasian}, Fabio},
       series = {Astronomical Society of the Pacific Conference Series},
       volume = {521},
        month = oct,
        pages = {531},
       adsurl = {https://ui.adsabs.harvard.edu/abs/2019ASPC..521..531M}
}
%
%

%


\begin{appendix}
\onecolumn 



\section{\label{app:expmaps}Exposure maps}

\begin{figure*}[h]
\centering
\begin{tabular}{lcc}
 & {\bf VIS} & {\bf NISP}  \\
 \noalign{\vskip 5pt}
 \hline
 \noalign{\vskip 5pt}
 {\bf EDF-N} & 
    \includegraphics[width=0.4\linewidth]{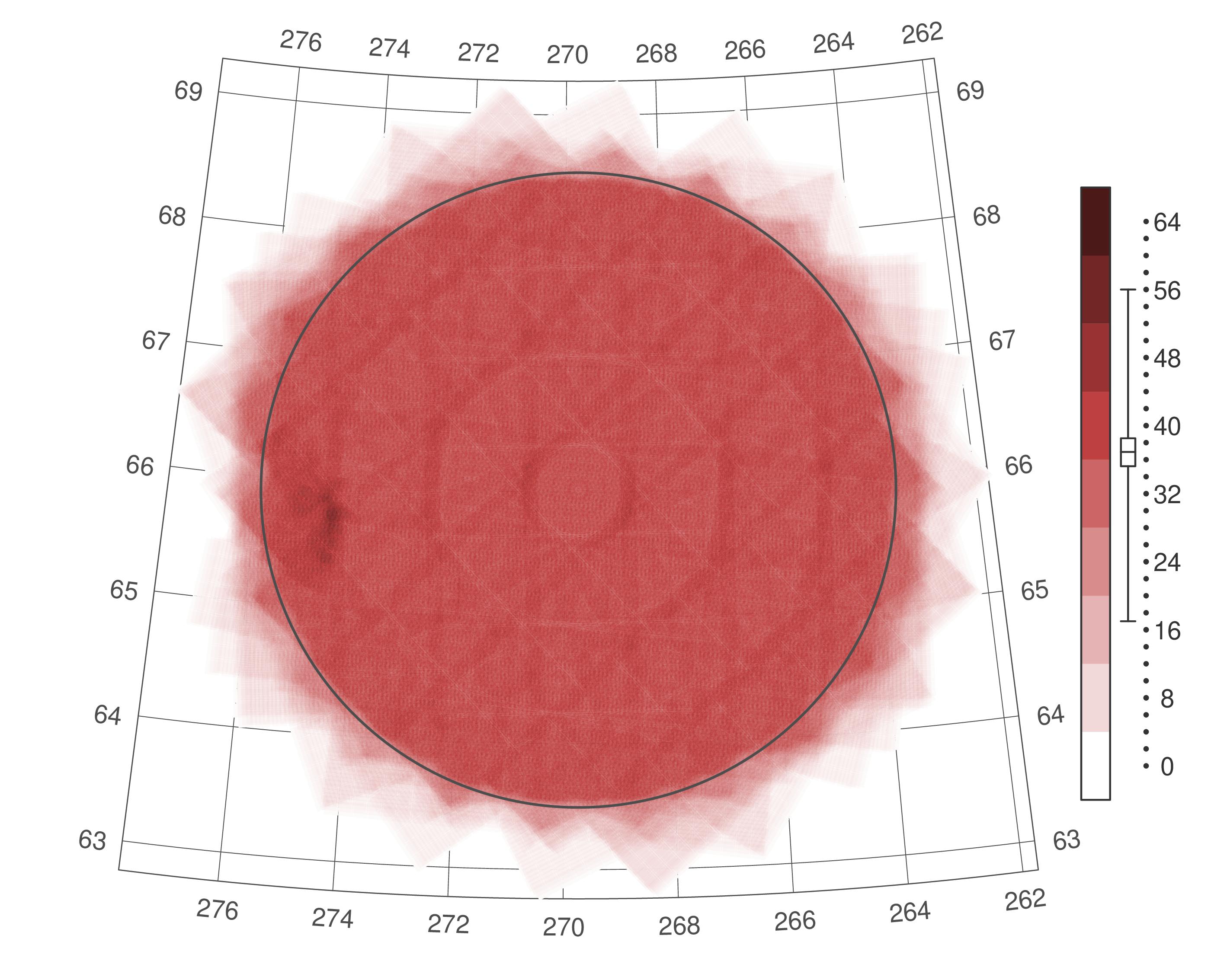}
    &  \includegraphics[width=0.4\linewidth] {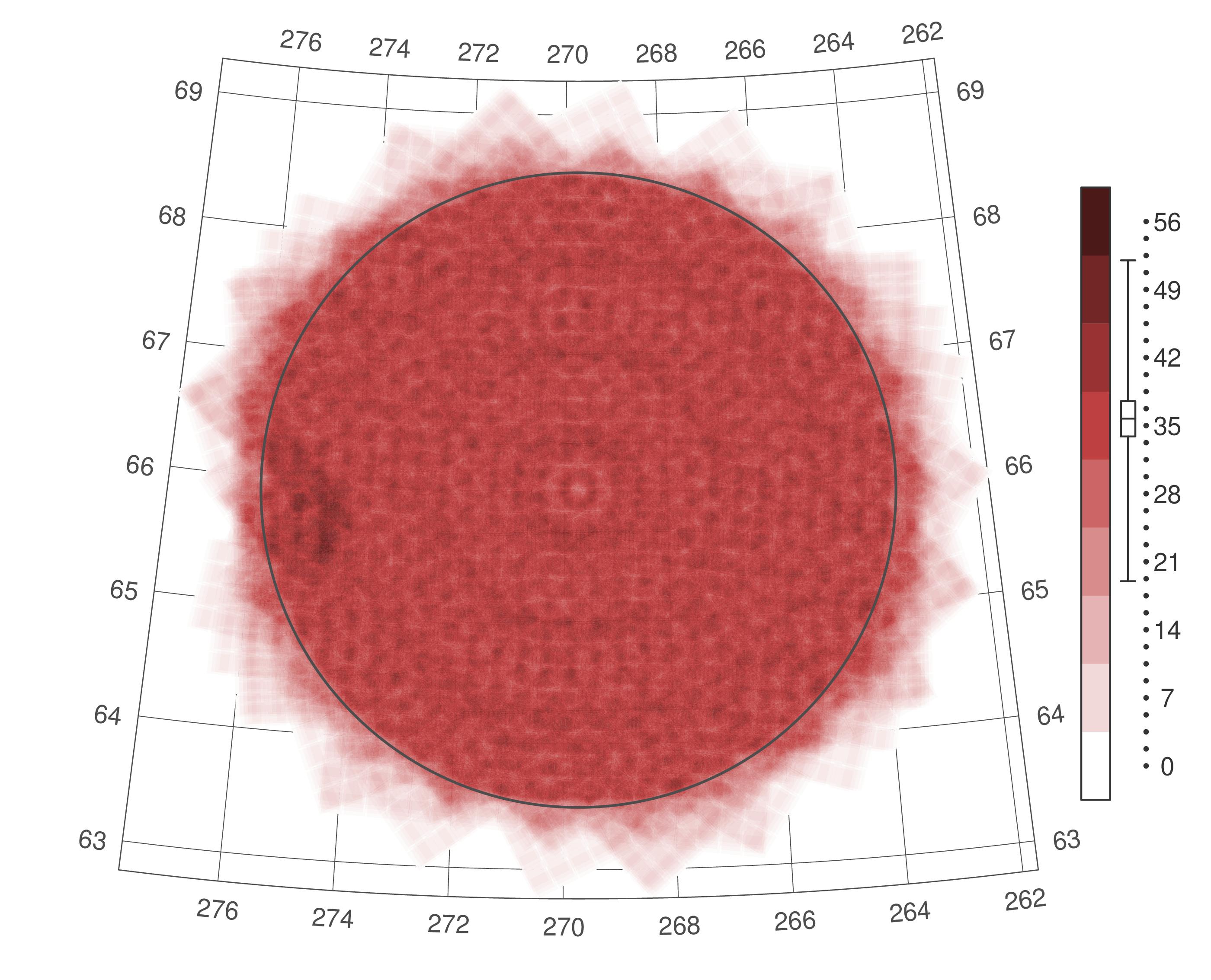}\\
 \noalign{\vskip 5pt}
 \hline
 \noalign{\vskip 5pt}
{\bf EDF-S} & 
    \includegraphics[width=0.4\linewidth]{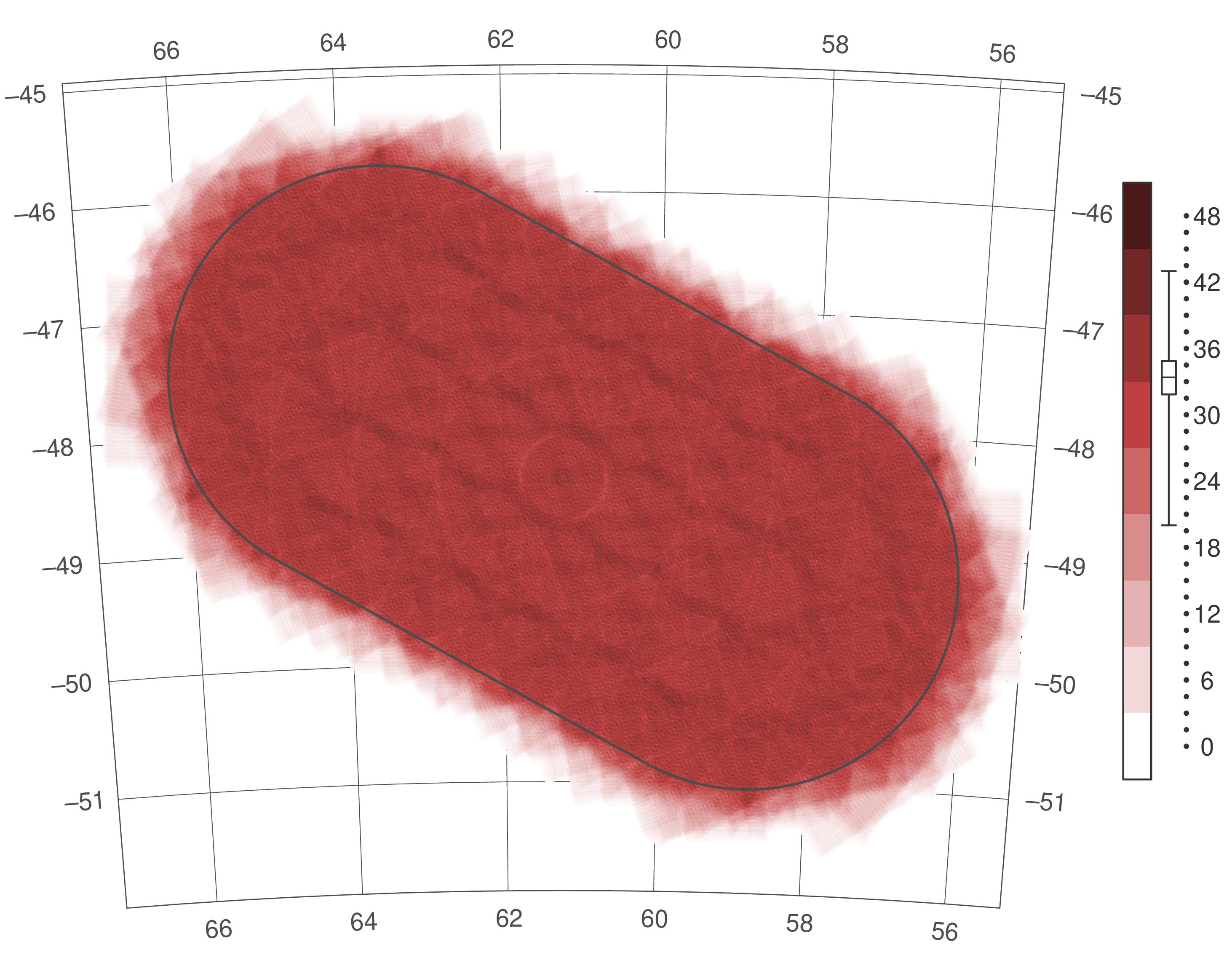}
    &  \includegraphics[width=0.4\linewidth] {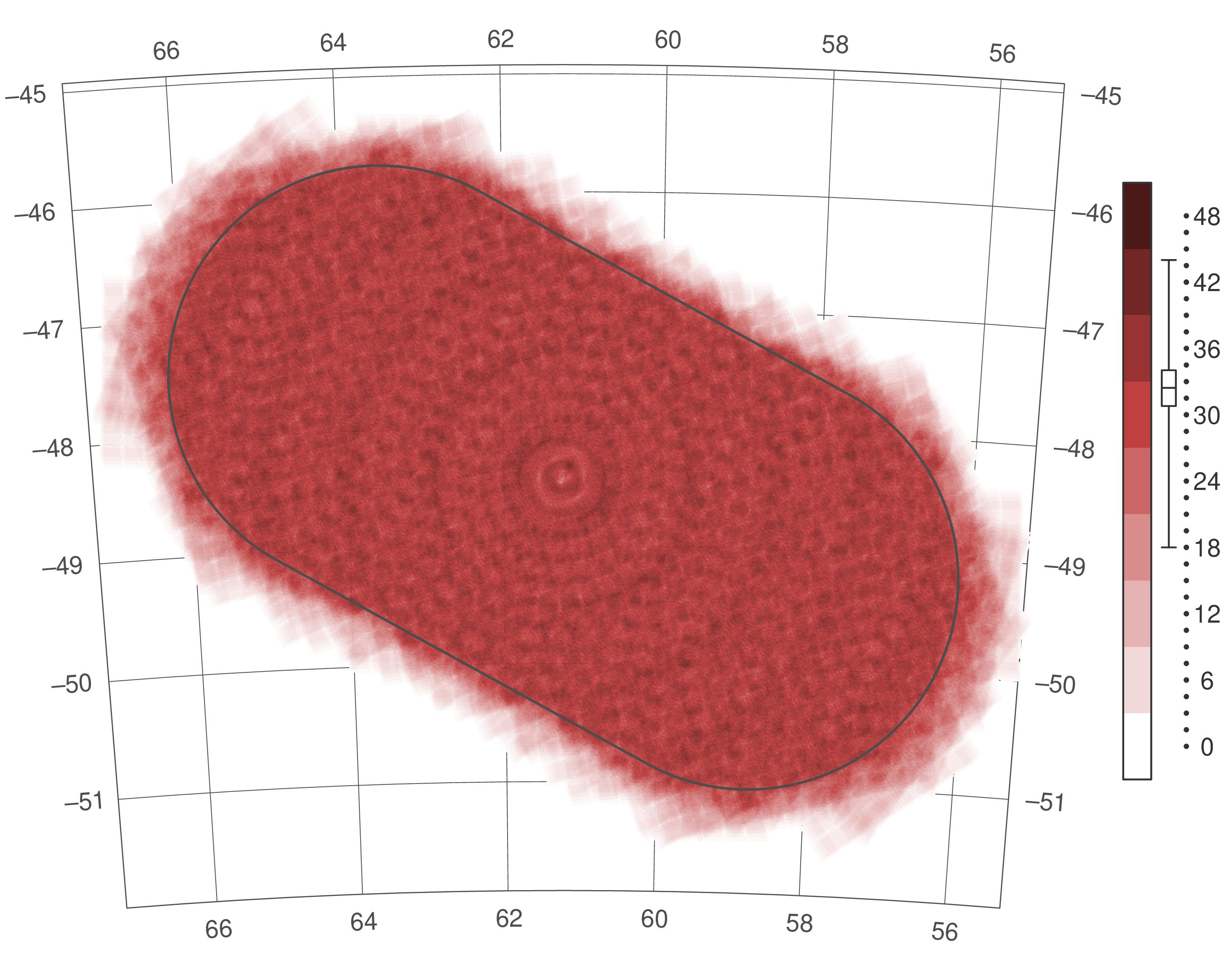}\\
 \noalign{\vskip 5pt}
 \hline
 \noalign{\vskip 5pt}
{\bf EDF-F} & 
    \includegraphics[width=0.4\linewidth]{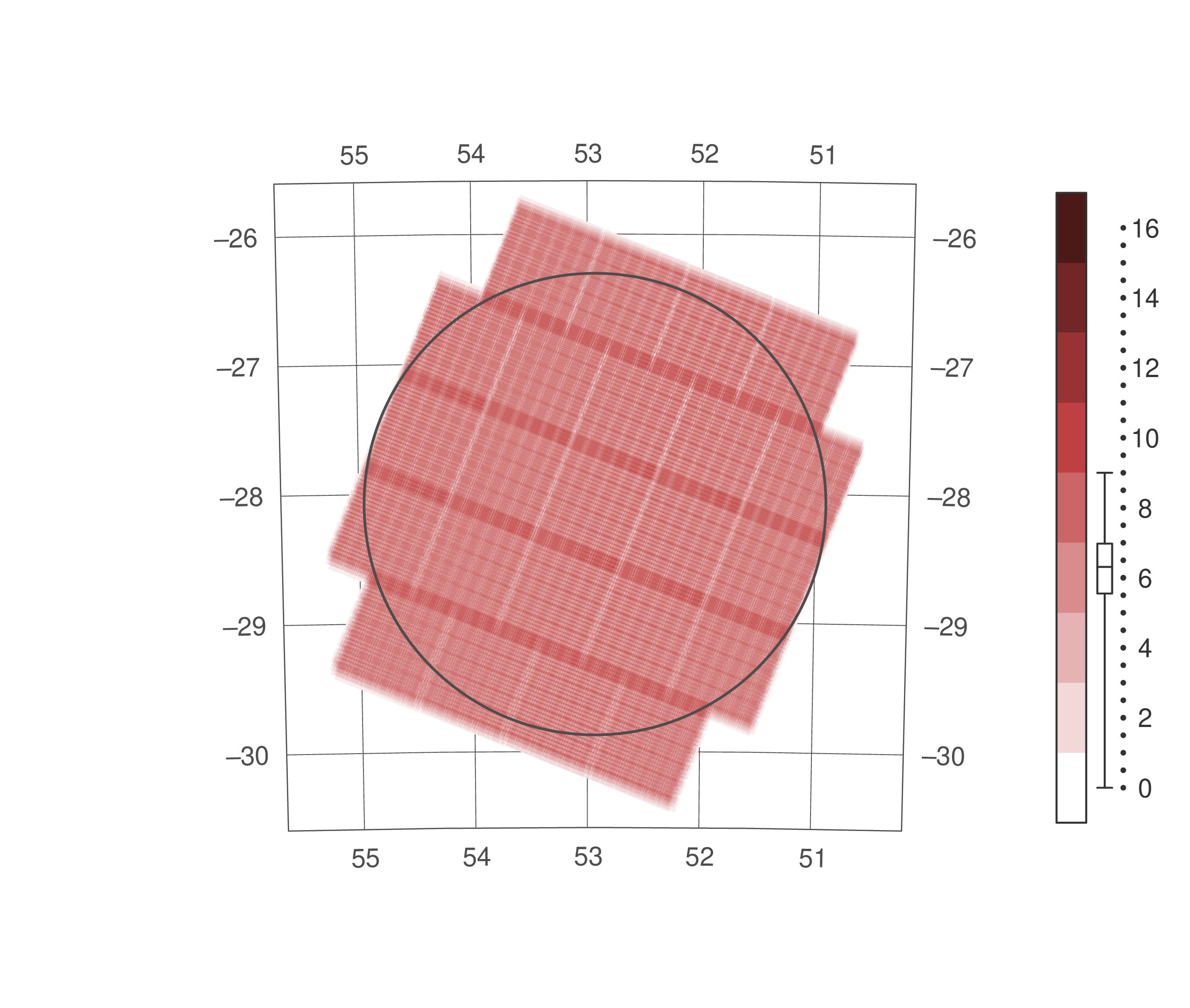}
    &  \includegraphics[width=0.4\linewidth] {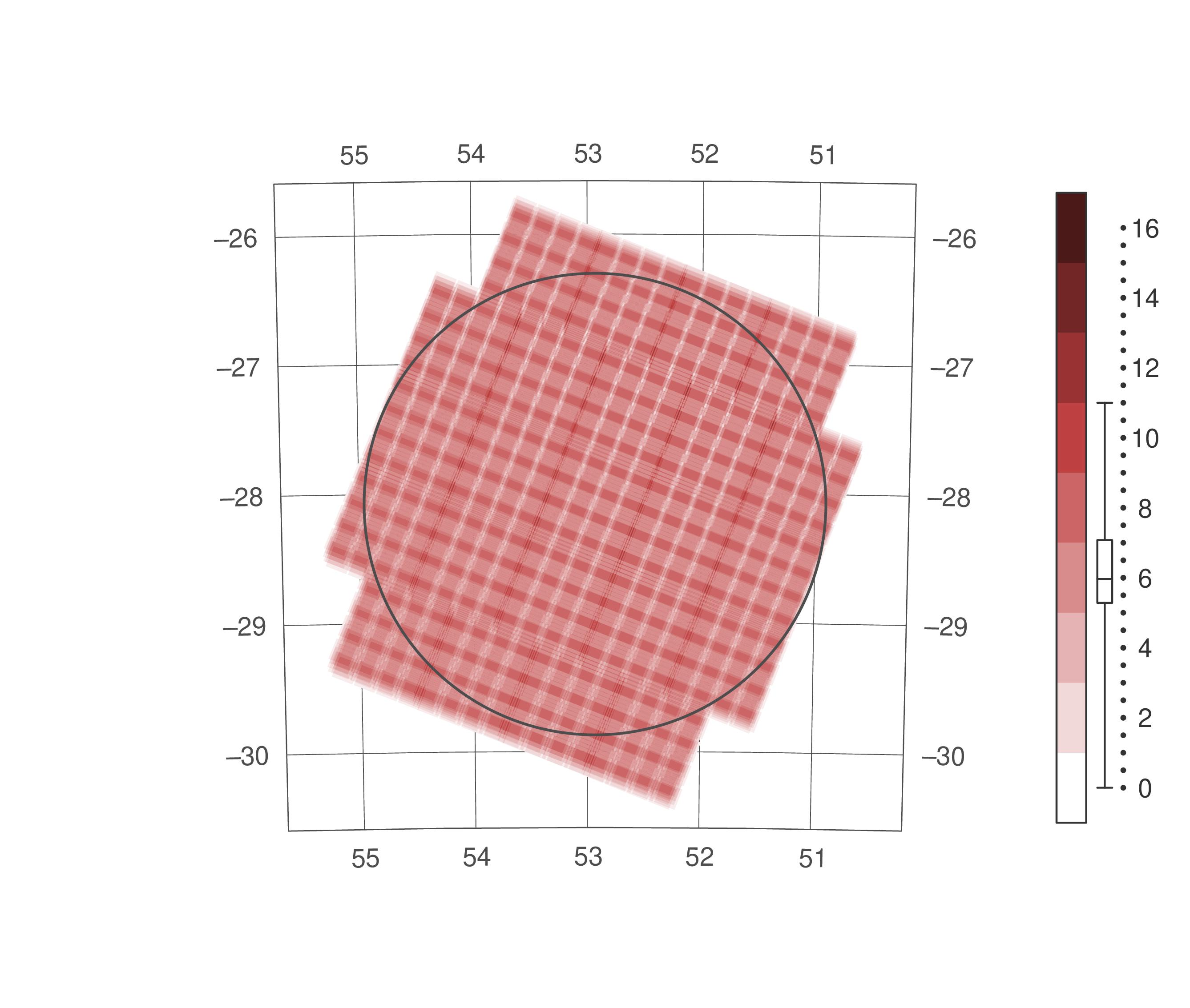}\\
     \noalign{\vskip 5pt}
 \hline
\end{tabular}
\caption{DR1 exposure maps of the Euclid Deep Fields per instrument. The depth is given by the number of single dither exposures per pixel. The scale includes a quartile bar, with the lower tick showing the first quartile (the median of the lower half of the exposure values), the middle tick showing the second quartile (the median of the full set of values), and the upper tick showing the third quartile (the median of the upper half of the exposure values). The box within the bar shows the inter-quartile range, i.e., the spread of the middle half of the points.}
\label{tab:appexpmapsedf}
\end{figure*} 


\begin{figure*}
\centering
\begin{tabular}{lcc}
 & {\bf VIS} & {\bf NISP}  \\
  \noalign{\vskip 5pt}
 \hline
 \noalign{\vskip 5pt}
 {\bf COSMOS} & 
    \includegraphics[width=0.4\linewidth]{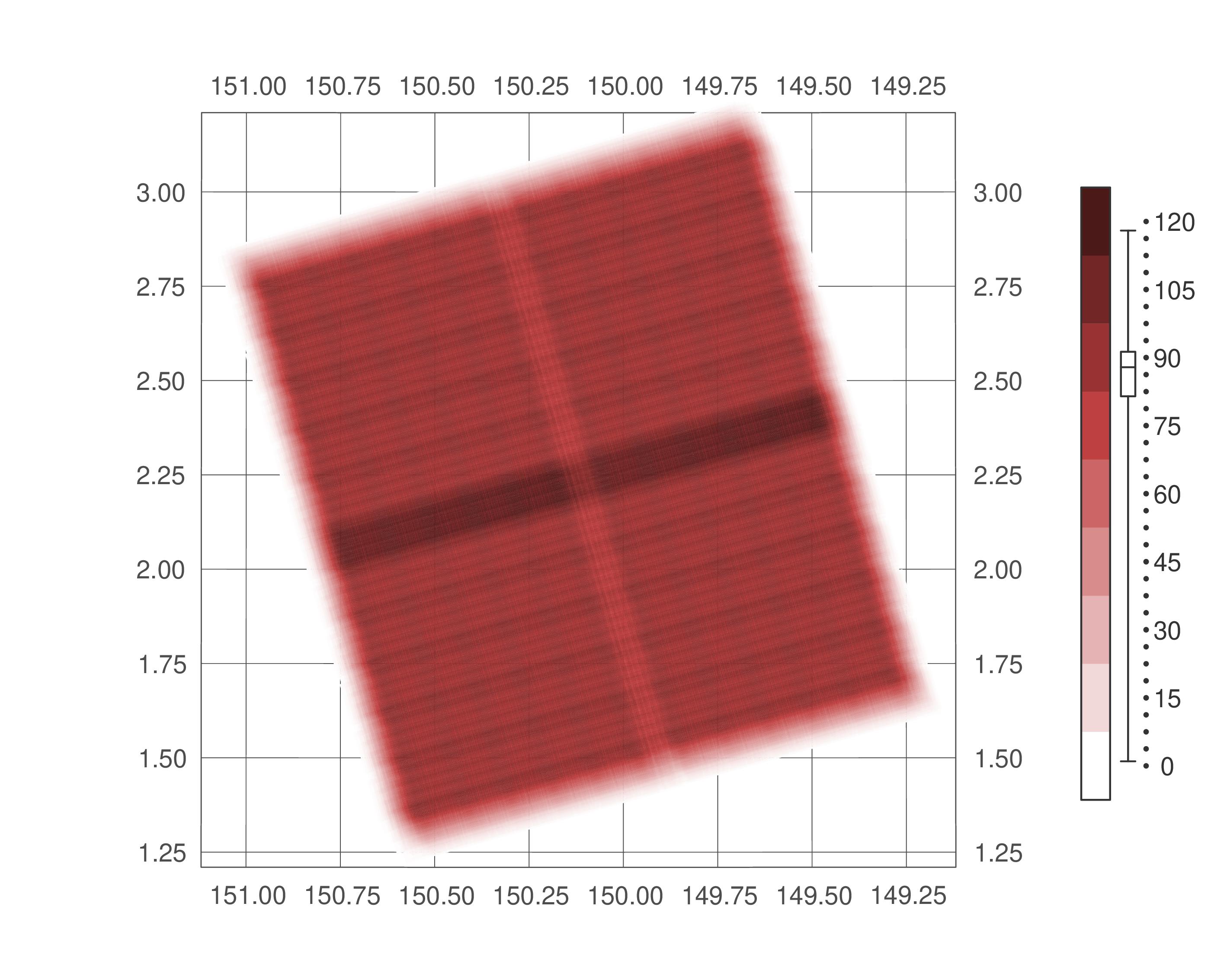}
    &  \includegraphics[width=0.4\linewidth] {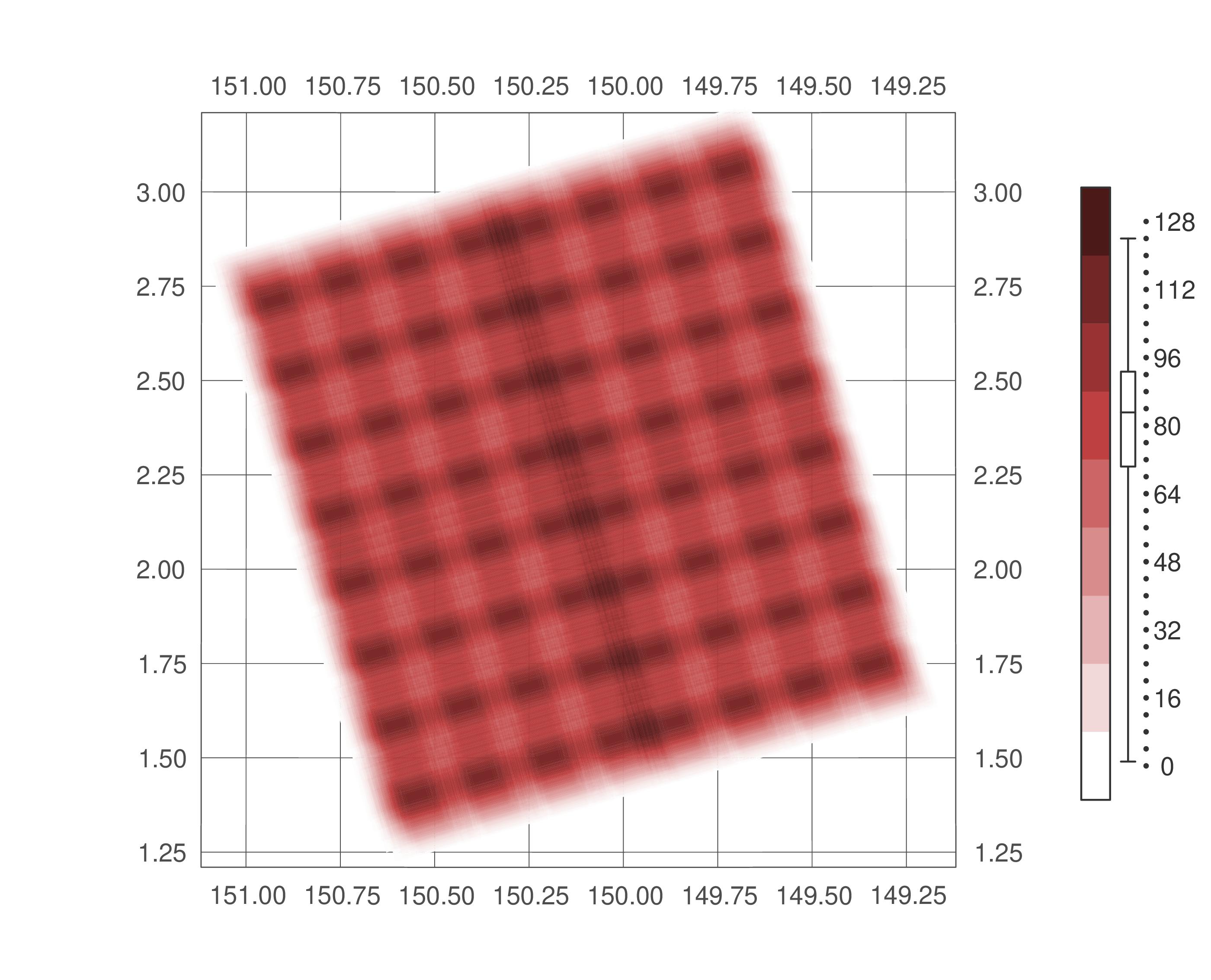}\\
 \noalign{\vskip 5pt}
 \hline
 \noalign{\vskip 5pt}
{\bf CDFS} & 
    \includegraphics[width=0.4\linewidth]{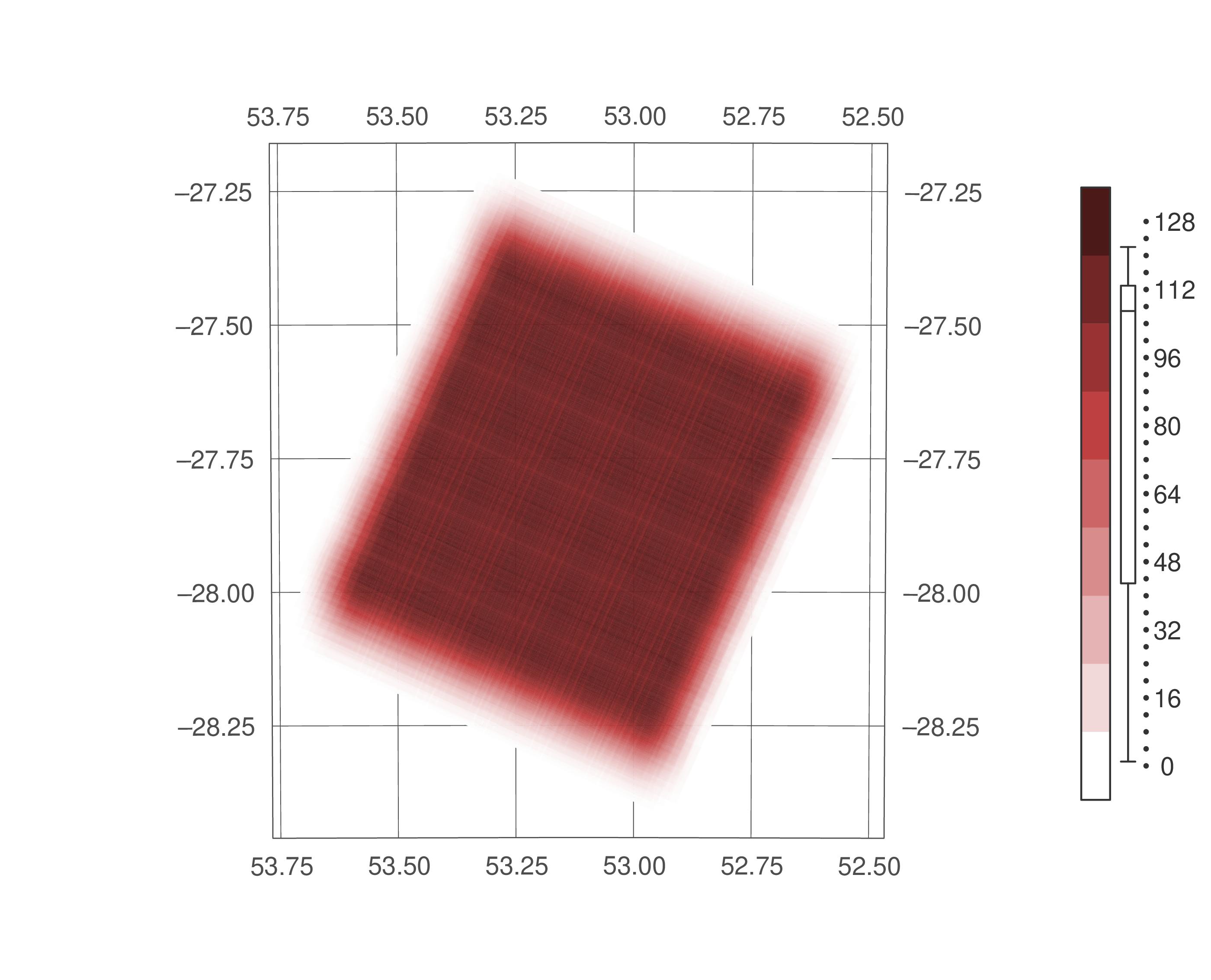}
    &  \includegraphics[width=0.4\linewidth] {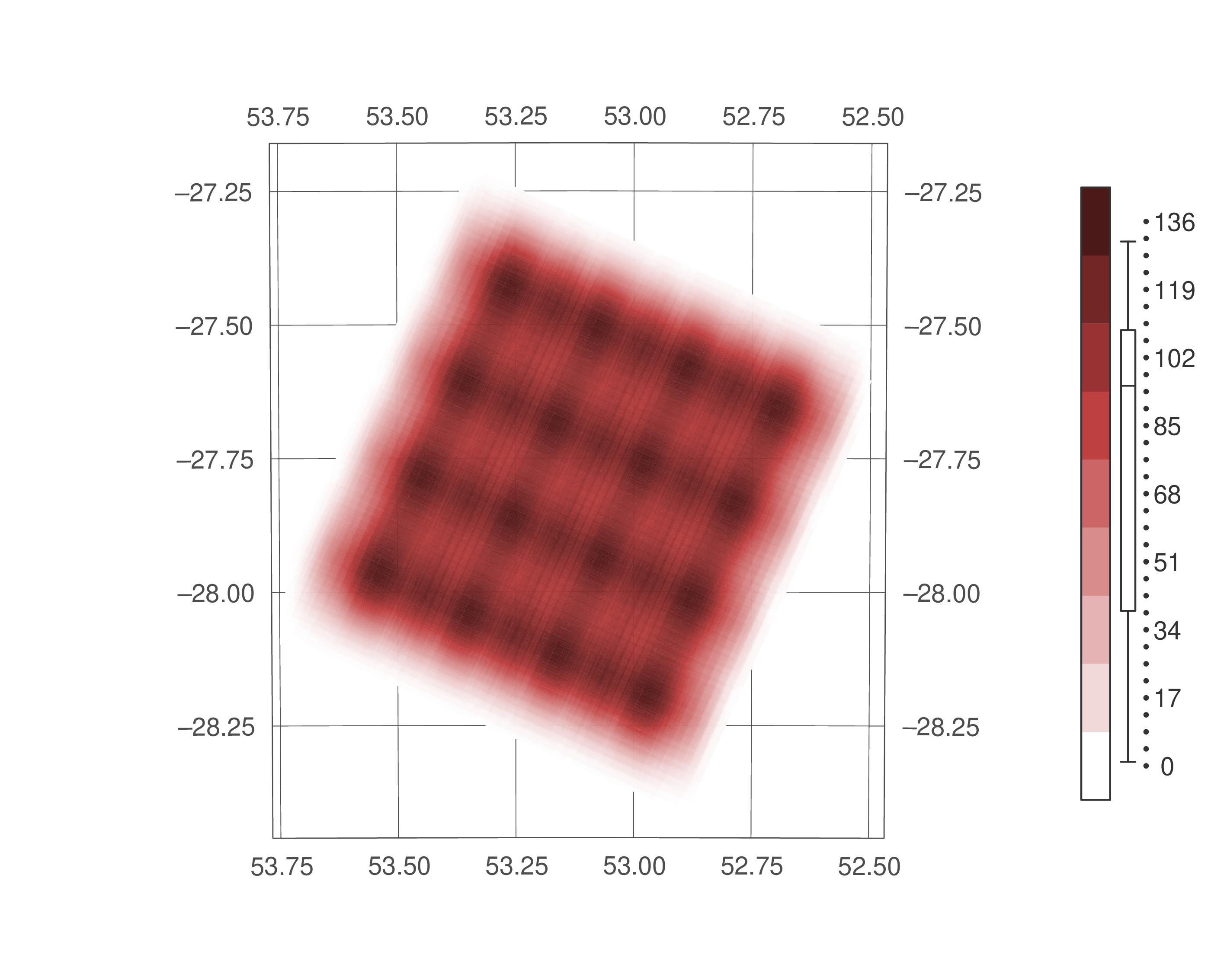}\\
 \noalign{\vskip 5pt}
 \hline
 \noalign{\vskip 5pt}
{\bf AEGIS} & 
    \includegraphics[width=0.4\linewidth]{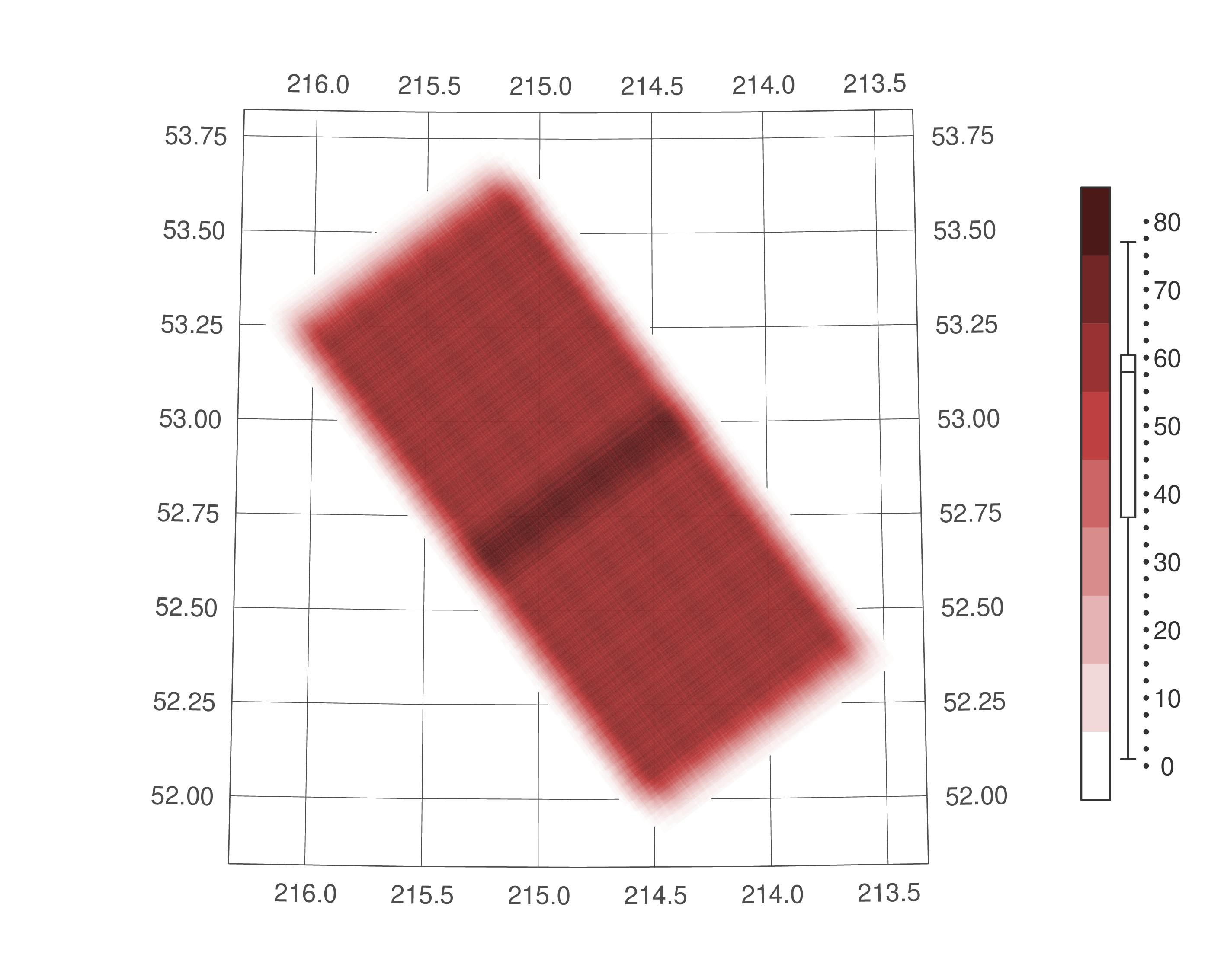}
    &  \includegraphics[width=0.4\linewidth] {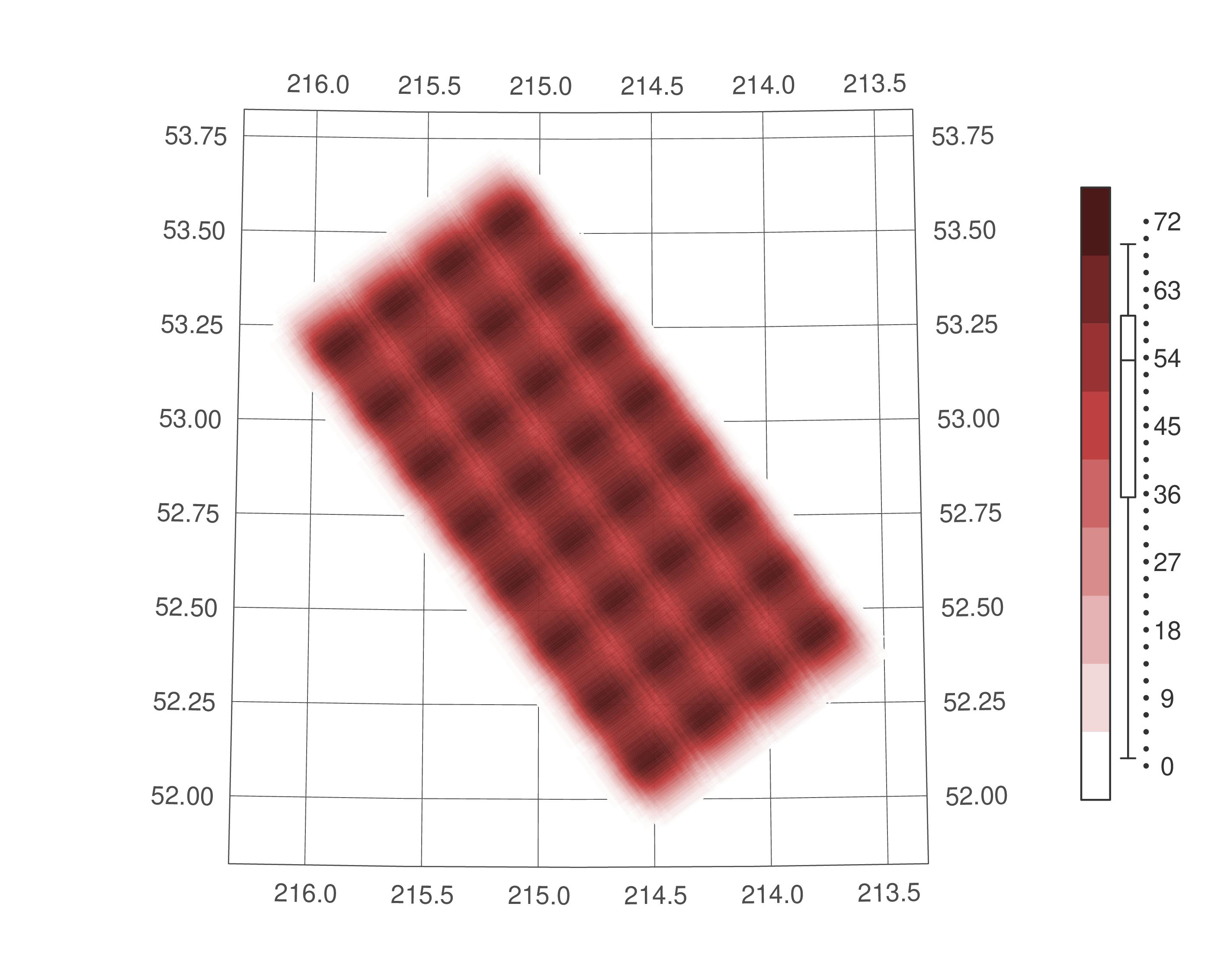}\\
 \noalign{\vskip 5pt}
 \hline
\end{tabular}
\caption{Same as Fig.~\ref{tab:appexpmapsedf} but for
the Euclid Auxiliary Fields. 
}
\label{tab:appexpmapseaf}
\end{figure*}


\begin{figure*}
\centering
\begin{tabular}{lcc}
 & {\bf VIS} & {\bf NISP}  \\
 \noalign{\vskip 5pt}
 \hline
 \noalign{\vskip 5pt}
 {\bf Self-calibration} & 
    \includegraphics[width=0.4\linewidth]{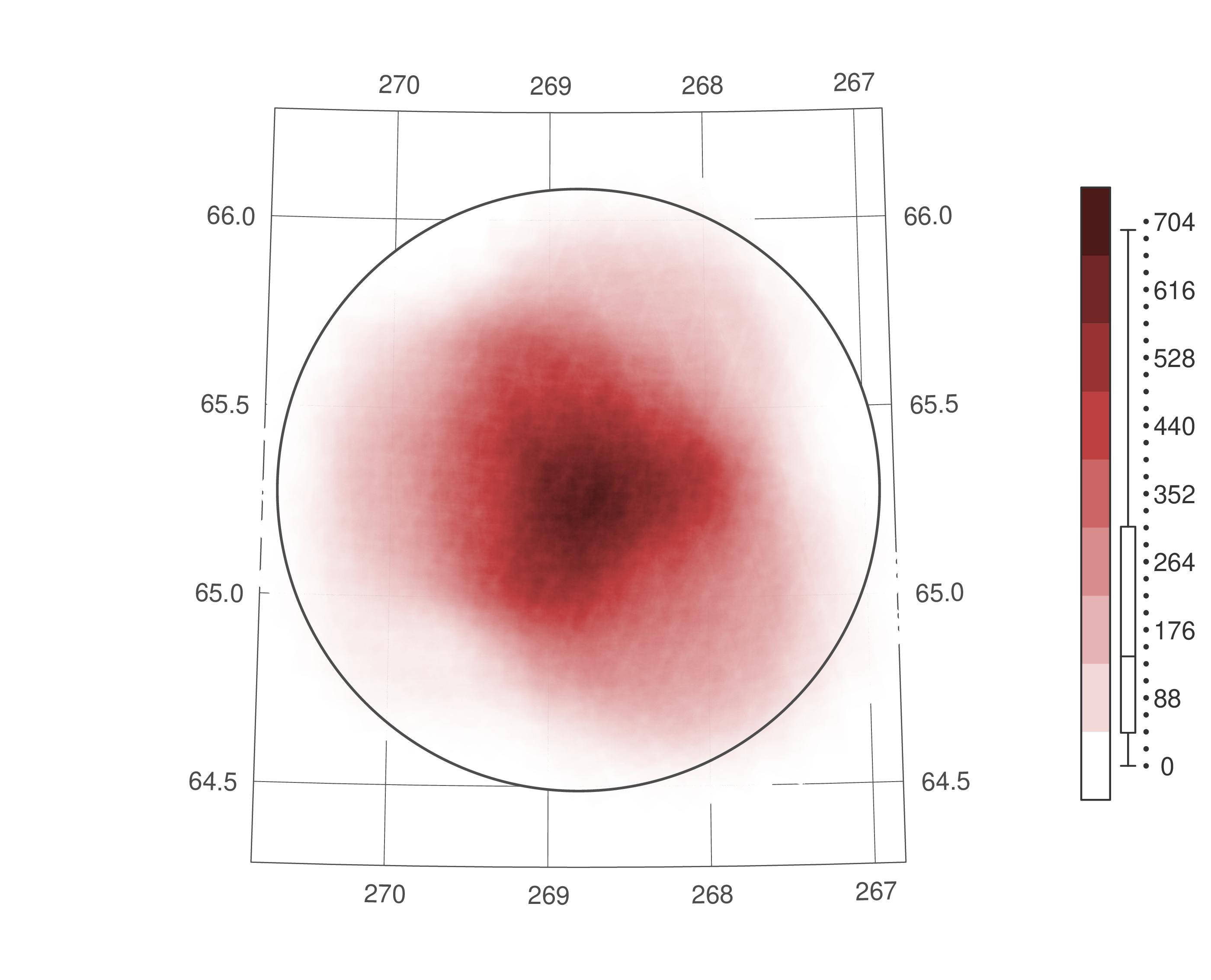}
    &  \includegraphics[width=0.4\linewidth] {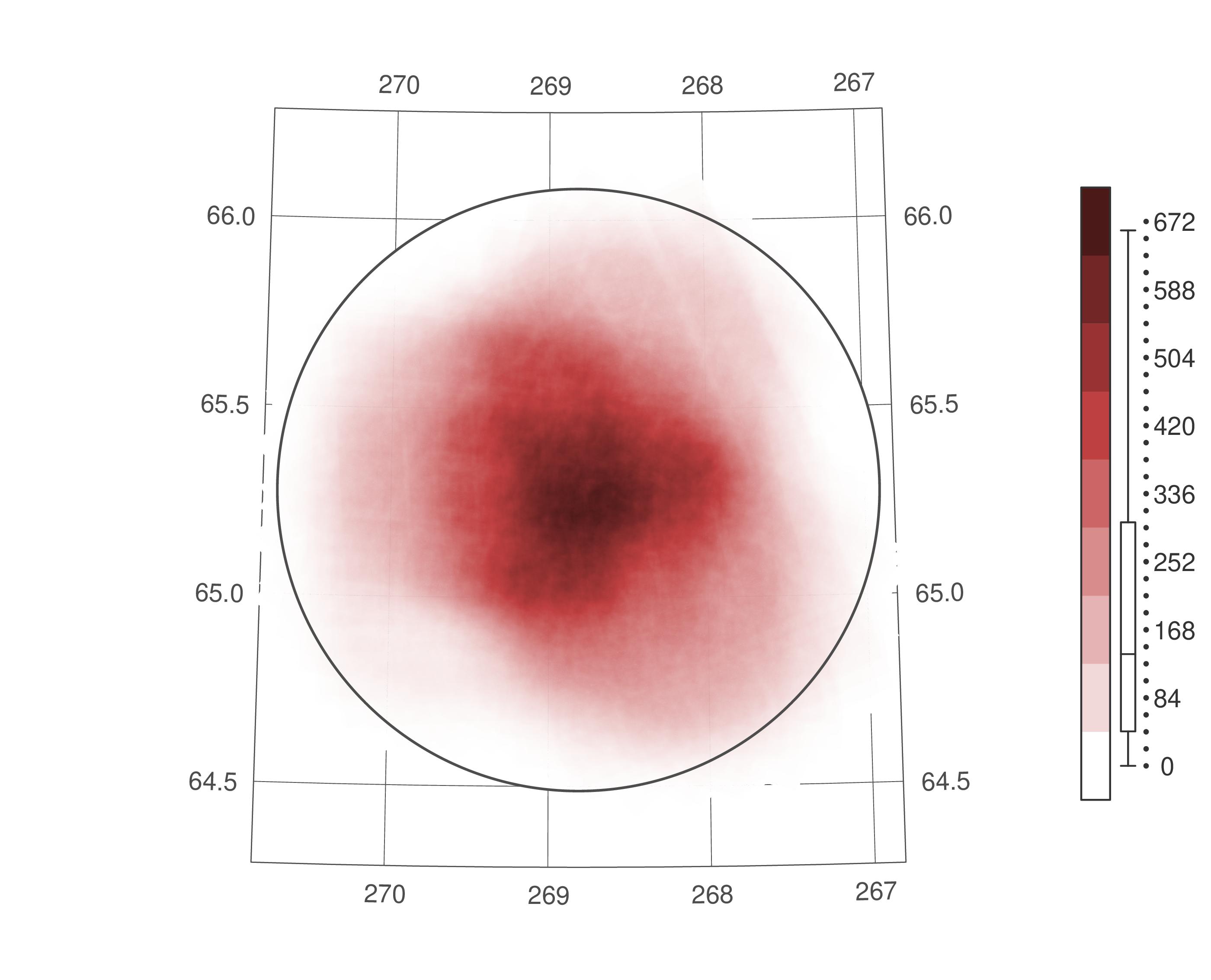}\\
\noalign{\vskip 5pt}
 \hline
 \noalign{\vskip 5pt}
 {\bf PN} & 
    \includegraphics[width=0.4\linewidth]{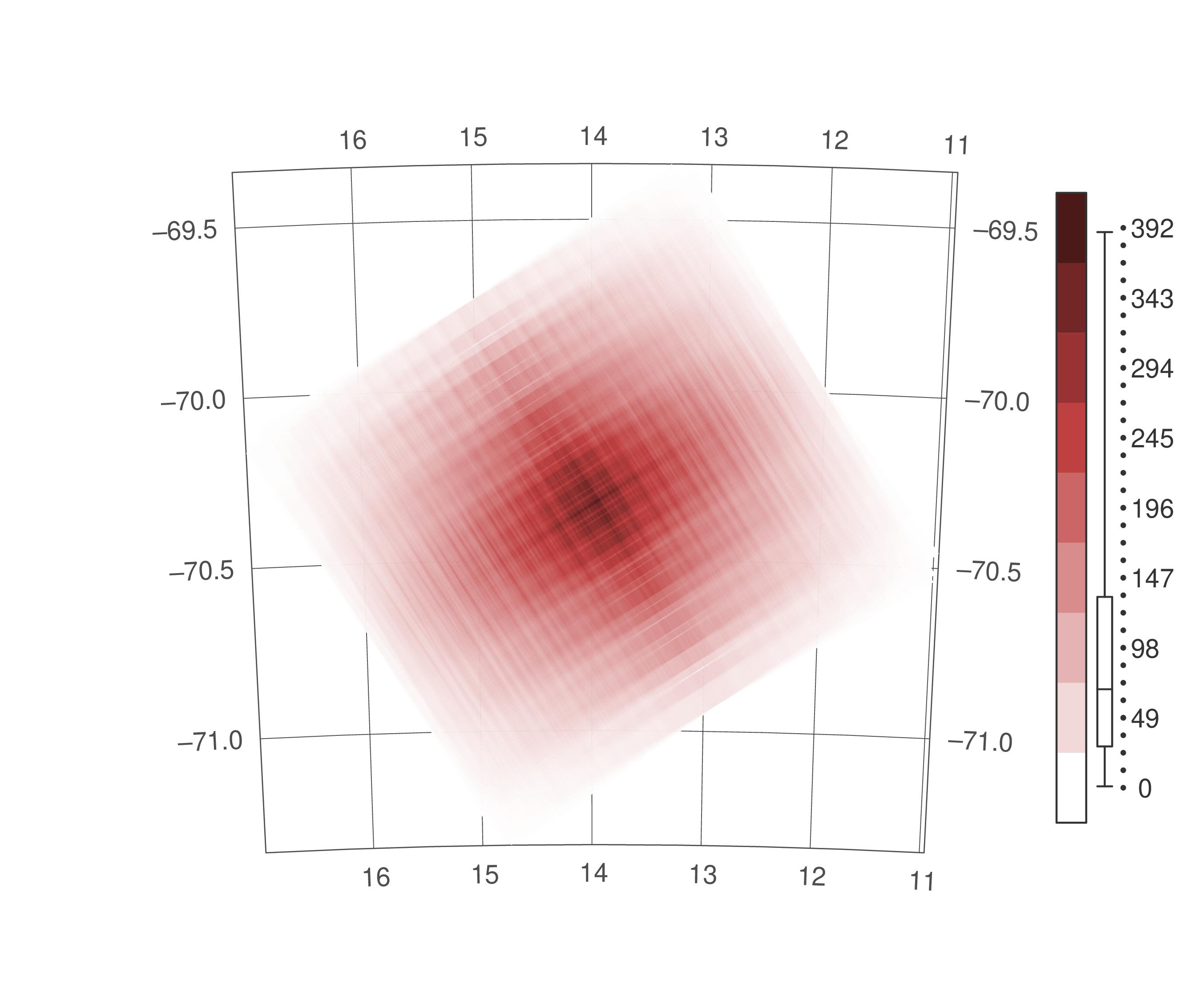}
    &  \includegraphics[width=0.4\linewidth] {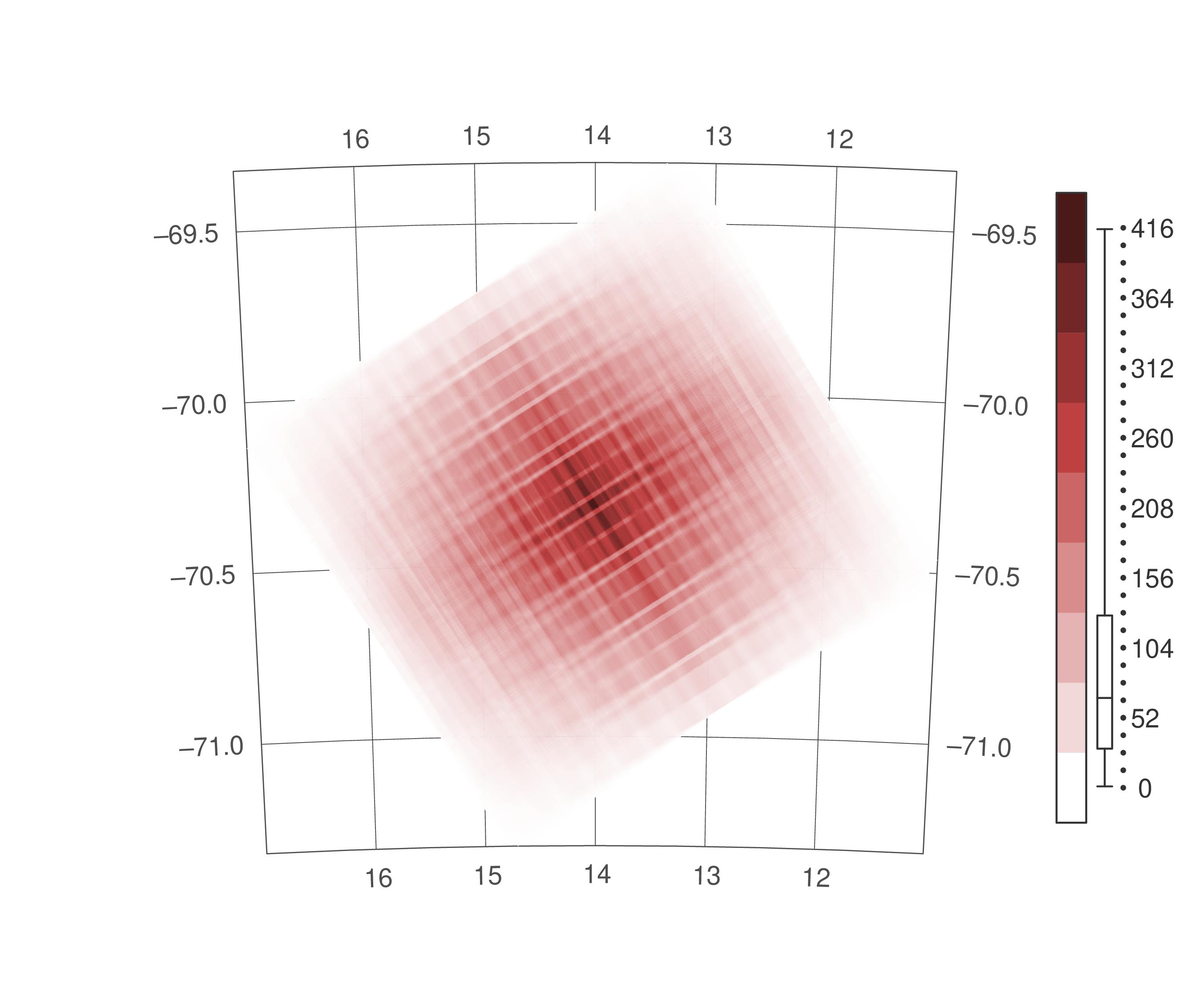}\\
\noalign{\vskip 5pt}
 \hline
 \noalign{\vskip 5pt}   
{\bf PSF} & 
    \includegraphics[width=0.4\linewidth]{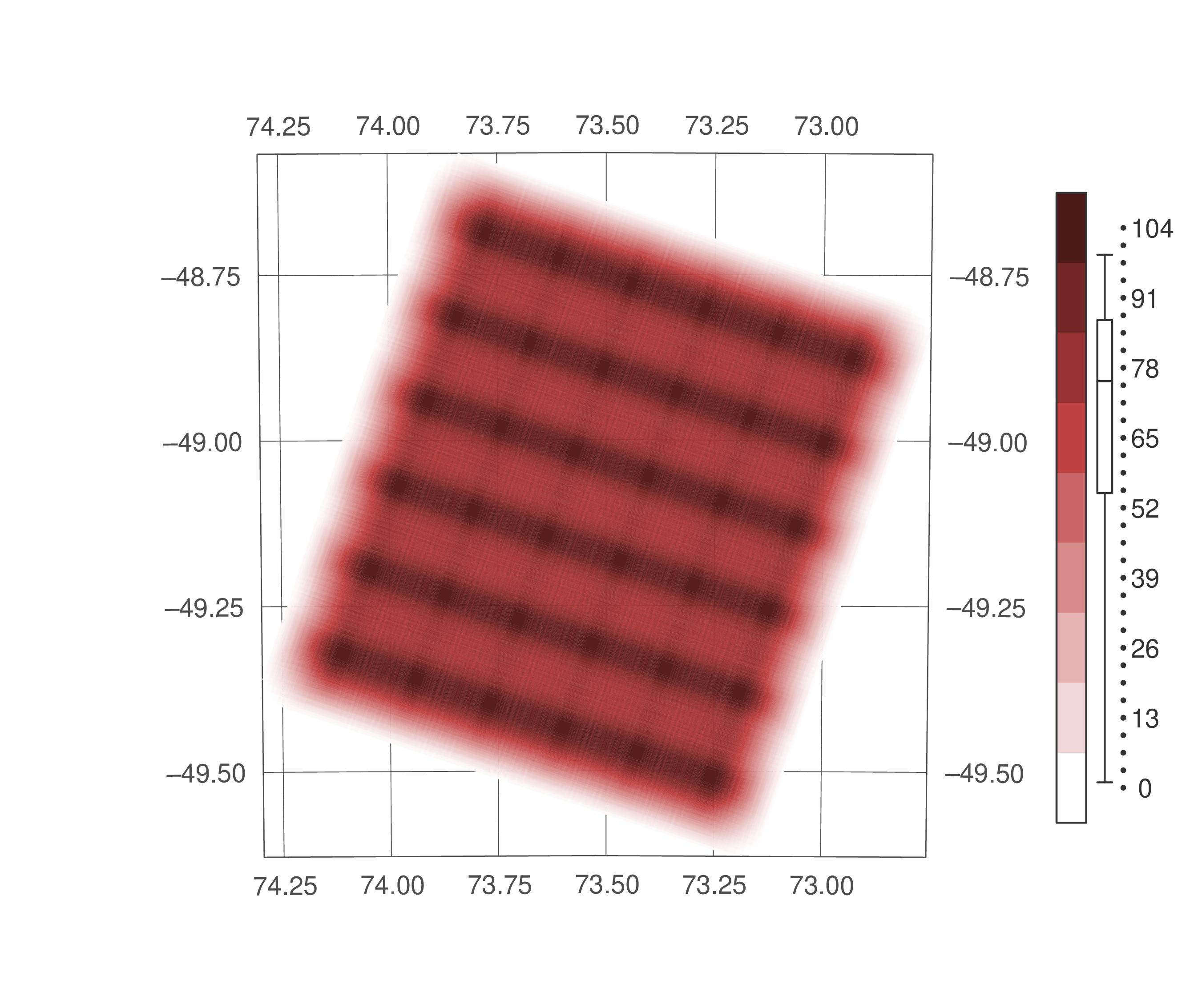}
    &  \\
\noalign{\vskip 5pt}
 \hline 
\end{tabular}
\caption{Same as Fig.~\ref{tab:appexpmapsedf} but for
the calibration fields. 
}
\label{tab:appexpmapsselfcal}
\end{figure*}


\begin{figure*}
\centering
\begin{tabular}{ccc}
 \includegraphics[width=0.33\linewidth]{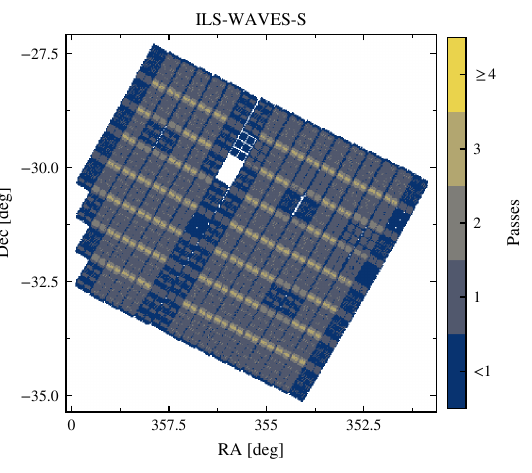}
    &  \includegraphics[width=0.33\linewidth]{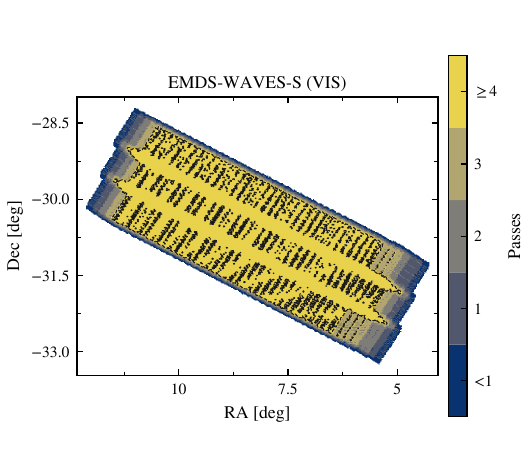} & \\
 \includegraphics[width=0.33\linewidth]{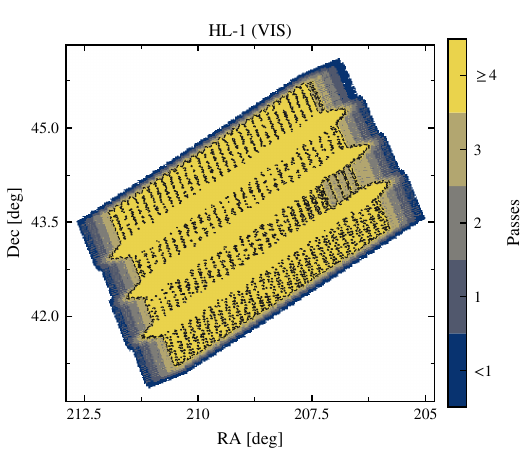} & 
    \includegraphics[width=0.33\linewidth]{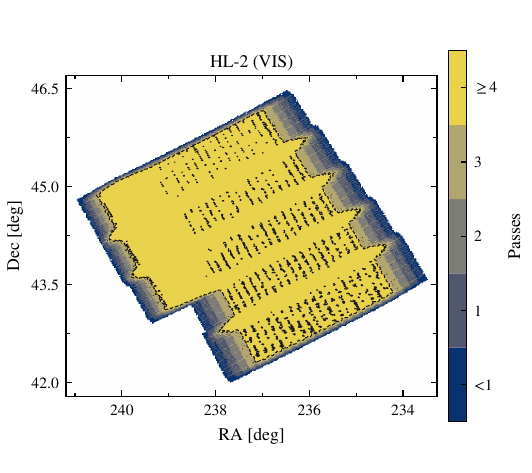}
    &  \\
    \includegraphics[width=0.33\linewidth]{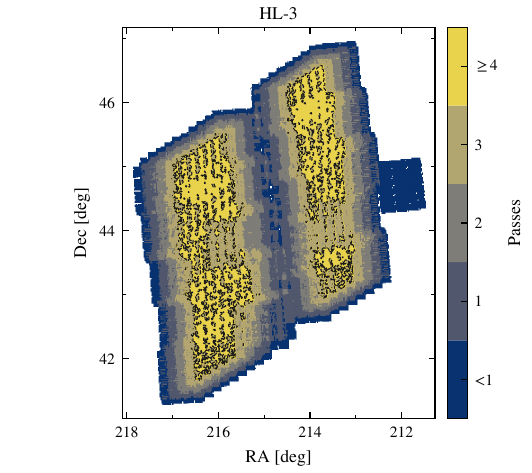}
    &  \includegraphics[width=0.3\linewidth] {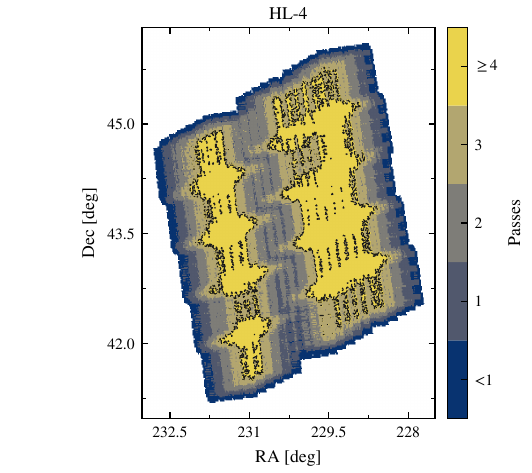} & \\
    \includegraphics[width=0.33\linewidth]{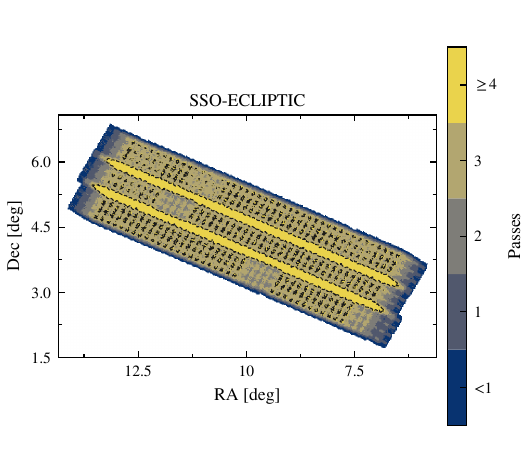}
    &&\\    
\end{tabular}
\caption{DR1 exposure maps of the PDC science fields. The depth in each pixel is given by the minimum number of 4-dither exposures between the VIS and NISP counts (for fields observed with ROS), or by the number of 4-dither VIS exposures (for fields only observed with VIS).}
\label{fig:appexpmappdc}
\end{figure*}

\clearpage

\section{\label{app:acronym}Acronyms and Glossary}

\begin{table*} [h!]
\centering
\caption{List of acronyms used in this paper with a  glossary of terms.}
\label{tab:acronyms}
\begin{tabular}{lll}
\hline\noalign{\vskip 2pt}
\multicolumn{1}{c}{Acronym} & \multicolumn{1}{c}{Definition} & \multicolumn{1}{c}{Glossary} \\
\hline
\noalign{\vskip 2pt}
\acs{AA} & $\alpha$ angle   & \Euclid's roll angle\\
AEGIS & AEGIS strip  &  EAF for colour gradient calibration\\
\acs{BGS} & Blue grism  & The shortest wavelength grism in NISP\\
\acs{CalF} & \acl{CalF}  & The repository of calibration specifications\\
\acs{CCB} & Configuration Control Board   &  A group that approves changes to a system\\
CDFS & Chandra Deep Field South  &  EAF for photometric redshift calibration, within EDF-F\\
COSMOS & Cosmos Evolution Survey  &  EAF for photo-$z$ and colour gradient calibrations, at the edge of the RoI\\
\acs{CPC} & Completeness and Purity calibration & Visits to EDFs with different orientations to decontaminate the spectra\\
\acs{DES} & \acl{DES}  &  Provides ground-based data in the southern sky\\
\acs{DR1} & Data Release 1  & The first large Euclid DR, including roughly $2000 \, {\rm deg}^2$ of EWS \\
\acs{EAF} & \acl{EAF}  & Well-known fields that \Euclid visits multiple times\\
EC & Euclid Consortium & 
Instruments building, scientific data processing, and exploitation \\
EDF-F & Euclid Deep Field Fornax   & The smallest field of the EDS\\
EDF-N & Euclid Deep Field North   & The most accessible field of the EDS, in the NEP\\
EDF-S & Euclid Deep Field South   & The largest field of the EDS\\
\acs{EDS} & \acl{EDS}   & The deep tier of the \Euclid surveys\\
\acs{EES} & \acl{EES}   & Survey of Solar System objects made in PDC thermalisation time\\
\acs{EGBS} & \acl{EGBS}   & A survey made expressly for Q2\\
\acs{EMDS} & \acl{EMDS}  & Tentative intermediate depth tier made in PDC thermalisation time\\
\acs{ERO} & \acl{ERO}  & The very first \Euclid data release \\
\acs{ESS} & Euclid Survey System   & SOC mission planning subsystem \\
ESSOT & ESS operational tool   &  ESS tool that generates an OSS from an RSD and also generates SSRs\\
\acs{EST} & \acl{EST}   & Team supporting the project scientist in  maximising the scientific return\\
\acs{EUDF} & \acl{EUDF}   & The deepest \Euclid field in preparation\\
\acs{EWS} & \acl{EWS}   & The main component of the Euclid survey\\
FGS & Fine-Guidance Sensor   & A dedicated instrument to find guide stars\\
FWA & Filter-wheel assembly   & Rotating device that holds multiple filters\\
GWA & Grism-wheel assembly & Rotating device that holds multiple grisms\\
\acs{ICR} & \acl{ICR}   & IOT input to specify low-level instrument commands on the OSS\\
\acs{ILS} & \acl{ILS}   & 
Two scientists external to the EC with selected rights to \Euclid data\\
\acs{IOT} & \acl{IOT}   & Two EC teams for operational support and monitoring of the instruments\\
LE1 & Level 1   &  Raw uncalibrated data, processed by SOC from raw spacecraft telemetry\\
LE2 & Level 2   & Calibrated and co-added images and spectra, processed by EC SGS PFs\\
\acs{MOC} & \acl{MOC}   & ESA, executes the operations and controls the spacecraft\\ 
\acs{NEP} & \acl{NEP} & Always visible to \Euclid, location of the EUDF\\
\acs{NISP} & NIR Spectrometer and Photometer   & Instrument, near-infrared wide-field imager and spectrometer\\
\acs{OSS} & Operational Sky Survey   &  Operational survey generated with ESSOT from an RSD and ICRs \\
\acs{PC} & Planning cycle   & The interval between two SOPS windows, duration of an SSR\\
\acs{PDC} & Phase-diversity calibration   & In-flight  calibration campaign for the wavefront model of the telescope\\
PF & Processing Function  &  Specialized EC SGS data processing pipelines\\
POR & Payload Operations Request   & Low-level operational product generated with SCS from OSS for MOC\\
\acs{PV} & Performance-Verification   & Early operations phase for in-flight characterisation of the instruments\\
Q1 & Quick Data Release 1 & DR not suited for cosmology with data from 1 pass on each EDF\\
Q2 & Quick Data Release 2 & DR not suited for cosmology with data from the EGBS\\
\acs{RGS} & Red grism   & The two longest wavelength grisms operating in NISP\\
\acs{RoI} & Region of Interest   & The extragalactic sky where the EWS is built\\
\acs{ROS} & Reference Observing Sequence   &  Observing sequence used in EWS, EDS, and EAFs  observations\\
\acs{RSD} & Reference Survey Definition   & Schedule of pointings and slews for science operations made by SOST\\
\acs{SAA} & \acl{SAA}   & \Euclid's pitch angle\\
\acs{SCS} & \acl{SCS}   & SOC subsystem that generates telecommands for a planning cycle\\
\acs{SGS} & Science Ground Segment (EC)  & Produces the science-ready calibrated images and source catalogues \\
\acs{SOC} & \acl{SOC}   & ESA, mission planning, monitoring, LE1 processing, and archive\\
SOPS & Special Operations   & Periodic time window allocated  spacecraft and orbit mainenance \\
\acs{SOST} & Survey Operations Support Team   & The survey scientist's EC team that produces the RSDs for SOC\\
\acs{SSR} & \acl{SSR}   & High-level product generated from an OSS with ESSOT for MOC\\
\acs{UNIONS} & UV NIR Optical Northern Survey  &  Provides ground-based data in the northern sky, from three observatories\\
\acs{VIS} & \acl{VIS}   & Instrument, wide-field optical imager\\

\hline
\end{tabular}
\end{table*}

\end{appendix}

\label{LastPage}
\end{document}